\documentclass[11pt,graphicx,amsmath]{article}
\usepackage{amsmath}
\usepackage{amssymb}
\usepackage{graphicx}
\usepackage{subfigure}
\usepackage{CJK}
\usepackage{bm}
\usepackage{xcolor}
\usepackage{ulem}
\usepackage[numbers,sort&compress]{natbib}
\usepackage{longtable}
\usepackage{seqsplit}
\numberwithin{equation}{section}

\def\be{\begin{equation}}
\def\ee{\end{equation}}
\def\ba{\begin{eqnarray}}
\def\ea{\end{eqnarray}}
\def\nn{\nonumber}
\def\bt{\bm{\theta}}
\def\cf{\mathcal{F}}
\def\ch{\mathcal{H}}
\def\bl#1\el{\begin{align}#1\end{align}}
\def\l{\left}
\def\r{\right}

\title{Second-order cosmological perturbations produced by scalar-scalar coupling during inflation stage}

\author{\small
   Bo Wang  \thanks{ymwangbo@ustc.edu.cn}
     \   \ and \ Yang  Zhang  \thanks{yzh@ustc.edu.cn}
           \\
 \small   CAS Key Laboratory for Researches in Galaxies and Cosmology/Department of Astronomy,\\
 \small   School of Astronomy and
Space Science, 
University of Science and Technology of China,   Hefei, Anhui, 230026,  China   }

 \date{}

\begin{document}
\maketitle

\def\bl#1\el{\begin{align}#1\end{align}}
\def\gsim{\;\rlap{\lower 2.5pt  \hbox{$\sim$}}\raise 1.5pt\hbox{$>$}\;}
\def\lsim{\;\rlap{\lower 2.5pt  \hbox{$\sim$}}\raise 1.5pt\hbox{$<$}\;}
\def\edth{\;\raise1.0pt\hbox{$'$}\hskip-6pt\partial\;}
\def\baredth{\;\overline{\raise1.0pt\hbox{$'$}\hskip-6pt \partial}\;}
\def\be{\begin{equation}}
\def\ee{\end{equation}}
\def\ba{\begin{eqnarray}}
\def\ea{\end{eqnarray}}
\def\nn{\nonumber}
\def\bt{\bm{\theta}}
\def\cf{\mathcal{F}}
\def\ch{\mathcal{H}}
\def\l{\left}
\def\r{\right}

\baselineskip=19truept


\Large

\allowdisplaybreaks

\begin{center}
\text{\large\bf Abstract}\\
\end{center}

We study the perturbations  up to the 2nd-order
for a power-law inflation driven by a scalar field
in synchronous coordinates.
We present the 1st-order solutions,
and analytically solve the 2nd-order perturbed Einstein equation and scalar field equation,
give the 2nd-order solutions for
all the scalar, vector, and tensor metric perturbations,
 as well as the perturbed scalar field.
During inflation, the 1st-order tensor perturbation is a wave and
is decoupled from other perturbations,
the scalar metric perturbation and the perturbed scalar field
are coupled waves,  propagating at the speed of light,
differing  from those in the dust and relativistic fluid models.
The 1st-order vector perturbation is not wave and just decreases during inflation.
The 2nd-order perturbed Einstein equation
is similar in structure to the 1st-order one,
but various  products of the 1st-order perturbations
occur as  the effective source,
among which the scalar-scalar coupling is considered in this paper.
The solutions of all the 2nd-order perturbations
consist of a homogeneous part similar to the 1st-order solutions,
and an inhomogeneous part
in a form of integrations of the effective source.
The 2nd-order vector perturbation is also a wave
since the effective source is composed of the 1st-order waves.
We perform the residual gauge transformations between synchronous coordinates
up to the 2nd-order,
and identify the 1st-order and 2nd-order gauge modes.

\section{Introduction}
\label{sec:introduction}

The spacetime  background of the expanding Universe is homogeneous and isotropic,
upon which there are perturbations
in both the spacetime metric and the cosmic energy-matter component.
The linear order of the cosmological perturbation
has been extensively studied \cite{Lifshitz1946,LifshitzKhalatnikov1963,
PressVishniac1980,Bardeen1980,KodamSasaki1984,
   BrandenbergerKahnPress1983,Grishchuk1994,Peebles1980},
which can account for
the cosmic microwave background radiation (CMB)
\cite{BaskoPolnarev1984,Polnarev1985, MaBertschinger1995,Bertschinger,
ZaldarriagaHarari1995,Kosowsky1996, Kamionkowski1997,KeatingTimbie1998,
ZhaoZhang2006,CaiZhang2012,Baskaran,Polnarevmiller2008}
and relic gravitational wave (RGW)
\cite{Grishchuk,Starobinsky,Allen1988,
zhangyang05,ZhangWang2016,ZhangWang2018b,
CaiLinWang2021,FordParker1977GW,Rubakov,
Fabbri,AbbottWise1984,Giovannini,Tashiro,Morais2014,BartoloBertaccaMatarrese2020},
and  to some extent the large-scale structure \cite{ Peebles1980}.
With the increasing accuracy of cosmological observations,
nonlinear perturbations
 are becoming an important part of
the cosmological studies
\cite{PyneCarroll1996,AcquavivaBartoloMatarrese2003,
JeongKomatsu2006,YangZhang2007,AnandaClarksonWands2007,Baumann2007,
Bartolo2010,Matarrese2007,Pietroni2008,Matsubara2008,Hwang2017GW}.

Ref.~\cite{PeresRosen1959} used  the 2nd-order perturbed Einstein equation
to study the gravitational radiation problem.
Ref.~\cite{Tomita1967} studied the 2nd-order perturbations
in synchronous coordinates,
and analyzed the 2nd-order density contrast for some special cases.
Refs.~\cite{MatarresePantanoSa'ez1994,Russ1996,Salopek}
studied gravitational instability in the 2nd-order perturbations in association
with large-scale structure.
Refs.~\cite{Bruni97,Matarrese98} studied the equations
of 2nd-order scalar and tensor perturbations for the scalar-scalar coupling
in Einstein-de Sitter model filled with a pressureless dust.
Refs.~\cite{MollerachHarariMatarrese2004,Lu2008}
studied the 2nd-order vector due to scalar-scalar coupling in Poisson gauge.
Ref.~\cite{Baumann2012} studied the cosmological effects due to small-scale nonlinearity.
{
In the Lambda cold dark matter ($\Lambda$CDM) framework,
Ref.~\cite{VillaRampf2016} calculated 2nd-order
scalar, vector and tensor perturbations as well as primordial non-Gaussianity in the synchronous-comoving
gauge, the Poisson gauge and the total matter gauge respectively;
}
and Ref.~\cite{Brilenkov&Eingorm2017} investigated 2nd-order
scalar, vector perturbations in the Poisson gauge and gave discussions of the cosmological backreaction from nonlinear effects.
The 2nd-order density perturbation was studied
with the squared RGW as the source
in the Arnowitt-Deser-Misner (ADM) framework
\cite{NohHwang2004}.
The 2nd-order perturbation equations were also
explored in a non-flat universe \cite{Noh2014,Sikora2023}.
Ref.~\cite{HwangNohPark2016} studied the non-linear equations of  perturbation
in the presence of multiple components of fluids and minimally coupled scalar fields.
Gauge-invariant 2nd-order perturbations
were studied in Refs.~\cite{Nakamura2003,MalikWands2004,
Domenech&Sasaki2017,AcquavivaBartoloMatarrese2003,Vernizzi2005,UgglaWainwright2019}.
Refs.~\cite{WangZhang2017,ZhangQinWang2017}
studied the 2nd-order perturbation in the matter-dominated (MD) stage,
including  the scalar-scalar, scalar-tensor, tensor-tensor couplings,
and  derived all the analytical 2nd-order solutions of
the  scalar, vector, tensor metric perturbations, and 2nd-order density contrast.
Refs.~\cite{WangZhang2018,WangZhang2019} studied the 2nd-order perturbation
in the radiation-dominated (RD) stage driven by
a relativistic perfect fluid,
and obtained the 2nd-order solutions of
the scalar, vector, tensor metric perturbations
and of the  density, pressure, velocity perturbations.
Ref.~\cite{ChoGongOh2020} investigated
the second-order effective energy-momentum tensor
due to the scalar metric perturbation
 with a perfect fluid in several gauges.

The inflation stage plays a special role in the studies of cosmology.
The cosmological perturbations are originated
during this stage \cite{Guth1981,Starobinsky1980,Linde1982,
LythRiotto1999,Riotto2002,Lucchin1985,JHwang1993},
and leave  the observational effects,
such as CMB anisotropy, non-Gaussianity of primordial perturbation
\cite{Bartolo2010,AcquavivaBartoloMatarrese2003,nongaussiancite},
primordial black hole formation \cite{primordialBH},
etc.
Ref.~\cite{AcquavivaBartoloMatarrese2003}
solved the 2nd-order  comoving curvature perturbation
in the longitudinal gauge and discussed the non-Gaussianity during inflation.
Rigopoulos {\it et al} \cite{Rigopoulos2004}
studied the second-order isocurvature and adiabatic perturbations
in long wavelength for multifield inflationary models.
Ref.~\cite{ChoGongOh2022} studied the second-order effective energy-momentum tensor
of the scalar metric perturbation during inflation.

In this paper, we shall study the 2nd-order perturbations
during inflation driven by a scalar field in synchronous coordinates.
At the 2nd-order level
there are six types of products of the linear perturbations
such as scalar-scalar, scalar-vector, scalar-tensor and so on.
In this paper we study the case of scalar-scalar couplings.
We shall derive the solutions of the scalar, vector, and tensor metric perturbations,
and the perturbed scalar field, up to the second order.
We shall also perform the residual gauge transformations
between synchronous coordinates and identify the gauge modes up to 2nd order.

In Sec. 2,
we give a basic setup for the metric perturbations
and the scalar field driving the power-law inflation
 in synchronous coordinates, and present the analytical solutions of all the 1st-order perturbations.

In Sec. 3,
we decompose the 2nd-order perturbed Einstein equation
into the separate equations of the scalar, vector, tensor metric perturbations,
and also give the 2nd-order perturbed equation of the scalar field.

In Sec. 4,
we derive the solutions of the 2nd-order metric perturbations
and perturbed scalar field.

In Sec. 5,  we perform the synchronous-to-synchronous transformation,
and identify the residual gauge modes in the 2nd-order solutions.

Sec. 6  gives the  conclusions and discussions.

Appendix \ref{sec:perturbedGT} gives
the  perturbed stress tensor  of the scalar field,
and its covariant conservation,  up to the second order.
Appendix \ref{sec:gauge_transform} gives  the  synchronous-to-synchronous
gauge transformations of the metric perturbations and
the scalar field up to 2nd order.
In Appendix \ref{sec:appendix_1stSolution},
we derive the solutions of the 1st-order perturbations in detail.
Appendix \ref{sec:appendix_gaugeInvariant} gives the primordial spectra of four gauge invariant scalar perturbations
and discuss theirs cosmological implications.

We use a unit with the speed of light $c=1$.

\section{ Basic setups and 1st-order perturbations}

A flat Robertson-Walker metric in synchronous coordinates is given by
\be \label{18q1}
ds^2=g_{\mu\nu}   dx^{\mu}dx^{\nu}
 = a^2(\tau)[-d\tau^2
        +\gamma_{ij}   dx^idx^j] ,
\ee
where $\tau$ is the  conformal time,
 the indices  $\mu,\nu = 0, 1, 2, 3$ and  $i,j= 1, 2, 3$.
We use the same notations as in
Refs.~\cite {Matarrese98,WangZhang2017,ZhangQinWang2017,WangZhang2018,WangZhang2019}.
As an advantage of the synchronous coordinates,
 $g_{00}$ and  $g_{0i}$ have no perturbations,  and
all possible type of metric perturbations are included in $\gamma_{ij}$,
which is expanded up to 2nd-order as
\be\label{eq1}
\gamma_{ij} =\delta_{ij} + \gamma_{ij}^{(1)} + \frac{1}{2} \gamma_{ij}^{(2)},
\ee
with $\gamma_{ij}^{(1)}$ and  $\gamma_{ij}^{(2)}$
being  the 1st- and 2nd-order metric perturbation, respectively.
Writing  $g^{ij}=a^{-2}\gamma^{ij}$,
one has
\be\label{metricUp2}
\gamma^{ij}=\delta^{ij} -\gamma^{(1)ij}
-\frac{1}{2}\gamma^{(2)ij}+\gamma^{(1)ik}\gamma^{(1)j}_{k} .
\ee
The determinant of the metric is expanded to 2nd order as
\be\label{determinantMetric}
g\equiv{\rm det} (g_{\mu\nu})= -a^8 \Big(
1+\gamma_{k}^{(1) k}+\frac{1}{2} \gamma_{k}^{(2) k}+\frac{1}{2}
\gamma_{k}^{(1) k} \gamma_{l}^{(1) l}
-\frac{1}{2} \gamma^{(1) kl} \gamma_{kl}^{(1)}
\Big)
,
\ee
which will be also used in this paper.
The 1st- and 2nd-order metric perturbation
can be further written as
\be  \label{gqamma1}
\gamma^{(A)}_{ij} = -2\phi^{(A)}\delta_{ij} +\chi_{ij}^{(A)},
~~\text{with}~~ A =1,2,
\ee
where  $\phi^{(A)}$ is the trace part of  scalar perturbation,
and $\chi_{ij}^{(A)}$ is  traceless and  decomposed as
\be \label{xqiij1}
\chi_{ij}^{(A)} =D_{ij}\chi^{||(A)}
               +\chi^{\perp(A)}_{ij}
               +\chi^{\top(A)}_{ij},
~~\text{with}~~ A=1,2,
\ee
where $D_{ij} \equiv  \partial_i\partial_j-\frac{1}{3}\delta_{ij}\nabla^2 $
and $\chi^{||(A)}$ is a scalar function,
and  $D_{ij}\chi^{||(A)}$ is the traceless part of the scalar perturbation.
The vector metric perturbation  $\chi^{\perp(A) }_{ij}$
satisfies a condition
$\partial^i\partial^j  \chi^{\perp(A) }_{ij}=0$
and can be written as
\be\label{chiVec0}
\chi^{\perp(A) }_{ij}= \partial_i B^{(A)}_j+\partial_j B^{(A)}_i,
\,\,\,\,\,
\text{with}~~  \partial^i B^{(A)}_i =0,
~~ A=1,2 ,
\ee
where $B^{(A)}_i$ is a curl  vector with two independent modes.
The tensor metric perturbation  $\chi^{\top(A)}_{ij}$
satisfies the traceless and transverse  condition:
$\chi^{\top(A)i}\, _i=0$, $\partial^i\chi^{\top(A)}_{ij}=0$,
having two independent modes.
(In the notation of  Ref.~\cite{Mukhanov1992},
 $\phi=0$, $B=0$, $E=\frac12 \chi^{(1)||}$,
 $\psi=  \phi^{(1)} + \frac{1}{6}\nabla^2 \chi^{(1)||}$.)

We consider a scalar field $\varphi$
driving the inflationary expansion \cite{Weinberg2008}
with the  Lagrangian density
\[
L= \sqrt{-g} \big( - \frac12  g^{\mu\nu}\varphi_{, \, \mu} \varphi_{, \, \nu}
    - V(\varphi)   \big),
\]
where $V(\varphi)$ is the potential of the  scalar field.
The field equation is
\be \label{fequvarphi}
\Box \varphi -  V(\varphi)_{,\, \varphi} =0,
\ee
where $\Box= \frac{1}{\sqrt{-g}}\partial_\mu (\sqrt{-g} g^{\mu\nu} \partial_\nu \, \,  )$.
The scalar field  can be expanded as \cite{AcquavivaBartoloMatarrese2003}
\be \label{expinflaton}
\varphi(\tau, {\bf x})
= \varphi^{(0)}(\tau) +\varphi^{(1)}(\tau, {\bf x})
        +\frac{1}{2}\varphi^{(2)}(\tau, {\bf x}),
\ee
where $\varphi^{(0)}$ is the background field,
$\varphi^{(1)}$ and $\varphi^{(2)}$ are
the 1st- and 2nd-order perturbations.
The potential   can be expanded   as
\be\label{Vexpand}
V(\varphi)=V(\varphi^{(0)})
+\varphi^{(1)}  V(\varphi^{(0)})_{,\, \varphi}
+\frac{1}{2}\varphi^{(1)}\varphi^{(1)}
 V(\varphi^{(0)})_{,\, \varphi \varphi }
+\frac{1}{2}\varphi^{(2)}V(\varphi^{(0)})_{,\, \varphi }
 \, .
\ee

The  Einstein equation is expanded up to 2nd-order of perturbations
\be\label{pertEinstein}
G^{(A)}_{\mu\nu}
\equiv R^{(A)}_{\mu\nu}-\frac{1}{2} \big[g_{\mu\nu}R\big]^{(A)}
=8\pi GT^{(A)}_{\mu\nu},
~~\text{with}~~ A= 0,1,2 ,
\ee
where $G^{(A)}_{\mu\nu}$ is the $A$th order perturbed Einstein tensor
(see Ref.~\cite{WangZhang2018} for the expressions),
and $T^{(A)}_{\mu\nu}$ is  the stress tensor,
listed  in Appendix \ref{sec:perturbedGT}.
For each  order,
the (00) component  of  (\ref{pertEinstein}) is the energy constraint,
 $(0i)$ components are the momentum constraints,
and  $(ij)$ components are the evolution equations.
The stress tensor is required to satisfy
the   covariant conservation
\be\label{covcons}
[T^{ \mu\nu}\,_{; \, \nu}]^{(A)}=0,
~~\text{with}~~ A= 0,1,2 .
\ee
As we have checked, to each order
the covariant conservation of energy
is equivalent to the field equation \eqref{fequvarphi}.
The dynamics of the  system is
determined by (\ref{pertEinstein}) and (\ref{covcons}).

The 0th-order Einstein equation gives the Friedmann equations
\be\label{Ein0th001}
 (\frac{a'}{a} )^2
=\frac{8\pi G}{3} a^2 \Big[
\frac{1}{2a^{2}} (\varphi^{(0)'} )^{2}
+V(\varphi^{(0)})
\Big],
\ee
\be\label{Ein0thij1}
-2\frac{a''}{a}
+  (\frac{a'}{a}  )^2
= 8\pi G   a^2  \Big[
 \frac{1}{2 a^{2}} (\varphi^{(0)'})^{2} - V(\varphi^{(0)})
 \Big].
\ee
with $'=d/d\tau$.
(\ref{Ein0th001}) and  (\ref{Ein0thij1}) lead to
$4\pi G \varphi^{(0)' 2} =-a H'$ with $H(\tau)\equiv a'/a^2$,
and  the background potential
\be\label{Vphi0}
V(\varphi^{(0)})=
 \frac{1}{8\pi G}\frac{1}{a^2}\Big[
 (\frac{a'}{a})^2
+\frac{a''}{a} \Big]
.
\ee
The 0th order  energy conservation is
\be\label{EnergyCons0th}
\varphi^{(0)''}
+\frac{2a'}{a}\varphi^{(0)'}
+a^2V(\varphi^{(0)})_{,\varphi}
 =0 .
\ee
In this paper we consider the power-law inflation with the scale factor
\cite{Grishchuk,zhangyang05,ZhangWang2018b}
\be\label{inflationscale}
a(\tau)=l_0|\tau|^{1+\beta}, ~~~ -\infty < \tau < \tau_1,
\ee
where $l_0$ and $\beta$ are two parameters of the model,
and $\tau_1$ is the ending time of inflation.
The expansion rate is $H=a'/a^2= -(\beta+1)/(l_0 |\tau|^{\beta+2})$,
and  $H= 1/l_0$ for the model $\beta = -2$,
which   corresponds to  de Sitter inflation.
The models $\beta >  -2$
would lead to a blue-tilted-up primordial spectrum,
disfavored by CMB observational data  \cite{Planck2018}
which give  the spectral index $n_s =0.961\sim0.969$
(corresponding to $\beta=-2.0195\sim-2.0155$).
So the models $\beta \lesssim  -2$ are more favored by observations.
The power-law inflation  \eqref{inflationscale} corresponds to
\be\label{phi0tau}
\varphi^{(0)}(\tau)
= M_{\rm Pl}\sqrt{2(\beta +1) (\beta +2)} \ln(-\tau)+\varphi^{(0)}_0,
\ee
\be\label{Vphi01}
V(\varphi^{(0)})=
V_0\exp\Big[-
\frac{2(\beta +2)}{\sqrt{2(\beta +1) (\beta +2) }}
\frac{\varphi^{(0)}}{M_{\rm Pl}}\Big]
,
\ee
with
\be
V_0\equiv
M_{\rm Pl}^2
\frac{(\beta +1) (2 \beta +1)}{l_0^2}
\exp\Big[
\frac{2(\beta +2)}{\sqrt{2(\beta +1) (\beta +2) }}
\frac{\varphi^{(0)}_0}{M_{\rm Pl}}\Big],
\ee
where $M_{\rm Pl}\equiv 1/\sqrt{8\pi G}$ is the Planck mass,
$\varphi^{(0)}_0$ is a constant to be determined by the initial condition.

The 1st-order perturbations have been   studied in literature,
and mostly expressed
in the Poisson coordinates with off-diagonal metric perturbations.
To study the 2nd-order perturbations  in synchronous coordinates,
we need  the 1st-order perturbations in synchronous coordinates.
Here we list the major results.
(Actually the treatment of the 1st-order perturbation
in synchronous coordinates is nontrivial  and more involved,
and the detailed derivation is given in Appendix \ref{sec:appendix_1stSolution}.)

From the 1st-order perturbed Einstein equation (\ref{Ein1st00}), (\ref{momentconstr1RDinf}) and (\ref{evoEq1stRD3}), and the 1st-order perturbed field equation (\ref{Enconsv1st}), one obtains the 1st-order solutions without residual gauge modes.
The 1st-order vector is vanishing,
\be\label{novect}
\chi^{\perp(1)}_{ij}=0 ,
\ee
like in the dust and fluid models.

The 1st-order tensor solution, i.e., the linear RGW propagating at the speed of light, is written as
\be  \label{Fourier}
\chi^{\top(1)}_{ij}  ( {\bf x},\tau)= \frac{1}{(2\pi)^{3/2}}
\int d^3k   e^{i \,\bf{k}\cdot\bf{x}}
\sum_{s={+,\times}} {\mathop \epsilon
\limits^s}_{ij}(k) ~ {\mathop h\limits^s}_k(\tau),
\ee
where ${\mathop \epsilon  \limits^s}_{ij}(k) $
are the polarization tensors,
${\mathop h\limits^s}_k(\tau)$ with  $s= {+,\times}$
are two modes of RGW
and can be assumed to be statistically equivalent,
so that the superscript $s$ can be dropped,
\bl \label{GWmode}
h_k(\tau ) = &
\frac{1}{a(\tau )}\sqrt{\frac{\pi }{2}} \sqrt{\frac{-\tau }{2}}
\left[b_1 H_{\beta +\frac{1}{2}}^{(1)}(-k \tau )
+b_2 H_{\beta +\frac{1}{2}}^{(2)}(-k \tau )\right]
,
\el
where  $H^{(1)}_{\beta +\frac{1}{2}}$, $H^{(2)}_{\beta +\frac{1}{2}}$
are the Hankel functions,
and  the constant coefficients $b_1$ and $b_2$
can be fixed by  the initial condition.
At high $k$,  $H_{\beta +\frac{1}{2}}^{(1)}(-k\tau ) \simeq
\sqrt{\frac{2}{\pi }} \frac{1}{\sqrt{- k\tau }}
 e^{ -i k\tau -\frac{i \pi (\beta+1) }{2}}$ and
\be\label{hikt}
h_k(\tau ) \sim \frac{1}{a(\tau )}
\big[ b_1 \frac{1}{ \sqrt{2k} } e^{ -i k\tau -\frac{i \pi  (\beta+1)  }{2}}
+  b_2 \frac{1}{ \sqrt{2k} } e^{ i k\tau +\frac{i \pi (\beta+1) }{2}} \big] .
\ee
For the Bunch-Davies vacuum state during inflation
and assuming the  quantum  normalization condition
that for  each $k$ mode and each polarization of tensor,
there is a zero point energy $\frac12 \hbar k$ in high frequency limit,
we obtain  $b_1=  \frac{2}{M_{\rm Pl}} e^{  \frac{i \pi (\beta+1) }{2}}$,
 $b_2=0$.
The primordial spectrum of tensor perturbation
is $\Delta^2_t \simeq 0.5 \frac{H^2}{M_{Pl}} k^{2\beta+4}$
at  low $k|\tau|\ll 1 $
\cite{zhangyang05,ZhangWang2018b,ZhangWang2016}.

And the 1st-order scalar solutions are
\bl\label{varphi1solv2}
\varphi^{(1) }_k
= &
k^{-1}(-\tau) ^{-\beta -\frac{1}{2}}
\Big[ d_1 H_{\beta +\frac{1}{2}}^{(1)}(-k\tau )
      +d_2 H_{\beta +\frac{1}{2}}^{(2)}(-k\tau ) \Big]
\nn\\
&
+(\beta +2)k^{-1} (-\tau) ^{-\beta -2}
 \int^\tau d\tau_1 \sqrt{-\tau_1 }
 \Big[d_1H_{\beta +\frac{1}{2}}^{(1)}(-k\tau_1 )
     +d_2 H_{\beta +\frac{1}{2}}^{(2)}(-k\tau_1 )  \Big],
\el
\bl\label{phi1kkfinal}
\phi^{(1)}_{ k}
=&
-\frac{\sqrt{2(\beta +1) (\beta +2)}}{(\beta +1)}
\frac{(-\tau)^{-\beta -\frac{1}{2}}}{6 M_{\rm Pl}k}
\Big[
d_1 H_{\beta +\frac{1}{2}}^{(1)}(-k\tau )
+d_2 H_{\beta +\frac{1}{2}}^{(2)}(-k\tau )
\Big]
\nn\\
&
-\frac{1}{M_{\rm Pl}}\frac{\sqrt{2(\beta +1) (\beta +2)}}{(\beta +1)}
\Big( \frac{\beta +1}{2 \,k} (-\tau) ^{-\beta -2}
 + \frac{k}{6 \beta} (- \tau)^{-\beta}  \Big)
\nn\\
&
\times\int^{\tau} d\tau_1 \sqrt{-\tau_1 }\Big[
d_1 H^{(1)}_{\beta +\frac{1}{2}}(-k \tau_1 )
+d_2 H^{(2)}_{\beta +\frac{1}{2}}(-k \tau_1 )   \Big]
\nn\\
&
+\frac{\sqrt{2(\beta +1) (\beta +2)}}{(\beta +1)\beta}
\frac{k}{6 M_{\rm Pl}}
\int^\tau d\tau_1  (-\tau_1)^{-\beta+ \frac12 }
\nn\\
&
\times
\Big[
d_1 H^{(1)}_{\beta +\frac{1}{2}}(-k \tau_1 )
+d_2 H^{(2)}_{\beta +\frac{1}{2}}(-k \tau_1 )   \Big]
 ,
\el
\bl\label{chi1kkfinal}
&
\chi^{||(1)}_{ k}
=
-\frac{\sqrt{2(\beta +1)(\beta +2)}}{ M_{\rm Pl}(\beta +1)
k^3(-\tau) ^{\beta +\frac{1}{2}}}
\Big[
d_1 H_{\beta +\frac{1}{2}}^{(1)}(-k\tau )
+d_2 H_{\beta +\frac{1}{2}}^{(2)}(-k\tau )
      \Big]
\nn\\
&
-\frac{\sqrt{2(\beta +1) (\beta +2)}}{ M_{\rm Pl}(\beta +1)k}
 \frac{1}{\beta}(-\tau )^{-\beta}
\int^\tau d\tau_1 (-\tau_1)^{\frac{1}{2}}
\Big[d_1H_{\beta +\frac{1}{2}}^{(1)}(-k\tau_1 )
     +d_2 H_{\beta +\frac{1}{2}}^{(2)}(-k\tau_1 ) \Big]
\nn\\
&
+\frac{\sqrt{2(\beta +1) (\beta +2)}}{ M_{\rm Pl}(\beta +1)k}
  \frac{1}{\beta}  \int^{\tau}d\tau_1
 (-\tau_1)^{-\beta+ \frac12  }
\Big[
d_1 H_{\beta +\frac{1}{2}}^{(1)}(-k\tau_1 )
+d_2 H_{\beta +\frac{1}{2}}^{(2)}(-k\tau_1 )
\Big]
,
\el
where
$(d_1, d_2)$ are the $k$-dependent integral constants.
{One  set of coefficients $(d_1, d_2)$
fixes all the scalar perturbations.}
To determine $(d_1, d_2)$, we shall adopt a gauge invariable
 \cite{Sasaki1986,Mukhanov1988,GordonWandsBassettMaartens2000}
\be\label{Qvarphi}
Q_\varphi
=\frac{\varphi^{(0)'}}{a H }{\cal R}
\equiv  \varphi^{(1)}  + \frac{\varphi^{(0)'} }{H a}
             \big(\phi^{(1)} +\frac16\nabla^2 \chi^{||(1)} \big) ,
\ee
where ${\cal R}$ is the comoving curvature perturbation.
${\cal R}$ measures the spatial curvature of constant-density hypersurfaces, and remains constant outside the horizon for adiabatic matter perturbations, which is often used in CMB observation analysis \cite{BassettTsuijikawa2006,Weinberg2008,Baumann2009}.
By the zero point energy of $Q_\varphi$ in the Bunch-Davies vacuum state \cite{BunchDavies1978} at high $k$
during inflation,
$d_1$ and $d_2$ are obtained as the following
\be \label{determd1}
d_1 =   \frac{ \sqrt{\pi} k}{2 l_0}  e^{\frac{i \pi  \beta }{2}+i\pi} ,
~~~
d_2 = 0.
\ee
Detailed calculation can be seen in Appendix \ref{sec:appendix_gaugeInvariant}.
Therefore, the 1st-order scalar solutions (\ref{varphi1solv2}), (\ref{phi1kkfinal}) and (\ref{chi1kkfinal}) become
\bl\label{varphi1solv2final}
\varphi^{(1) }_k
= &
\frac{ \sqrt{\pi}}{2 l_0}  e^{\frac{i \pi  \beta }{2}+i\pi} (-\tau) ^{-\beta -\frac{1}{2}}
H_{\beta +\frac{1}{2}}^{(1)}(-k\tau )
\nn\\
&
+\frac{ \sqrt{\pi}}{2 l_0}  e^{\frac{i \pi  \beta }{2}+i\pi}
(\beta +2) (-\tau) ^{-\beta -2}
 \int^\tau d\tau_1 \sqrt{-\tau_1 }
 H_{\beta +\frac{1}{2}}^{(1)}(-k\tau_1 ) ,
\el
\bl\label{phi1kkfinalfinal}
\phi^{(1)}_{ k}
=&
-\frac{\sqrt{2(\beta +1) (\beta +2)}}{(\beta +1)}
\frac{ \sqrt{\pi} }{2 l_0}  e^{\frac{i \pi  \beta }{2}+i\pi}
\frac{(-\tau)^{-\beta -\frac{1}{2}}}{6 M_{\rm Pl}}
H_{\beta +\frac{1}{2}}^{(1)}(-k\tau )
\nn\\
&
-\frac{1}{M_{\rm Pl}}\frac{\sqrt{2(\beta +1) (\beta +2)}}{(\beta +1)}
\frac{ \sqrt{\pi} }{2 l_0}  e^{\frac{i \pi  \beta }{2}+i\pi}
\Big( \frac{\beta +1}{2 } (-\tau) ^{-\beta -2}
 + \frac{k^2}{6 \beta} (- \tau)^{-\beta}  \Big)
\nn\\
&
\times\int^{\tau} d\tau_1 \sqrt{-\tau_1 }
H^{(1)}_{\beta +\frac{1}{2}}(-k \tau_1 )
\nn\\
&
+\frac{\sqrt{2(\beta +1) (\beta +2)}}{(\beta +1)\beta}
\frac{k^2}{6 M_{\rm Pl}}
\frac{ \sqrt{\pi}}{2 l_0}  e^{\frac{i \pi  \beta }{2}+i\pi}
\int^\tau d\tau_1  (-\tau_1)^{-\beta+ \frac12 }
H^{(1)}_{\beta +\frac{1}{2}}(-k \tau_1 )
 ,
\el
\bl\label{chi1kkfinalfinal}
\chi^{||(1)}_{ k}
=
&
-\frac{\sqrt{2(\beta +1)(\beta +2)}}{ M_{\rm Pl}(\beta +1)
(-\tau) ^{\beta +\frac{1}{2}}}
\frac{ \sqrt{\pi}}{2 l_0 k^2}  e^{\frac{i \pi  \beta }{2}+i\pi}
 H_{\beta +\frac{1}{2}}^{(1)}(-k\tau )
\nn\\
&
-\frac{\sqrt{2(\beta +1) (\beta +2)}}{ M_{\rm Pl}(\beta +1)}
\frac{ \sqrt{\pi} }{2 l_0}  e^{\frac{i \pi  \beta }{2}+i\pi}
 \frac{1}{\beta}(-\tau )^{-\beta}
\int^\tau d\tau_1 (-\tau_1)^{\frac{1}{2}}
H_{\beta +\frac{1}{2}}^{(1)}(-k\tau_1 )
\nn\\
&
+\frac{\sqrt{2(\beta +1) (\beta +2)}}{ M_{\rm Pl}(\beta +1)}\frac{ \sqrt{\pi}}{2 l_0}  e^{\frac{i \pi  \beta }{2}+i\pi}
  \frac{1}{\beta}  \int^{\tau}d\tau_1
 (-\tau_1)^{-\beta+ \frac12  }
H_{\beta +\frac{1}{2}}^{(1)}(-k\tau_1 )
.
\el
The above scalar solutions are also waves propagating with the speed of light,
like the tensor modes.
It seems that $\chi^{||(1)}$ suffers $k^{-2}$ divergence at $k\rightarrow0$.
However, since the operator $D_{ij}\propto k^2$,
the scalar part of metric $D_{ij}\chi^{||(1)}$
is actually convergent at $k\rightarrow0$.
The above 1st-order scalar metric  perturbations and
perturbed scalar field are generally subject to
changes under the transformations
between synchronous coordinates
(see \eqref{gaugetrphi}--\eqref{deltarho} in Appendix \ref{sec:gauge_transform}.)
Therefore,  the gauge-invariant 1st-order scalar perturbations
are often used in cosmology
 \cite{Bardeen1980,KodamaSasaki1984,Mukhanov1992,Hwang1993,Sasaki1986,
Mukhanov1988,Mukhanov2005}, which are reviewed and discussed in Appendix \ref{sec:appendix_gaugeInvariant}.

\section{The 2nd-order Einstein equation by scalar-scalar coupling}

The 2nd-order perturbed Einstein equation
is similar to the 1st-order one,
but contains various products of the 1sr-order metric perturbations
as  part of the source.
Since the  1sr-order vector modes
are  vanishing as (\ref{novect}),
the product terms consist of the scalar-scalar, scalar-tensor, tensor-tensor.
We consider the scalar-scalar coupling
 for the inflation stage  in this paper.
The perturbed stress tensor of the scalar field up to 2nd-order
is listed in Appendix \ref{sec:perturbedGT}.

The  $(00)$ component of 2nd-order  Einstein equation
is   the energy constraint
\bl  \label{Ein2th003RD}
&
-\frac{6(\beta +1)}{\tau } \phi^{(2)'}_S
+2\nabla^2\phi^{(2)}_S
+\frac{1}{3}\nabla^2\nabla^2\chi^{||(2)}_{S}
\nn\\
=&
\frac{1}{M_{\rm Pl}}\frac{\sqrt{2(\beta +1) (\beta +2)}}{\tau } \varphi^{(2)'}_{S}
-\frac{1}{M_{\rm Pl}}\frac{\sqrt{2(\beta +1) (\beta +2)} (2 \beta +1)}{\tau^2}\varphi^{(2)}_{S}
+E_S
,
\el
where the subindex $`` S"$ indicates the scalar-scalar coupling, and
\bl\label{ES1RD}
E_S
=
&
\frac{1}{M_{\rm Pl}^2} \Big[
\varphi^{(1)'}\varphi^{(1)'}
+\varphi^{(1),l}\varphi^{(1)}_{,l}
+\frac{2 (\beta +2) (2 \beta +1)}{\tau^{2}}
\varphi^{(1)}\varphi^{(1)}
\Big]
\nn\\
&
+\frac{24(\beta +1)}{\tau }\phi^{(1)'}\phi^{(1)}
-6\phi^{(1) '}\phi^{(1) '}
-6\phi^{(1)}_{,\,k}\phi^{(1),\,k}
-16\phi^{(1)}\nabla^2\phi^{(1)}
-\frac{8}{3}\phi^{(1)}\nabla^2\nabla^2\chi^{||(1)}
\nn\\
    &
+2\phi^{(1),\,kl}\chi^{||(1)}_{,\,kl}
-\frac{2}{3}\nabla^2\phi^{(1)}\nabla^2\chi^{||(1)}
+\frac{1}{4} \chi^{||(1)',\,kl}\chi^{||(1)'}_{,\,kl}
-\frac{1}{12} \nabla^2\chi^{||(1)'}\nabla^2\chi^{||(1)'}
            \nn\\
            &
+ \frac{2(\beta +1)}{\tau} \chi^{||(1),\,kl}\chi^{||(1)'}_{,\,kl}
-\frac{2(\beta +1)}{3\tau} \nabla^2\chi^{||(1)}\nabla^2\chi^{||(1)'}
+\frac{1}{3}\chi^{||(1),\,kl}\nabla^2\chi^{||(1)}_{,\,kl}
\nn\\
&
-\frac{1}{9}\nabla^2\chi^{||(1)}\nabla^2\nabla^2\chi^{||(1)}
-\frac{1}{4}\chi^{||(1),\,klm}\chi^{||(1)}_{,\,klm}
+\frac{5}{12}\nabla^2\chi^{||(1)}_{,\,k}\nabla^2\chi^{||(1),\,k}
,
\el
which  consists of the products of
$\phi^{(1)}$, $\chi^{||(1)}$ and  $\varphi^{(1)}$.
\eqref{Ein2th003RD} has a form similar to the 1st order \eqref{Ein1st00},
and has the inhomogeneous term $E_S$ as part of the source.

The  $(0i)$ component of the 2nd-order
Einstein equation is  the momentum constraint
{
\be\label{MoConstr2ndv3RD}
2 \phi^{(2)'} _{S,\,i}
+\frac{1}{2}D_{ij}\chi^{||(2)',\,j}_S
+\frac{1}{2}\chi^{\perp(2)',\,j}_{S\,ij}
=
\frac{1}{M_{\rm Pl}}\frac{\sqrt{2(\beta +1) (\beta+2)}}{\tau}\varphi^{(2)}_{S, i}
+M_{S\,i},
\ee
}
where
\bl\label{MSi1}
M_{S\,i}
\equiv&
\frac{2}{M_{\rm Pl}^2}\varphi^{(1)}_{, i}\varphi^{(1)'}
-8\phi^{(1)'} \phi^{(1) } _{,\,i}
-8\phi^{(1) } \phi^{(1) '} _{,\,i}
-\frac{4}{3}\phi^{(1) }\nabla^2\chi^{||(1)'}_{,\,i}
-\frac{4}{3}\phi^{(1)'}\nabla^2\chi^{||(1)}_{ ,\,i}
\nn\\
&
+ \phi^{(1) ,\,k }\chi^{||(1)' }_{,\,ki}
-\frac{1}{3}\phi^{(1) }_{,\,i }\nabla^2\chi^{||(1)' }
-2\phi^{(1)' ,\,k }\chi^{||(1) }_{,\,ki}
+\frac{2}{3}\phi^{(1)'}_{ ,\,i }\nabla^2\chi^{||(1) }
\nn\\
&
+\frac{2}{3}\chi^{||(1)' }_{,\,ki}\nabla^2\chi^{||(1),\,k}
-\frac{1}{18}\nabla^2\chi^{||(1)}_{,\,i}\nabla^2\chi^{||(1)' }
-\frac{1}{3}\chi^{||(1)  }_{,\,ki}\nabla^2\chi^{||(1)',\,k}
\nn\\
&
+\frac{1}{9}\nabla^2\chi^{||(1)  }\nabla^2\chi^{||(1)'}_{,\,i}
- \frac{1}{2}\chi^{||(1)',\,kl }\chi^{||(1) }_{,\,kli}
.
\el
Equation (\ref{MoConstr2ndv3RD}) can be further decomposed into two parts.
The longitudinal part is
\be \label{MoConstr2ndv3RD2}
2\phi^{(2)'}_{S}
+\frac{1}{3}\nabla^2\chi^{||(2)'}_S
=
\frac{1}{M_{\rm Pl}}\frac{\sqrt{2(\beta +1) (\beta+2)}}{\tau}\varphi^{(2)}_{S}
+\nabla^{-2}M_{S\,l}^{,\,l}
.
\ee
where
{\allowdisplaybreaks
\bl\label{MSkkScalar}
\nabla^{-2}M_{S\,l}^{,l}
&=
-8\phi^{(1)'} \phi^{(1) }
-\frac{1}{3}\phi^{(1) }\nabla^2\chi^{||(1)' }
-\frac{4}{3}\phi^{(1)'}\nabla^2\chi^{||(1)}
-\frac{1}{18}\nabla^2\chi^{||(1)}\nabla^2\chi^{||(1)' }
\nn\\
&
-\frac{1}{6}\chi^{||(1)}_{,\,k}\nabla^2\chi^{||(1)',\,k}
+\nabla^{-2}\Big[
\frac{2}{M_{\rm Pl}^2}\varphi^{(1), k}\varphi^{(1)'}_{, k}
+\frac{2}{M_{\rm Pl}^2}\varphi^{(1)'}\nabla^2\varphi^{(1)}
\nn\\
&
+ \phi^{(1) ,\,kl }\chi^{||(1)' }_{,\,kl}
-\phi^{(1) }\nabla^2\nabla^2\chi^{||(1)'}
+2\nabla^2\phi^{(1)'}\nabla^2\chi^{||(1) }
-2\phi^{(1)' ,\,kl}\chi^{||(1) }_{,\,kl}
\nn\\
&
+\frac{1}{6}\chi^{||(1)}_{,\,k}\nabla^2\nabla^2\chi^{||(1)',\,k}
+\frac{1}{6}\nabla^2\chi^{||(1)  }\nabla^2\nabla^2\chi^{||(1)'}
+\frac{1}{6}\chi^{||(1)'}_{,\,kl}\nabla^2\chi^{||(1),\,kl}
\nn\\
&
+\frac{2}{3}\nabla^2\chi^{||(1)' }_{,\,k}\nabla^2\chi^{||(1),\,k}
- \frac{1}{2}\chi^{||(1)',\,klm }\chi^{||(1) }_{,\,klm}
\Big]
.
\el
}
The transverse part  is
\be \label{MoCons2ndCurlRD1}
\frac{1}{2}\chi^{\perp(2)',\,j}_{S\,ij}
=
\l(
M_{S\,i}
-\partial_i\nabla^{-2}M_{S\,l}^{,\,l}
\r)
\ ,
\ee
where
{
\bl\label{MSiCurl}
&
M_{S\,i}-\partial_i\nabla^{-2}M_{S\,k}^{,k}
\nn\\
=&
\frac{2}{M_{\rm Pl}^2}\varphi^{(1)}_{, i}\varphi^{(1)'}
-\phi^{(1) }\nabla^2\chi^{||(1)' }_{,\,i}
+2\phi^{(1)'}_{ ,\,i }\nabla^2\chi^{||(1) }
+ \phi^{(1) ,\,k }\chi^{||(1)' }_{,\,ki}
\nn\\
&
-2\phi^{(1)' ,\,k }\chi^{||(1) }_{,\,ki}
+\frac{2}{3}\chi^{||(1)' }_{,\,ki}\nabla^2\chi^{||(1),\,k}
-\frac{1}{6}\chi^{||(1)}_{,\,ki}\nabla^2\chi^{||(1)',\,k}
\nn\\
&
+\frac{1}{6}\chi^{||(1)}_{,\,k}\nabla^2\chi^{||(1)',\,k}_{,\,i}
+\frac{1}{6}\nabla^2\chi^{||(1)  }\nabla^2\chi^{||(1)'}_{,\,i}
- \frac{1}{2}\chi^{||(1)',\,kl }\chi^{||(1) }_{,\,kli}
\nn\\
&
+\partial_i\nabla^{-2}\Big[
-\frac{2}{M_{\rm Pl}^2}\varphi^{(1), k}\varphi^{(1)'}_{, k}
-\frac{2}{M_{\rm Pl}^2}\varphi^{(1)'}\nabla^2\varphi^{(1)}
- \phi^{(1) ,\,kl }\chi^{||(1)' }_{,\,kl}
+\phi^{(1) }\nabla^2\nabla^2\chi^{||(1)'}
\nn\\
&
-2\nabla^2\phi^{(1)'}\nabla^2\chi^{||(1) }
+2\phi^{(1)' ,\,kl}\chi^{||(1) }_{,\,kl}
 -\frac{1}{6}\chi^{||(1)}_{,\,k}\nabla^2\nabla^2\chi^{||(1)',\,k}
-\frac{1}{6}\nabla^2\chi^{||(1)  }\nabla^2\nabla^2\chi^{||(1)'}
\nn\\
&
-\frac{1}{6}\chi^{||(1)'}_{,\,kl}\nabla^2\chi^{||(1),\,kl}
-\frac{2}{3}\nabla^2\chi^{||(1)' }_{,\,k}\nabla^2\chi^{||(1),\,k}
+ \frac{1}{2}\chi^{||(1)',\,klm }\chi^{||(1) }_{,\,klm}
\Big]  .
\el
}

The $(ij)$ component of 2nd-order Einstein equation is given by
\bl\label{Evo2ndSs1RD}
&
2\phi^{(2)''}_S \delta_{ij}
+4\frac{\beta +1}{\tau }\phi^{(2)'}_S \delta_{ij}
+\phi^{(2)}_{S,\,ij}
-\nabla^2\phi^{(2)}_S \delta_{ij}
\nn\\
&
+\frac{1}{2}D_{ij}\chi^{||(2)''}_S
+\frac{\beta +1}{\tau } D_{ij}\chi^{||(2)'}_S
+\frac{1}{6}\nabla^2D_{ij}\chi^{||(2)}_S
-\frac{1}{9}\delta_{ij}\nabla^2\nabla^2\chi^{||(2) }_S
\nn\\
&
+\frac{1}{2}\chi^{\perp(2)''}_{S\,ij}
+\frac{\beta +1}{\tau } \chi^{\perp(2)'}_{S\,ij}
\nn\\
&
+\frac{1}{2}\chi^{\top(2)''}_{S\,ij}
+\frac{\beta +1}{\tau }\chi^{\top(2)'}_{S\,ij}
-\frac{1}{2}\nabla^2\chi^{\top(2)}_{S\,ij}
\nn\\
=&
\frac{1}{M_{\text{Pl}}}\frac{ \sqrt{2(\beta +1) (\beta +2)}}{\tau  }\varphi^{(2)'}_S\delta_{ij}
+\frac{1}{M_{\text{Pl}}}\frac{ \sqrt{2(\beta +1) (\beta +2)} (2 \beta +1)}{\tau ^2 }\varphi^{(2)}_S\delta_{ij}
+S_{S\,ij}
,
\el
where
\bl\label{Ss2ndij2RD}
S_{S\,ij}
&=
\frac{1}{M_{\rm Pl}^2}\varphi^{(1)'}\varphi^{(1)'}\delta_{ij}
-\frac{1}{M_{\rm Pl}^2}\varphi^{(1),k}\varphi^{(1)}_{, k}\delta_{ij}
-\frac{1}{M_{\rm Pl}^2}\frac{2 (\beta +2) (2 \beta +1)}{\tau ^2}\varphi^{(1)}\varphi^{(1)}\delta_{ij}
\nn\\
&
-\frac{4}{M_{\rm Pl}} \frac{  \sqrt{2(\beta +1) (\beta +2)} }{\tau }\varphi^{(1)'}\phi^{(1)}\delta_{ij}
-\frac{4}{M_{\rm Pl}}\frac{\sqrt{2(\beta +1) (\beta +2)} (2 \beta +1)}{\tau ^2}\varphi^{(1)}\phi^{(1)}\delta_{ij}
\nn\\
&
+\frac{2}{M_{\rm Pl}^2}\varphi^{(1)}_{, i}\varphi^{(1)}_{, j}
+\frac{2}{M_{\rm Pl}}\frac{\sqrt{2(\beta +1) (\beta +2)} (2 \beta +1)}{\tau ^2}\chi^{||(1)}_{,ij}\varphi^{(1)}
\nn\\
&
-\frac{2}{3M_{\rm Pl}}\frac{\sqrt{2(\beta +1) (\beta +2)} (2 \beta +1) }{\tau ^2}\varphi^{(1)}\nabla^2\chi^{||(1)}\delta_{ij}
\nn\\
&
+\frac{2}{M_{\rm Pl}}\frac{\sqrt{2(\beta +1) (\beta +2)} }{\tau }\chi^{||(1)}_{,ij}\varphi^{(1)'}
-\frac{2}{3M_{\rm Pl}}\frac{ \sqrt{2(\beta +1) (\beta +2)} }{\tau }\varphi^{(1)'}\nabla^2\chi^{||(1)}\delta_{ij}
\nn\\
&
-6\phi^{(1)}_{,\,i}\phi^{(1)}_{,\,j}
+4\phi^{(1),\,k}\phi^{(1)}_{,\,k}\delta_{ij}
+4\phi^{(1)}\nabla^2\phi^{(1)}\delta_{ij}
-4\phi^{(1)}\phi^{(1)}_{,\,ij}
-2\phi^{(1)'}\phi^{(1)'}\delta_{ij}
\nn\\
&
-\frac{12(\beta +1)}{\tau }\phi^{(1)'}\chi^{||(1)}_{,\,ij}
+\frac{4(\beta+1)}{\tau }\phi^{(1)'}\nabla^2\chi^{||(1)}\delta_{ij}
-\phi^{(1)'}\chi^{||(1)'}_{,\,ij}
+\frac{1}{3}\phi^{(1)'}\nabla^2\chi^{||(1)'}\delta_{ij}
\nn\\
&
-6\phi^{(1)''}\chi^{||(1)}_{,\,ij}
+2\phi^{(1)''}\nabla^2\chi^{||(1)}\delta_{ij}
-\phi^{(1)}_{,\,j}\nabla^2\chi^{||(1)}_{,\,i}
-\phi^{(1)}_{,\,i}\nabla^2\chi^{||(1)}_{,\,j}
+\phi^{(1)}_{,\,k}\chi^{||(1),\,k}_{,\,ij}
\nn\\
&
-\frac{2}{3}\phi^{(1)}\nabla^2\chi^{||(1)}_{,\,ij}
+\frac{1}{3}\phi^{(1)}_{,\,k}\nabla^2\chi^{||(1),\,k}\delta_{ij}
+\frac{2}{3}\phi^{(1)}\nabla^2\nabla^2\chi^{||(1)}\delta_{ij}
+4\chi^{||(1)}_{,\,ij}\nabla^2\phi^{(1)}
\nn\\
&
-\frac{4}{3}\nabla^2\phi^{(1)}\nabla^2\chi^{||(1)}\delta_{ij}
-2\phi^{(1)}_{,\,ki}\chi^{||(1),\,k}_{,\,j}
-2\phi^{(1)}_{,\,kj}\chi^{||(1),\,k}_{,\,i}
+\frac{4}{3}\phi^{(1)}_{,\,ij}\nabla^2\chi^{||(1)}
\nn\\
&
+\frac{1}{4}\chi^{||(1),\,klm}\chi^{||(1)}_{,\,klm}\delta_{ij}
-\frac{11}{36}\nabla^2\chi^{||(1),\,k}\nabla^2\chi^{||(1)}_{,\,k}\delta_{ij}
-\frac{2}{9}\nabla^2\chi^{||(1)}\nabla^2\nabla^2\chi^{||(1)}\delta_{ij}
\nn\\
&
-\frac{1}{6}\nabla^2\chi^{||(1)}_{,\,i}\nabla^2\chi^{||(1)}_{,\,j}
-\frac{1}{3}\chi^{||(1),\,k}_{,\,j}\nabla^2\chi^{||(1)}_{,\,ik}
-\frac{1}{3}\chi^{||(1),\,k}_{,\,i}\nabla^2\chi^{||(1)}_{,\,jk}
+\frac{2}{3}\chi^{||(1)}_{,\,ij}\nabla^2\nabla^2\chi^{||(1)}
\nn\\
&
+\frac{2}{9}\nabla^2\chi^{||(1)}\nabla^2\chi^{||(1)}_{,\,ij}
-\frac{1}{2}\chi^{||(1),\,kl}_{,\,i}\chi^{||(1)}_{,\,klj}
+\frac{2}{3}\chi^{||(1)}_{,\,kij}\nabla^2\chi^{||(1),\,k}
-\frac{2(\beta +1)}{\tau }\chi^{||(1),\,kl}\chi^{||(1)'}_{,\,kl}\delta_{ij}
\nn\\
&
+\frac{2(\beta +1)}{3\tau }\nabla^2\chi^{||(1)}\nabla^2\chi^{||(1)'}\delta_{ij}
-\frac{3}{4}\chi^{||(1)',\,kl}\chi^{||(1)'}_{,\,kl}\delta_{ij}
-\chi^{||(1),\,kl}\chi^{||(1)''}_{,\,kl}\delta_{ij}
\nn\\
&
+\frac{1}{3}\nabla^2\chi^{||(1)}\nabla^2\chi^{||(1)''}\delta_{ij}
+\frac{13}{36}\nabla^2\chi^{||(1)'}\nabla^2\chi^{||(1)'}\delta_{ij}
+\chi^{||(1)',\,k}_{,\,i}\chi^{||(1)'}_{,\,kj}
-\frac{2}{3}\chi^{||(1)'}_{,\,ij}\nabla^2\chi^{||(1)'}
,
\el
By decomposition,  the trace part of  \eqref{Evo2ndSs1RD}  is
{
\bl\label{Evo2ndSsTr2RD}
&
\phi^{(2)''}_S
+2\frac{\beta +1}{\tau }\phi^{(2)'}_S
-\frac{1}{3}\nabla^2\phi^{(2)}_S
-\frac{1}{18}\nabla^2\nabla^2\chi^{||(2) }_S
\nn\\
=&
\frac{1}{M_{\rm Pl}}\frac{\sqrt{2(\beta +1) (\beta +2)} }{2\tau }\varphi^{(2)'}_S
+\frac{1}{M_{\rm Pl}}\frac{\sqrt{2(\beta +1)
(\beta +2)} (2 \beta +1) }{2\tau ^2}\varphi^{(2)}_S
+\frac{1}{6}S_{S\,k}^k
\,,
\el
}
where
{
\bl\label{SijTraceRD}
S_{S\,k}^k
=&\frac{3}{M_{\rm Pl}^2}\varphi^{(1)'}\varphi^{(1)'}
-\frac{1}{M_{\rm Pl}^2}\varphi^{(1),k}\varphi^{(1)}_{, k}
-\frac{6}{M_{\rm Pl}^2}\frac{(\beta +2) (2 \beta +1)}{\tau ^2}\varphi^{(1)}\varphi^{(1)}
\nn\\
&
-\frac{12}{M_{\rm Pl}} \frac{ \sqrt{2(\beta +1) (\beta +2)} }{\tau }\varphi^{(1)'}\phi^{(1)}
-\frac{12}{M_{\rm Pl}}\frac{ \sqrt{2(\beta +1) (\beta +2)} (2 \beta +1) }{\tau ^2}\varphi^{(1)}\phi^{(1)}
\nn\\
&
+6\phi^{(1)}_{,\,k}\phi^{(1),\,k}
+8\phi^{(1)}\nabla^2\phi^{(1)}
-6\phi^{(1)'}\phi^{(1)'}
-4\phi^{(1)}_{,\,kl}\chi^{||(1),\,kl}
+\frac{4}{3}\phi^{(1)}\nabla^2\nabla^2\chi^{||(1)}
\nn\\
&
+\frac{4}{3}\nabla^2\phi^{(1)}\nabla^2\chi^{||(1)}
+\frac{2}{9}\nabla^2\chi^{||(1)}\nabla^2\nabla^2\chi^{||(1)}
-\frac{2}{3}\chi^{||(1),\,kl}\nabla^2\chi^{||(1)}_{,\,kl}
-\frac{5}{12}\nabla^2\chi^{||(1),\,k}\nabla^2\chi^{||(1)}_{,\,k}
\nn\\
&
+\frac{1}{4}\chi^{||(1),\,klm}\chi^{||(1)}_{,\,klm}
-\frac{6(\beta +1)}{\tau }\chi^{||(1),\,kl}\chi^{||(1)'}_{,\,kl}
+\frac{2(\beta +1)}{\tau }\nabla^2\chi^{||(1)}\nabla^2\chi^{||(1)'}
\nn\\
&
-\frac{5}{4}\chi^{||(1)',\,kl}\chi^{||(1)'}_{,\,kl}
-3\chi^{||(1),\,kl}\chi^{||(1)''}_{,\,kl}
+\nabla^2\chi^{||(1)}\nabla^2\chi^{||(1)''}
+\frac{5}{12}\nabla^2\chi^{||(1)'}\nabla^2\chi^{||(1)'}
 .
\el
}
The traceless part of  \eqref{Evo2ndSs1RD}  is
{
\bl\label{Evo2ndSsNoTr1RD}
&
D_{ij}\phi^{(2)}_{S}
+\frac{1}{2}D_{ij}\chi^{||(2)''}_S
+\frac{\beta +1}{\tau } D_{ij}\chi^{||(2)'}_S
+\frac{1}{6}\nabla^2D_{ij}\chi^{||(2)}_S
\nn\\
&
+\frac{1}{2}\chi^{\perp(2)''}_{S\,ij}
+\frac{\beta +1}{\tau } \chi^{\perp(2)'}_{S\,ij}
\nn\\
&
+\frac{1}{2}\chi^{\top(2)''}_{S\,ij}
+\frac{\beta +1}{\tau } \chi^{\top(2)'}_{S\,ij}
-\frac{1}{2}\nabla^2\chi^{\top(2)}_{S\,ij}
=
\bar S_{S\,ij}
,
\el
}
where
{ \allowdisplaybreaks
\bl\label{SijTracelessRD}
\bar S_{S\,ij}
\equiv&
S_{S\,ij}-\frac{1}{3}S_{S\,k}^k\delta_{ij}
\nn\\
=&
\frac{2}{M_{\rm Pl}^2}\varphi^{(1)}_{, i}\varphi^{(1)}_{, j}
-\frac{2}{3M_{\rm Pl}^2}\varphi^{(1), k}\varphi^{(1)}_{, k}\delta_{ij}
+\frac{2}{M_{\rm Pl}}\frac{ \sqrt{2(\beta +1) (\beta +2)} (2 \beta +1)}{\tau ^2}\chi^{||(1)}_{,ij}\varphi^{(1)}
\nn\\
&
-\frac{2}{3M_{\rm Pl}}\frac{\sqrt{2(\beta +1) (\beta +2)} (2 \beta +1) }{\tau ^2}\varphi^{(1)}\nabla^2\chi^{||(1)}\delta_{ij}
\nn\\
&
+\frac{2}{M_{\rm Pl}}\frac{\sqrt{2(\beta +1) (\beta +2)} }{\tau }\chi^{||(1)}_{,ij}\varphi^{(1)'}
-\frac{2}{3M_{\rm Pl}}\frac{\sqrt{2(\beta +1) (\beta +2)} }{\tau }\varphi^{(1)'}\nabla^2\chi^{||(1)}\delta_{ij}
\nn\\
&
-6\phi^{(1)}_{,\,i}\phi^{(1)}_{,\,j}
+2\phi^{(1),\,k}\phi^{(1)}_{,\,k}\delta_{ij}
-4\phi^{(1)}\phi^{(1)}_{,\,ij}
+\frac{4}{3}\phi^{(1)}\nabla^2\phi^{(1)}\delta_{ij}
-\frac{12(\beta +1)}{\tau }\phi^{(1)'}\chi^{||(1)}_{,\,ij}
\nn\\
&
+\frac{4(\beta +1)}{\tau }\phi^{(1)'}\nabla^2\chi^{||(1)}\delta_{ij}
-\phi^{(1)'}\chi^{||(1)'}_{,\,ij}
+\frac{1}{3}\phi^{(1)'}\nabla^2\chi^{||(1)'}\delta_{ij}
-6\phi^{(1)''}\chi^{||(1)}_{,\,ij}
\nn\\
&
+2\phi^{(1)''}\nabla^2\chi^{||(1)}\delta_{ij}
-\phi^{(1)}_{,\,j}\nabla^2\chi^{||(1)}_{,\,i}
-\phi^{(1)}_{,\,i}\nabla^2\chi^{||(1)}_{,\,j}
+\phi^{(1)}_{,\,k}\chi^{||(1),\,k}_{,\,ij}
+\frac{1}{3}\phi^{(1)}_{,\,k}\nabla^2\chi^{||(1),\,k}\delta_{ij}
\nn\\
&
-\frac{2}{3}\phi^{(1)}\nabla^2\chi^{||(1)}_{,\,ij}
+\frac{2}{9}\phi^{(1)}\nabla^2\nabla^2\chi^{||(1)}\delta_{ij}
+4\chi^{||(1)}_{,\,ij}\nabla^2\phi^{(1)}
-\frac{4}{3}\nabla^2\phi^{(1)}\nabla^2\chi^{||(1)}\delta_{ij}
\nn\\
&
-2\phi^{(1)}_{,\,ki}\chi^{||(1),\,k}_{,\,j}
-2\phi^{(1)}_{,\,kj}\chi^{||(1),\,k}_{,\,i}
+\frac{4}{3}\phi^{(1)}_{,\,kl}\chi^{||(1),\,kl}\delta_{ij}
+\frac{4}{3}\phi^{(1)}_{,\,ij}\nabla^2\chi^{||(1)}
-\frac{4}{9}\nabla^2\phi^{(1)}\nabla^2\chi^{||(1)}\delta_{ij}
\nn\\
&
+\frac{2}{3}\chi^{||(1)}_{,\,kij}\nabla^2\chi^{||(1),\,k}
-\frac{1}{6}\nabla^2\chi^{||(1)}_{,\,i}\nabla^2\chi^{||(1)}_{,\,j}
-\frac{1}{6}\nabla^2\chi^{||(1),\,k}\nabla^2\chi^{||(1)}_{,\,k}\delta_{ij}
-\frac{1}{3}\chi^{||(1),\,k}_{,\,j}\nabla^2\chi^{||(1)}_{,\,ik}
\nn\\
&
-\frac{1}{3}\chi^{||(1),\,k}_{,\,i}\nabla^2\chi^{||(1)}_{,\,jk}
+\frac{2}{9}\chi^{||(1),\,kl}\nabla^2\chi^{||(1)}_{,\,kl}\delta_{ij}
+\frac{2}{3}\chi^{||(1)}_{,\,ij}\nabla^2\nabla^2\chi^{||(1)}
+\frac{2}{9}\nabla^2\chi^{||(1)}\nabla^2\chi^{||(1)}_{,\,ij}
\nn\\
&
-\frac{8}{27}\nabla^2\chi^{||(1)}\nabla^2\nabla^2\chi^{||(1)}\delta_{ij}
-\frac{1}{2}\chi^{||(1),\,kl}_{,\,i}\chi^{||(1)}_{,\,klj}
+\frac{1}{6}\chi^{||(1),\,klm}\chi^{||(1)}_{,\,klm}\delta_{ij}
+\chi^{||(1)',\,k}_{,\,i}\chi^{||(1)'}_{,\,kj}
\nn\\
&
-\frac{1}{3}\chi^{||(1)',\,kl}\chi^{||(1)'}_{,\,kl}\delta_{ij}
-\frac{2}{3}\chi^{||(1)'}_{,\,ij}\nabla^2\chi^{||(1)'}
+\frac{2}{9}\nabla^2\chi^{||(1)'}\nabla^2\chi^{||(1)'}\delta_{ij}
\ .
\el }
Applying $\partial_i\partial_j$ on (\ref{Evo2ndSsNoTr1RD})
gives the evolution equation for the scalar $\chi^{||(2)}_S$
as the following:
{
\bl\label{Evo2ndSsChi1RD}
&
\frac{1}{2}\chi^{||(2)''}_S
+\frac{\beta +1}{\tau }\chi^{||(2)'}_S
+\frac{1}{6}\nabla^2\chi^{||(2)}_S
= -  \phi^{(2)}_{S}
 + \frac{3}{2}\nabla^{-2}\nabla^{-2}\bar S_{S\,kl}^{\, ,\,kl}
,
\el
}
where
{ \allowdisplaybreaks
\bl\label{SijPijbarRD}
&\bar S_{S\,ij}^{,\,ij}
=
\frac{2}{3}\nabla^2\nabla^2\Big[
-\frac{9}{4}\phi^{(1)}\phi^{(1)}
-\frac{2}{3}\phi^{(1)}\nabla^2\chi^{||(1)}
-\frac{1}{18}\nabla^2\chi^{||(1)}\nabla^2\chi^{||(1)}
-\frac{1}{16}\chi^{||(1)}_{,\,kl}\chi^{||(1),\,kl}
\Big]
\nn\\
&
+\nabla^2\Big[
\frac{1}{3M_{\rm Pl}^2}\varphi^{(1), k}\varphi^{(1)}_{, k}
-\frac{2}{3M_{\rm Pl}}\frac{\sqrt{2(\beta +1) (\beta +2)} (2 \beta +1)}{\tau ^2}\varphi^{(1)}\nabla^2\chi^{||(1)}
\nn\\
&
-\frac{2}{3M_{\rm Pl}}\frac{\sqrt{2(\beta +1) (\beta +2)} }{\tau }\varphi^{(1)'}\nabla^2\chi^{||(1)}
-\frac{5}{3}\phi^{(1)}\nabla^2\phi^{(1)}
-\frac{8(\beta +1)}{\tau }\phi^{(1)'}\nabla^2\chi^{||(1)}
\nn\\
&
-\frac{2}{3}\phi^{(1)'}\nabla^2\chi^{||(1)'}
-4\phi^{(1)''}\nabla^2\chi^{||(1)}
-\frac{4}{9}\phi^{(1),k}\nabla^2\chi^{||(1)}_{,\,k}
+\frac{4}{3}\phi^{(1)}_{,\,kl}\chi^{||(1),kl}
\nn\\
&
+\frac{1}{54}\nabla^2\chi^{||(1),k}\nabla^2\chi^{||(1)}_{,\,k}
+\frac{11}{36}\chi^{||(1)}_{,\,kl}\nabla^2\chi^{||(1),kl}
+\frac{1}{6}\chi^{||(1)',kl}\chi^{||(1)'}_{,\,kl}
-\frac{1}{9}\nabla^2\chi^{||(1)'}\nabla^2\chi^{||(1)'}
\Big]
\nn\\
&
+\frac{2}{M_{\rm Pl}^2}\varphi^{(1)}_{, k}\nabla^2\varphi^{(1),k}
+\frac{2}{M_{\rm Pl}^2}\nabla^2\varphi^{(1)}\nabla^2\varphi^{(1)}
+\frac{2}{M_{\rm Pl}}\frac{\sqrt{2(\beta +1) (\beta +2)} (2 \beta +1) }{\tau ^2}\varphi^{(1)}\nabla^2\nabla^2\chi^{||(1)}
\nn\\
&
+\frac{4}{M_{\rm Pl}}\frac{\sqrt{2(\beta +1) (\beta +2)} (2 \beta +1) }{\tau ^2}\varphi^{(1),k}\nabla^2\chi^{||(1)}_{,k}
\nn\\
&
+\frac{2}{M_{\rm Pl}}\frac{\sqrt{2(\beta +1) (\beta +2)} (2 \beta +1) }{\tau ^2}\chi^{||(1)}_{,kl}\varphi^{(1),kl}
+\frac{2}{M_{\rm Pl}}\frac{\sqrt{2(\beta +1) (\beta +2)}}{\tau }\varphi^{(1)'}\nabla^2\nabla^2\chi^{||(1)}
\nn\\
&
+\frac{4}{M_{\rm Pl}}\frac{\sqrt{2 (\beta +1) (\beta +2)} }{\tau }\varphi^{(1)',k}\nabla^2\chi^{||(1)}_{,k}
+\frac{2}{M_{\rm Pl}}\frac{\sqrt{2 (\beta +1) (\beta +2)} }{\tau }\chi^{||(1)}_{,kl}\varphi^{(1)',kl}
\nn\\
&
+2\phi^{(1),\,k}\nabla^2\phi^{(1)}_{,\,k}
+2\phi^{(1)}\nabla^2\nabla^2\phi^{(1)}
-12\frac{\beta +1}{\tau }\phi^{(1)',\,kl}\chi^{||(1)}_{,\,kl}
+12\frac{\beta +1}{\tau }\nabla^2\phi^{(1)'}\nabla^2\chi^{||(1)}
\nn\\
&
-\phi^{(1)',\,kl}\chi^{||(1)'}_{,\,kl}
+\nabla^2\phi^{(1)'}\nabla^2\chi^{||(1)'}
-6\phi^{(1)'',\,kl}\chi^{||(1)}_{,\,kl}
+6\nabla^2\phi^{(1)''}\nabla^2\chi^{||(1)}
\nn\\
&
+\frac{4}{3}\nabla^2\nabla^2\chi^{||(1)}\nabla^2\phi^{(1)}
+\frac{11}{3}\nabla^2\chi^{||(1)}_{,\,k}\nabla^2\phi^{(1),k}
+\frac{2}{3}\phi^{(1),\,k}\nabla^2\nabla^2\chi^{||(1)}_{,\,k}
-3\phi^{(1)}_{,\,klm}\chi^{||(1),klm}
\nn\\
&
+\frac{7}{9}\nabla^2\chi^{||(1)}_{,\,k}\nabla^2\nabla^2\chi^{||(1),k}
+\frac{5}{18}\nabla^2\nabla^2\chi^{||(1)}\nabla^2\nabla^2\chi^{||(1)}
-\frac{1}{2}\chi^{||(1)}_{,\,klm}\nabla^2\chi^{||(1),\,klm}
\nn\\
&
+\frac{1}{3}\chi^{||(1)'}_{,\,kl}\nabla^2\chi^{||(1)',\,kl}
+\frac{1}{3}\nabla^2\chi^{||(1)'}_{,\,k}\nabla^2\chi^{||(1)',\,k}
\ .
\el }
A combination $\partial^i$[(\ref{Evo2ndSsNoTr1RD})$-D_{ij}$(\ref{Evo2ndSsChi1RD})],
gives the evolution equation for the 2nd-order vector
\be\label{Evo2ndSsVec2RD}
\chi^{\perp(2)''}_{S\,ij}
  +2 \frac{\beta+1}{\tau} \chi^{\perp(2)'}_{S\,ij}
   = 2 V_{S\,ij} \, ,
\ee
where the  inhomogeneous term
{ \allowdisplaybreaks
\bl\label{SourceCurl1RD}
V_{S\,ij}
\equiv
&
\nabla^{-2}\bar S_{S\,kj,\,i}^{,\,k}
+\nabla^{-2}\bar S_{S\,ki,j}^{,\,k}
-2\nabla^{-2}\nabla^{-2}\bar S_{S\,kl,\,ij}^{\, ,\,kl}
\nn\\
=&
\partial_i\nabla^{-2}\l(
\bar S_{S\,kj}^{,\,k}
-\nabla^{-2}\bar S_{S\,kl,\,j}^{\, ,\,kl}
\r)
+(i\leftrightarrow j)
\nn\\
=&
\partial_i\Big[\frac{1}{3}\phi^{(1)}\nabla^2\chi^{||(1)}_{,\,j}
+2\phi^{(1),\,k}\chi^{||(1)}_{,\,kj}
+\frac{1}{3}\chi^{||(1)}_{,\,kj}\nabla^2\chi^{||(1),\,k}
\Big]
\nn\\
&
-\partial_i\partial_j\nabla^{-2}\Big[
-12\frac{\beta +1}{\tau }\phi^{(1)'}\nabla^2\chi^{||(1)}
-\phi^{(1)'}\nabla^2\chi^{||(1)'}
-6\phi^{(1)''}\nabla^2\chi^{||(1)}
\nn\\
&
+\frac{1}{3}\phi^{(1)}\nabla^2\nabla^2\chi^{||(1)}
+2\phi^{(1),\,kl}\chi^{||(1)}_{,\,kl}
+\frac{1}{12}\nabla^2\chi^{||(1)}_{,\,k}\nabla^2\chi^{||(1),\,k}
+\frac{1}{3}\chi^{||(1)}_{,\,kl}\nabla^2\chi^{||(1),\,kl}
\Big]
\nn\\
&
+\partial_i\nabla^{-2}\Big[
\frac{2}{M_{\rm Pl}^2}\varphi^{(1)}_{, j}\nabla^2\varphi^{(1)}
+\frac{2}{M_{\rm Pl}}\frac{\sqrt{2(\beta +1) (\beta +2)} (2 \beta +1)}{\tau ^2}
    \Big(
    \varphi^{(1)}\nabla^2\chi^{||(1)}_{,j}
\nn\\
&
    +\chi^{||(1)}_{,kj}\varphi^{(1),k}
    \Big)
+\frac{2}{M_{\rm Pl}}\frac{\sqrt{2(\beta +1) (\beta +2)} }{\tau }
    \Big(
    \varphi^{(1)'}\nabla^2\chi^{||(1)}_{,j}
    +\chi^{||(1)}_{,kj}\varphi^{(1)',k}
    \Big)
\nn\\
&
+2\phi^{(1)}\nabla^2\phi^{(1)}_{,\,j}
-12\frac{\beta +1}{\tau }\phi^{(1)',\,k}\chi^{||(1)}_{,\,kj}
-12\frac{\beta +1}{\tau }\phi^{(1)'}\nabla^2\chi^{||(1)}_{,\,j}
\nn\\
&
-\phi^{(1)'}\nabla^2\chi^{||(1)'}_{,\,j}
-\phi^{(1)',\,k}\chi^{||(1)'}_{,\,kj}
-6\phi^{(1)''}\nabla^2\chi^{||(1)}_{,\,j}
-6\phi^{(1)'',\,k}\chi^{||(1)}_{,\,kj}
\nn\\
&
+\frac{4}{3}\nabla^2\chi^{||(1)}_{,\,j}\nabla^2\phi^{(1)}
-\frac{5}{3}\phi^{(1),\,k}\nabla^2\chi^{||(1)}_{,\,kj}
-3\phi^{(1)}_{,\,kl}\chi^{||(1),\,kl}_{,\,j}
\nn\\
&
+\frac{5}{18}\nabla^2\chi^{||(1)}_{,\,j}\nabla^2\nabla^2\chi^{||(1)}
-\frac{1}{2}\chi^{||(1)}_{,\,klj}\nabla^2\chi^{||(1),\,kl}
+\frac{1}{3}\chi^{||(1)'}_{,\,kj}\nabla^2\chi^{||(1)',\,k}
\Big]
\nn\\
&
-\partial_i\partial_j\nabla^{-2}\nabla^{-2}\Big[
\frac{2}{M_{\rm Pl}^2}\varphi^{(1)}_{, k}\nabla^2\varphi^{(1),k}
+\frac{2}{M_{\rm Pl}^2}\nabla^2\varphi^{(1)}\nabla^2\varphi^{(1)}
\nn\\
&
+\frac{2}{M_{\rm Pl}}\frac{\sqrt{2(\beta +1) (\beta +2)} (2 \beta +1) }{\tau ^2}
\Big(
\varphi^{(1)}\nabla^2\nabla^2\chi^{||(1)}
+2\varphi^{(1),k}\nabla^2\chi^{||(1)}_{,k}
\nn\\
&
+\chi^{||(1)}_{,kl}\varphi^{(1),kl}
\Big)
+\frac{2}{M_{\rm Pl}}\frac{\sqrt{2(\beta +1) (\beta +2)} }{\tau }
\Big(
\varphi^{(1)'}\nabla^2\nabla^2\chi^{||(1)}
+2\varphi^{(1)',k}\nabla^2\chi^{||(1)}_{,k}
\nn\\
&
+\chi^{||(1)}_{,kl}\varphi^{(1)',kl}
\Big)
+2\phi^{(1),\,k}\nabla^2\phi^{(1)}_{,\,k}
+2\phi^{(1)}\nabla^2\nabla^2\phi^{(1)}
-12\frac{\beta +1}{\tau }\phi^{(1)',\,kl}\chi^{||(1)}_{,\,kl}
\nn\\
&
+12\frac{\beta +1}{\tau }\nabla^2\phi^{(1)'}\nabla^2\chi^{||(1)}
-\phi^{(1)',\,kl}\chi^{||(1)'}_{,\,kl}
+\nabla^2\phi^{(1)'}\nabla^2\chi^{||(1)'}
\nn\\
&
-6\phi^{(1)'',\,kl}\chi^{||(1)}_{,\,kl}
+6\nabla^2\phi^{(1)''}\nabla^2\chi^{||(1)}
+\frac{4}{3}\nabla^2\nabla^2\chi^{||(1)}\nabla^2\phi^{(1)}
\nn\\
&
+\frac{11}{3}\nabla^2\chi^{||(1)}_{,\,k}\nabla^2\phi^{(1),\,k}
+\frac{2}{3}\phi^{(1),\,k}\nabla^2\nabla^2\chi^{||(1)}_{,\,k}
-3\phi^{(1)}_{,\,klm}\chi^{||(1),\,klm}
\nn\\
&
+\frac{7}{9}\nabla^2\chi^{||(1)}_{,\,k}\nabla^2\nabla^2\chi^{||(1),\,k}
-\frac{1}{2}\chi^{||(1)}_{,\,klm}\nabla^2\chi^{||(1),\,klm}
+\frac{5}{18}\nabla^2\nabla^2\chi^{||(1)}\nabla^2\nabla^2\chi^{||(1)}
\nn\\
&
+\frac{1}{3}\chi^{||(1)'}_{,\,kl}\nabla^2\chi^{||(1)',\,kl}
+\frac{1}{3}\nabla^2\chi^{||(1)'}_{,\,k}\nabla^2\chi^{||(1)',\,k}
\Big]
+(i\leftrightarrow j)
\ .
\el
}
We have checked that  $V_{S\, ij}$
satisfies the vector condition (\ref{chiVec0}).

Finally, a combination
[(\ref{Evo2ndSsNoTr1RD})$-D_{ij}$(\ref{Evo2ndSsChi1RD})
$-$(\ref{Evo2ndSsVec2RD})] gives the  equation of 2nd-order tensor
\be \label{Evo2ndSsTen1RD}
\chi^{\top(2)''}_{S\,ij}
+2\frac{\beta +1}{\tau } \chi^{\top(2)'}_{S\,ij}
-\nabla^2\chi^{\top(2)}_{S\,ij}
= 2 J_{S\,ij} ,
\ee
where  the   inhomogeneous term
{\allowdisplaybreaks
\bl\label{2ndTensorSourceRD}
J_{S\,ij}
&
\equiv
\bar S_{S\,ij}
-\frac{3}{2}D_{ij}\nabla^{-2}\nabla^{-2}\bar S_{S\,kl}^{\, ,\,kl}
-\nabla^{-2}\bar S_{S\,kj,\,i}^{,\,k}
-\nabla^{-2}\bar S_{S\,ki,j}^{,\,k}
+2\nabla^{-2}\nabla^{-2}\bar S_{S\,kl,\,ij}^{\, ,\,kl}
\nn\\
&=
D_{ij}\Big[
-\frac{3}{4}\phi^{(1)}\phi^{(1)}
-\frac{1}{3}\phi^{(1)}\nabla^2\chi^{||(1)}
-2\phi^{(1)}_{,\,k}\chi^{||(1),\,k}
+\frac{1}{6}\nabla^2\chi^{||(1)}\nabla^2\chi^{||(1)}
\nn\\
&
 +\frac{1}{16}\chi^{||(1)}_{,\,kl}\chi^{||(1),\,kl}
-\frac{1}{3}\chi^{||(1)}_{,\,k}\nabla^2\chi^{||(1),\,k}
\Big]
+\frac{2}{M_{\rm Pl}^2}\varphi^{(1)}_{, i}\varphi^{(1)}_{, j}
-\frac{2}{3M_{\rm Pl}^2}\varphi^{(1), k}\varphi^{(1)}_{, k}\delta_{ij}
\nn\\
&
+\frac{2}{M_{\rm Pl}}\frac{\sqrt{2 (\beta +1) (\beta +2)} (2 \beta +1) }{\tau ^2}
\Big(
\chi^{||(1)}_{,ij}\varphi^{(1)}
-\frac{1}{3}\varphi^{(1)}\nabla^2\chi^{||(1)}\delta_{ij}
\Big)
\nn\\
&
+\frac{2}{M_{\rm Pl}}\frac{\sqrt{2 (\beta +1) (\beta +2)} }{\tau }
\Big(
\chi^{||(1)}_{,ij}\varphi^{(1)'}
-\frac{1}{3}\varphi^{(1)'}\nabla^2\chi^{||(1)}\delta_{ij}
\Big)
+2\phi^{(1)}\phi^{(1)}_{,\,ij}
\nn\\
&
-\frac{2}{3}\phi^{(1)}\nabla^2\phi^{(1)}\delta_{ij}
-12\frac{\beta +1}{\tau }\phi^{(1)'}\chi^{||(1)}_{,\,ij}
+4\frac{\beta +1}{\tau }\phi^{(1)'}\nabla^2\chi^{||(1)}\delta_{ij}
-\phi^{(1)'}\chi^{||(1)'}_{,\,ij}
\nn\\
&
+\frac{1}{3}\phi^{(1)'}\nabla^2\chi^{||(1)'}\delta_{ij}
-6\phi^{(1)''}\chi^{||(1)}_{,\,ij}
+2\phi^{(1)''}\nabla^2\chi^{||(1)}\delta_{ij}
+\frac{7}{3}\phi^{(1)}_{,\,ij}\nabla^2\chi^{||(1)}
\nn\\
&
-\frac{7}{9}\nabla^2\phi^{(1)}\nabla^2\chi^{||(1)}\delta_{ij}
+\frac{1}{3}\phi^{(1)}\nabla^2\chi^{||(1)}_{,\,ij}
-\frac{1}{9}\phi^{(1)}\nabla^2\nabla^2\chi^{||(1)}\delta_{ij}
+3\phi^{(1)}_{,\,k}\chi^{||(1),\,k}_{,\,ij}
\nn\\
&
-\phi^{(1)}_{,\,k}\nabla^2\chi^{||(1),\,k}\delta_{ij}
+2\chi^{||(1),\,k}\phi^{(1)}_{,\,kij}
-\frac{2}{3}\chi^{||(1),\,k}\nabla^2\phi^{(1)}_{,\,k}\delta_{ij}
+4\chi^{||(1)}_{,\,ij}\nabla^2\phi^{(1)}
\nn\\
&
-\frac{4}{3}\nabla^2\phi^{(1)}\nabla^2\chi^{||(1)}\delta_{ij}
+\frac{2}{3}\chi^{||(1)}_{,\,ij}\nabla^2\nabla^2\chi^{||(1)}
-\frac{2}{9}\nabla^2\chi^{||(1)}\nabla^2\nabla^2\chi^{||(1)}\delta_{ij}
\nn\\
&
-\frac{7}{18}\nabla^2\chi^{||(1)}_{,\,i}\nabla^2\chi^{||(1)}_{,\,j}
+\chi^{||(1)}_{,\,kij}\nabla^2\chi^{||(1),\,k}
-\frac{11}{54}\nabla^2\chi^{||(1),\,k}\nabla^2\chi^{||(1)}_{,\,k}\delta_{ij}
\nn\\
&
-\frac{1}{9}\chi^{||(1),\,k}\nabla^2\nabla^2\chi^{||(1)}_{,\,k}\delta_{ij}
+\frac{1}{3}\chi^{||(1)}_{,\,k}\nabla^2\chi^{||(1),\,k}_{,\,ij}
-\frac{1}{2}\chi^{||(1),\,kl}_{,\,i}\chi^{||(1)}_{,\,klj}
+\frac{1}{6}\chi^{||(1),\,klm}\chi^{||(1)}_{,\,klm}\delta_{ij}
\nn\\
&
+\chi^{||(1)',\,k}_{,\,i}\chi^{||(1)'}_{,\,kj}
-\frac{1}{3}\chi^{||(1)',\,kl}\chi^{||(1)'}_{,\,kl}\delta_{ij}
-\frac{2}{3}\chi^{||(1)'}_{,\,ij}\nabla^2\chi^{||(1)'}
+\frac{2}{9}\nabla^2\chi^{||(1)'}\nabla^2\chi^{||(1)'}\delta_{ij}
    \nn\\
    &
-\frac{3}{2}D_{ij}\nabla^{-2}\Big[
\frac{1}{3M_{\rm Pl}^2}\varphi^{(1), k}\varphi^{(1)}_{, k}
-\frac{2}{3M_{\rm Pl}}\frac{\sqrt{2 (\beta +1) (\beta +2)} (2 \beta +1) }{\tau ^2}\varphi^{(1)}\nabla^2\chi^{||(1)}
\nn\\
&
-\frac{2}{3M_{\rm Pl}}\frac{\sqrt{2 (\beta +1) (\beta +2)} }{\tau }\varphi^{(1)'}\nabla^2\chi^{||(1)}
-\frac{5}{3}\phi^{(1)}\nabla^2\phi^{(1)}
-8\frac{\beta +1}{\tau }\phi^{(1)'}\nabla^2\chi^{||(1)}
\nn\\
&
-\frac{2}{3}\phi^{(1)'}\nabla^2\chi^{||(1)'}
-4\phi^{(1)''}\nabla^2\chi^{||(1)}
-\frac{4}{9}\phi^{(1),k}\nabla^2\chi^{||(1)}_{,\,k}
+\frac{4}{3}\phi^{(1)}_{,\,kl}\chi^{||(1),kl}
\nn\\
&
+\frac{1}{54} \nabla^2\chi^{||(1),k}\nabla^2\chi^{||(1)}_{,\,k}
+\frac{11}{36}\chi^{||(1)}_{,\,kl}\nabla^2\chi^{||(1),kl}
+\frac{1}{6}\chi^{||(1)',kl}\chi^{||(1)'}_{,\,kl}
-\frac{1}{9}\nabla^2\chi^{||(1)'}\nabla^2\chi^{||(1)'}
\Big]
\nn\\
&
-\frac{3}{2}D_{ij}\nabla^{-2}\nabla^{-2}
\Big[
\frac{2}{M_{\rm Pl}^2}\varphi^{(1)}_{, k}\nabla^2\varphi^{(1),k}
+\frac{2}{M_{\rm Pl}^2}\nabla^2\varphi^{(1)}\nabla^2\varphi^{(1)}
\nn\\
&
+\frac{2}{M_{\rm Pl}}\frac{\sqrt{2 (\beta +1) (\beta +2)} (2 \beta +1) }{\tau ^2}
\Big(
\varphi^{(1)}\nabla^2\nabla^2\chi^{||(1)}
+2\varphi^{(1),k}\nabla^2\chi^{||(1)}_{,k}
+\chi^{||(1)}_{,kl}\varphi^{(1),kl}
\Big)
\nn\\
&
+\frac{2}{M_{\rm Pl}}\frac{\sqrt{2 (\beta +1) (\beta +2)} }{\tau }
\Big(
\varphi^{(1)'}\nabla^2\nabla^2\chi^{||(1)}
+2\varphi^{(1)',k}\nabla^2\chi^{||(1)}_{,k}
+\chi^{||(1)}_{,kl}\varphi^{(1)',kl}
\Big)
\nn\\
&
+2\phi^{(1),\,k}\nabla^2\phi^{(1)}_{,\,k}
+2\phi^{(1)}\nabla^2\nabla^2\phi^{(1)}
-12\frac{\beta +1}{\tau }\phi^{(1)',\,kl}\chi^{||(1)}_{,\,kl}
\nn\\
&
+12\frac{\beta +1}{\tau }\nabla^2\phi^{(1)'}\nabla^2\chi^{||(1)}
-\phi^{(1)',\,kl}\chi^{||(1)'}_{,\,kl}
+\nabla^2\phi^{(1)'}\nabla^2\chi^{||(1)'}
-6\phi^{(1)'',\,kl}\chi^{||(1)}_{,\,kl}
\nn\\
&
+6\nabla^2\phi^{(1)''}\nabla^2\chi^{||(1)}
+\frac{4}{3}\nabla^2\nabla^2\chi^{||(1)}\nabla^2\phi^{(1)}
+\frac{11}{3}\nabla^2\chi^{||(1)}_{,\,k}\nabla^2\phi^{(1),k}
+\frac{2}{3}\phi^{(1),\,k}\nabla^2\nabla^2\chi^{||(1)}_{,\,k}
\nn\\
&
-3\phi^{(1)}_{,\,klm}\chi^{||(1),klm}
+\frac{7}{9}\nabla^2\chi^{||(1)}_{,\,k}\nabla^2\nabla^2\chi^{||(1),k}
+\frac{5}{18}\nabla^2\nabla^2\chi^{||(1)}\nabla^2\nabla^2\chi^{||(1)}
\nn\\
&
-\frac{1}{2}\chi^{||(1)}_{,\,klm}\nabla^2\chi^{||(1),\,klm}
+\frac{1}{3}\chi^{||(1)'}_{,\,kl}\nabla^2\chi^{||(1)',\,kl}
+\frac{1}{3}\nabla^2\chi^{||(1)'}_{,\,k}\nabla^2\chi^{||(1)',\,k}
\Big]
\nn
\\
    &
-\partial_i\Big[
\frac{1}{3}\phi^{(1)}\nabla^2\chi^{||(1)}_{,\,j}
+2\phi^{(1),\,k}\chi^{||(1)}_{,\,kj}
+\frac{1}{3}\chi^{||(1)}_{,\,kj}\nabla^2\chi^{||(1),\,k}
\Big]
\nn\\
&
-\partial_j\Big[
\frac{1}{3}\phi^{(1)}\nabla^2\chi^{||(1)}_{,\,i}
+2\phi^{(1),\,k}\chi^{||(1)}_{,\,ki}
+\frac{1}{3}\chi^{||(1)}_{,\,ki}\nabla^2\chi^{||(1),\,k}
\Big]
\nn\\
&
+\partial_i\partial_j\nabla^{-2}\Big[
-24\frac{\beta +1}{\tau }\phi^{(1)'}\nabla^2\chi^{||(1)}
-2\phi^{(1)'}\nabla^2\chi^{||(1)'}
-12\phi^{(1)''}\nabla^2\chi^{||(1)}
\nn\\
&
+\frac{2}{3}\phi^{(1)}\nabla^2\nabla^2\chi^{||(1)}
+4\phi^{(1),\,kl}\chi^{||(1)}_{,\,kl}
+\frac{1}{6}\nabla^2\chi^{||(1)}_{,\,k}\nabla^2\chi^{||(1),\,k}
+\frac{2}{3}\chi^{||(1)}_{,\,kl}\nabla^2\chi^{||(1),\,kl}
\Big]
\nn\\
&
-\partial_i\nabla^{-2}\Big[
\frac{2}{M_{\rm Pl}^2}\varphi^{(1)}_{, j}\nabla^2\varphi^{(1)}
+\frac{2}{M_{\rm Pl}}\frac{\sqrt{2 (\beta +1) (\beta +2)} (2 \beta +1) }{\tau ^2}
\Big(
\varphi^{(1)}\nabla^2\chi^{||(1)}_{,j}
\nn\\
&
+\chi^{||(1)}_{,kj}\varphi^{(1),k}
\Big)
+\frac{2}{M_{\rm Pl}}\frac{\sqrt{2 (\beta +1) (\beta +2)} }{\tau }
\Big(
\varphi^{(1)'}\nabla^2\chi^{||(1)}_{,j}
+\chi^{||(1)}_{,kj}\varphi^{(1)',k}
\Big)
\nn\\
&
+2\phi^{(1)}\nabla^2\phi^{(1)}_{,\,j}
-12\frac{\beta +1}{\tau }\phi^{(1)',\,k}\chi^{||(1)}_{,\,kj}
-12\frac{\beta +1}{\tau }\phi^{(1)'}\nabla^2\chi^{||(1)}_{,\,j}
\nn\\
&
-\phi^{(1)'}\nabla^2\chi^{||(1)'}_{,\,j}
-\phi^{(1)',\,k}\chi^{||(1)'}_{,\,kj}
-6\phi^{(1)''}\nabla^2\chi^{||(1)}_{,\,j}
-6\phi^{(1)'',\,k}\chi^{||(1)}_{,\,kj}
\nn\\
&
+\frac{4}{3}\nabla^2\chi^{||(1)}_{,\,j}\nabla^2\phi^{(1)}
-\frac{5}{3}\phi^{(1),\,k}\nabla^2\chi^{||(1)}_{,\,kj}
-3\phi^{(1)}_{,\,kl}\chi^{||(1),\,kl}_{,\,j}
\nn\\
&
+\frac{5}{18}\nabla^2\chi^{||(1)}_{,\,j}\nabla^2\nabla^2\chi^{||(1)}
-\frac{1}{2}\chi^{||(1)}_{,\,klj}\nabla^2\chi^{||(1),\,kl}
+\frac{1}{3}\chi^{||(1)'}_{,\,kj}\nabla^2\chi^{||(1)',\,k}
\Big]
\nn\\
&
-\partial_j\nabla^{-2}\Big[
\frac{2}{M_{\rm Pl}^2}\varphi^{(1)}_{, i}\nabla^2\varphi^{(1)}
+\frac{2}{M_{\rm Pl}}\frac{\sqrt{2 (\beta +1) (\beta +2)} (2 \beta +1) }{\tau ^2}
\Big(
\varphi^{(1)}\nabla^2\chi^{||(1)}_{,i}
\nn\\
&
+\chi^{||(1)}_{,ki}\varphi^{(1),k}
\Big)
+\frac{2}{M_{\rm Pl}}\frac{\sqrt{2 (\beta +1) (\beta +2)} }{\tau }
\Big(
\varphi^{(1)'}\nabla^2\chi^{||(1)}_{,i}
+\chi^{||(1)}_{,ki}\varphi^{(1)',k}
\Big)
\nn\\
&
+2\phi^{(1)}\nabla^2\phi^{(1)}_{,\,i}
-12\frac{\beta +1}{\tau }\phi^{(1)',\,k}\chi^{||(1)}_{,\,ki}
-12\frac{\beta +1}{\tau }\phi^{(1)'}\nabla^2\chi^{||(1)}_{,\,i}
-\phi^{(1)'}\nabla^2\chi^{||(1)'}_{,\,i}
\nn\\
&
-\phi^{(1)',\,k}\chi^{||(1)'}_{,\,ki}
-6\phi^{(1)''}\nabla^2\chi^{||(1)}_{,\,i}
-6\phi^{(1)'',\,k}\chi^{||(1)}_{,\,ki}
+\frac{4}{3}\nabla^2\chi^{||(1)}_{,\,i}\nabla^2\phi^{(1)}
-\frac{5}{3}\phi^{(1),\,k}\nabla^2\chi^{||(1)}_{,\,ki}
\nn\\
&
-3\phi^{(1)}_{,\,kl}\chi^{||(1),\,kl}_{,\,i}
+\frac{5}{18}\nabla^2\chi^{||(1)}_{,\,i}\nabla^2\nabla^2\chi^{||(1)}
-\frac{1}{2}\chi^{||(1)}_{,\,kli}\nabla^2\chi^{||(1),\,kl}
+\frac{1}{3}\chi^{||(1)'}_{,\,ki}\nabla^2\chi^{||(1)',\,k}
\Big]
\nn\\
&
+\partial_i\partial_j\nabla^{-2}\nabla^{-2}\Big[
\frac{4}{M_{\rm Pl}^2}\varphi^{(1)}_{, k}\nabla^2\varphi^{(1),k}
+\frac{4}{M_{\rm Pl}^2}\nabla^2\varphi^{(1)}\nabla^2\varphi^{(1)}
\nn\\
&
+\frac{4}{M_{\rm Pl}}\frac{\sqrt{2 (\beta +1) (\beta +2)} (2 \beta +1) }{\tau ^2}
\Big(
\varphi^{(1)}\nabla^2\nabla^2\chi^{||(1)}
+2\varphi^{(1),k}\nabla^2\chi^{||(1)}_{,k}
+\chi^{||(1)}_{,kl}\varphi^{(1),kl}
\Big)
\nn\\
&
+\frac{4}{M_{\rm Pl}}\frac{\sqrt{2 (\beta +1) (\beta +2)} }{\tau }
\Big(
\varphi^{(1)'}\nabla^2\nabla^2\chi^{||(1)}
+2\varphi^{(1)',k}\nabla^2\chi^{||(1)}_{,k}
+\chi^{||(1)}_{,kl}\varphi^{(1)',kl}
\Big)
\nn\\
&
+4\phi^{(1),\,k}\nabla^2\phi^{(1)}_{,\,k}
+4\phi^{(1)}\nabla^2\nabla^2\phi^{(1)}
-24\frac{\beta +1}{\tau }\phi^{(1)',\,kl}\chi^{||(1)}_{,\,kl}
+24\frac{\beta +1}{\tau }\nabla^2\phi^{(1)'}\nabla^2\chi^{||(1)}
\nn\\
&
-2\phi^{(1)',\,kl}\chi^{||(1)'}_{,\,kl}
+2\nabla^2\phi^{(1)'}\nabla^2\chi^{||(1)'}
-12\phi^{(1)'',\,kl}\chi^{||(1)}_{,\,kl}
+12\nabla^2\phi^{(1)''}\nabla^2\chi^{||(1)}
\nn\\
&
+\frac{8}{3}\nabla^2\nabla^2\chi^{||(1)}\nabla^2\phi^{(1)}
+\frac{22}{3}\nabla^2\chi^{||(1)}_{,\,k}\nabla^2\phi^{(1),\,k}
+\frac{4}{3}\phi^{(1),\,k}\nabla^2\nabla^2\chi^{||(1)}_{,\,k}
\nn\\
&
-6\phi^{(1)}_{,\,klm}\chi^{||(1),\,klm}
+\frac{14}{9}\nabla^2\chi^{||(1)}_{,\,k}\nabla^2\nabla^2\chi^{||(1),\,k}
-\chi^{||(1)}_{,\,klm}\nabla^2\chi^{||(1),\,klm}
\nn\\
&
+\frac{5}{9}\nabla^2\nabla^2\chi^{||(1)}\nabla^2\nabla^2\chi^{||(1)}
+\frac{2}{3}\chi^{||(1)'}_{,\,kl}\nabla^2\chi^{||(1)',\,kl}
+\frac{2}{3}\nabla^2\chi^{||(1)'}_{,\,k}\nabla^2\chi^{||(1)',\,k}
\Big] .
\el
}

Beside  the 2nd-order perturbed Einstein equation,
we also need the equation of the 2nd-order perturbed scalar field,
 which follows from the expansion of \eqref{fequvarphi}  to the 2nd-order
\bl \label{2ndeqinflation}
\varphi^{(2)''}
 &  + 2 \frac{a'}{a}\varphi^{(2)'}
- \nabla^2\varphi^{(2)}
+   a^2 V_{,\varphi \varphi} \varphi^{(2)}
+\frac{1}{2}  \varphi^{(0)'}\gamma_{k}^{(2)' k}
\nn\\
&
+ \gamma_{k}^{(1)' k}\varphi^{(1)'}
-  \varphi^{(0)'}\gamma^{(1)' kl} \gamma_{kl}^{(1)}
+ 2 \gamma^{(1)kl}_{,k}\varphi^{(1)}_{,l}
\nn\\
&
+ 2 \gamma^{(1)kl}\varphi^{(1)}_{,kl}
- \gamma_{m,k}^{(1) m}\varphi^{(1),k}
+   a^2 V_{,\varphi \varphi\varphi} \varphi^{(1)}\varphi^{(1)}
=0.
\el
This is equivalent to the 2nd-order energy conservation \eqref{Enconsv2nd0}.
(See \eqref{enerConsev2} in  Appendix \ref{sec:perturbedGT}.)
For  the power-law inflation with the scalar-scalar coupling,
\eqref{2ndeqinflation}  becomes
\bl\label{enerConsev2inflation}
\varphi^{(2)''}_S
& +\frac{2\beta+ 2}{\tau }\varphi^{(2)'}_S
+\frac{2 (\beta +2) (2 \beta +1)}{\tau ^2}\varphi^{(2)}_S
-\nabla^2\varphi^{(2)}_S
\nn \\
& = -A_S  +3M_{\rm Pl}\frac{\sqrt{2(\beta +1) (\beta+2)}}{\tau}\phi^{(2)'}_S ,
\el
where
\bl\label{AS2inflation}
A_S
\equiv &
-\frac{2 \sqrt{2} (\beta +2)^2 (2 \beta +1)}{\sqrt{(\beta +1)
 (\beta +2)} \tau ^2 M_{\text{Pl}}}\varphi^{(1)}\varphi^{(1)}
-6\varphi^{(1)'}\phi^{(1)'}
+2\varphi_{,k}^{(1)}\phi^{(1),k}
-4\phi^{(1)}\nabla^2\varphi^{(1)}
\nn\\
&
-12\frac{\sqrt{2} \sqrt{(\beta +1) (\beta +2)} M_{\text{Pl}}}{\tau }\phi^{(1)} \phi^{(1)'}
+\frac{4}{3}\varphi_{,k }^{(1)}\nabla^2\chi^{||(1),k}
+2\varphi_{,kl}^{(1)}\chi^{||(1),kl}
\nn\\
&
-\frac{2}{3}\nabla^2\varphi^{(1)}\nabla^2\chi^{||(1)}
-\frac{\sqrt{2} \sqrt{(\beta +1) (\beta +2)}
M_{\text{Pl}}}{\tau }\chi^{||(1),kl}\chi^{||(1)'}_{,kl}
\nn\\
&
+\frac{1}{3}\frac{\sqrt{2} \sqrt{(\beta +1) (\beta +2)}
 M_{\text{Pl}}}{\tau }\nabla^2\chi^{||(1)}\nabla^2\chi^{||(1)'}.
\el

We  find
that the 2nd-order trace evolution equation (\ref{Evo2ndSsTr2RD})
can be written as
\bl\label{relation12nd}
(\ref{Evo2ndSsTr2RD})
=&
-\frac{1}{6}(\ref{Ein2th003RD})
-\frac{\tau }{6 (\beta +1)}\frac{d}{d\tau}(\ref{Ein2th003RD})
+\frac{\tau }{6 (\beta +1)}\nabla^2(\ref{MoConstr2ndv3RD2})
\nn\\
&
+\frac{\beta +2}{3 M_{\rm Pl}\sqrt{2 (\beta +1) (\beta +2)} }
(\ref{enerConsev2inflation})  ,
\el
and the traceless scalar equation (\ref{Evo2ndSsChi1RD})  can be written as
\bl\label{relation22nd}
(\ref{Evo2ndSsChi1RD})
=& \nabla^{-2}\Big[\frac{1}{2}(\ref{Ein2th003RD})
+\frac{\tau}{2(\beta +1)} \frac{d}{d \tau}(\ref{Ein2th003RD})
-\frac{\tau}{2(\beta +1)} \nabla^{2}(\ref{MoConstr2ndv3RD2})
+\frac{3}{2} \frac{d}{d \tau}(\ref{MoConstr2ndv3RD2})
\nn\\
&
+\frac{3(\beta +1)}{\tau}(\ref{MoConstr2ndv3RD2})
-\frac{\beta +2}{M_{\rm Pl}\sqrt{2 (\beta +1)
(\beta +2)} }(\ref{enerConsev2inflation})
\Big] .
\el
Thus,  we can also use the conservation equation \eqref{enerConsev2inflation}
and the constraints \eqref{Ein2th003RD} \eqref{MoConstr2ndv3RD2}
to solve the scalar perturbations,
and the solutions will satisfy the evolution equations
of the scalars automatically.

\section{Solutions of the 2nd-order perturbations}

Now we solve the 2nd-order equations given in the last section.
First,
consider the 2nd-order tensor equation (\ref{Evo2ndSsTen1RD}) which is inhomogeneous.
The homogeneous solution has the same form of
the 1st-order solution \eqref{Fourier} and can be absorbed by it,
and the inhomogeneous solution  is
{
\bl  \label{Fourier2nd2nd}
\chi^{\top(2)}_{S\,ij}({\bf x},\tau)
= &
\frac{1}{(2\pi)^{3/2}}
\int d^3k   e^{i \,\bf{k}\cdot\bf{x}}
\bar I_{S\, ij}(k, \tau) ,
\el
}
where
\bl
\bar I_{S\, ij}(k, \tau)
&=
-\frac{\pi}{2 i}(-\tau )^{-\beta -\frac{1}{2}}
\bigg[
H_{\beta +\frac{1}{2}}^{(1)}(-k \tau ) \int^{\tau }
 J_{S\,ij \, k}(\tau_1) (-\tau_1)^{\beta +\frac{3}{2}}
 H_{\beta +\frac{1}{2}}^{(2)}(-k \tau_1) \, d\tau_1
\nn\\
&
-H_{\beta +\frac{1}{2}}^{(2)}(-k \tau ) \int^{\tau }
 J_{S\,ij \, k}(\tau_1) (-\tau_1)^{\beta +\frac{3}{2}}
 H_{\beta +\frac{1}{2}}^{(1)}(-k \tau_1) \, d\tau_1
\bigg],
\el
where $J_{S\,ij \, k}$ is the $k$ mode of the source (\ref{2ndTensorSourceRD}),
consisting of products of 1st-order solutions.

Next,
the 2nd-order vector equation \eqref{Evo2ndSsVec2RD} is also  inhomogeneous.
{In fact this can be replaced by the vector constraint equation  \eqref{MoCons2ndCurlRD1}}
which has the  solution
\bl\label{vect2ndsol}
\chi^{\perp(2) }_{S\,ij} =  2\int^\tau d\tau_1
 \Big[ \nabla^{-2} (M_{S\,i,j}(\tau_1)  + M_{S\,j,i}(\tau_1))
 -2\partial_i\partial_j\nabla^{-2}\nabla^{-2} M_{S\,l}^{,\,l}(\tau_1) \Big] ,
\el
where the effective source $M_{S\,i}$ is given by \eqref{MSi1}.
As can be checked,
this solution is the   inhomogeneous solution of \eqref{Evo2ndSsVec2RD}.
The 2nd-order vector modes  are  nonvanishing,
generated by the products of the 1st-order perturbations
during the inflation.
The 2nd order vector  $\chi^{\perp(2) }_{S\,ij}$ is a wave
since  $M_{S\,i}$ in  \eqref{vect2ndsol}
are products of the 1st-order perturbation waves.

Then, we solve the 2nd-order scalars.
The 2nd-order scalar field equation (\ref{enerConsev2inflation})
of  $\varphi^{(2)}_S$ is not closed
as it still contains the scalar metric perturbation $\phi^{(2)}_S $.
With the help of the constraints (\ref{Ein2th003RD})  \eqref{MoConstr2ndv3RD2}
we put  \eqref{enerConsev2inflation} into a  closed,  third-order equation
\bl\label{Sephi1d3}
\varphi^{(2)'''}_S &
+\frac{ 3\beta +4}{\tau}\varphi^{(2)''}_S
+\frac{ 2 \beta^2+2 \beta-2}{\tau^{2}}\varphi^{(2)'}_S
-\frac{4 \beta ^2+10 \beta +4}{\tau^{3}}\varphi^{(2)}_S
\nn\\
& -\nabla^2(\varphi^{(2)'}_S+\frac{\beta+2}{\tau}\varphi^{(2)}_{S})
 = M_{\rm Pl} \frac{\sqrt{ (\beta +2)(\beta +1)}}{\sqrt2(\beta +1)}
\Big(M_{S\,l}^{,\,l}-E_S^{\,\prime}\Big) ,
\el
which is similar to the 1st-order equation \eqref{phi1d3},
but contains the inhomogeneous source terms,
with $M_{S\, l}$ in \eqref{MSi1} and $E_S$ in \eqref{ES1RD}.
The homogeneous solutions of \eqref{Sephi1d3}
can be  absorbed into  the 1st-order solution \eqref{varphi1solv2},
and the inhomogeneous solution is obtained as the following
\bl\label{Sevarphi1solv2}
&
\varphi^{(2) }_{S\, k } (\tau)
= -\frac{\pi}{4 i}
M_{\rm Pl}
\frac{\sqrt{ (\beta +2)(\beta +1)}}{\sqrt2(\beta +1)}
(-\tau)^{-\beta -2}
\nn\\
&
\times
\bigg\{
\int^{\tau }  d\tau_2
(-\tau_2)^{\frac{3}{2}}
H_{\beta +\frac{3}{2}}^{(1)}(-k\tau_2 )
\int^{\tau_2}  d\tau_1
(-\tau_1) ^{\beta +\frac{3}{2}}
\Big(
\big[M_{S\,l}^{,\,l}(\tau_1)\big]_{ k}
-E_{S\, k}^{\,\prime}(\tau_1)
\Big)
H_{\beta +\frac{3}{2}}^{(2)}(-k \tau_1)
\nn\\
&
-\int^{\tau }  d\tau_2
(-\tau_2)^{\frac{3}{2}}
H_{\beta +\frac{3}{2}}^{(2)}(-k\tau_2 )
\int^{\tau_2}  d\tau_1
(-\tau_1) ^{\beta +\frac{3}{2}}
\Big(
\big[M_{S\,l}^{,\,l}(\tau_1)\big]_{ k}
-E_{S\,  k}^{\,\prime}(\tau_1)
\Big)
 H_{\beta +\frac{3}{2}}^{(1)}(-k \tau_1)
\bigg\} ,
\el
which  contains integrations of the source terms,
and contains no residual gauge mode.

Now we proceed to derive the  scalar metric perturbations
 $\phi^{(2)}$ and $\chi^{||(2)}$.
Analogous to \eqref{phiChiZeta} \eqref{2zeta3rd},
we denote
\be\label{SephiChiZeta}
\zeta^{(2)}\equiv
-2\phi^{(2)}
-\frac{1}{3}\nabla^2\chi^{||(2)} .
\ee
By combination of the conservation  (\ref{enerConsev2inflation})
and the constraints (\ref{Ein2th003RD}) (\ref{MoConstr2ndv3RD2}),
we get   the $k$ mode equation
\bl\label{Sezeta3rd}
\zeta^{(2)'''}_{S \, k }
 + & 3\frac{(\beta   +2  )}{\tau}\zeta^{(2)''}_{S \, k }
+\frac{ 2 (\beta +1) (\beta +3)}{\tau^2} \zeta^{(2)'}_{S \, k }
   +k^2  \zeta^{(2)'}_{S \, k }
   +  \frac{(\beta +2)}{\tau} k^2  \zeta^{(2)}_{S \, k }
\nn\\
=&
\frac{1}{ k^2}M_{S\,l}^{'',\,l}
+\frac{3(\beta +2)}{k^2\tau}M_{S\,l}^{',\,l}
+\Big(\frac{2 (\beta +1)(\beta +3)}{k^2 \tau^2 }+1 \Big) M_{S\,l}^{,\,l}
\nn\\
&
+\frac{\sqrt{2} \sqrt{\beta ^2+3 \beta +2}}{ M_{\rm Pl}\, \tau}A_S
+\frac{(\beta +2)}{\tau}E_S,
\el
where the rhs denote the $k$ mode of the source,
absent in the 1st-order  \eqref{2zeta3rd}.
In fact, the moment constraint \eqref{MoConstr2ndv3RD2} gives
a simple equation
\[
-\zeta^{(2)'}_S
 =
\frac{1}{M_{\rm Pl}}\frac{\sqrt{2(\beta +1) (\beta +2)}}{\tau} \varphi^{(2)}_S
+\nabla^{-2}M_{S\,l}^{,\,l} \, .
\]
With $\varphi^{(2)}$ given in \eqref{Sevarphi1solv2}
and $\nabla^{-2}M_{S\,l}^{,\,l}$ given by (\ref{MSkkScalar}),
integrating the above  directly,  we obtain
\bl \label{SeSereclpch222}
&
\zeta^{(2)}_{S \, k}
= -\frac{\pi}{4 i}(\beta +2)
\Big[
\int^\tau
(-\tau_3)^{-\beta -3}
d\tau_3
\Big]
\times
\nn\\
&
\times
\bigg[
\int^{\tau }
  d\tau_2
(-\tau_2)^{\frac{3}{2}}
H_{\beta +\frac{3}{2}}^{(1)}(-k\tau_2 )
\int^{\tau_2}
 d\tau_1
(-\tau_1) ^{\beta +\frac{3}{2}}
\Big(
\big[M_{S\,l}^{,\,l}(\tau_1)\big]_{k}
-E_{S\,  k}^{\,\prime}(\tau_1)
\Big)
H_{\beta +\frac{3}{2}}^{(2)}(-k \tau_1)
\nn\\
&
-\int^{\tau }
 d\tau_2
(-\tau_2)^{\frac{3}{2}}
H_{\beta +\frac{3}{2}}^{(2)}(-k\tau_2 )
\int^{\tau_2}
 d\tau_1
(-\tau_1)^{\beta +\frac{3}{2}}
\Big(
\big[M_{S\,l}^{,\,l}(\tau_1)\big]_{ k}
-E_{S \,  k}^{\,\prime}(\tau_1)
\Big)
 H_{\beta +\frac{3}{2}}^{(1)}(-k \tau_1)
\bigg]
\nn\\
&
+\frac{\pi}{4 i}(\beta +2)\int^{\tau } d\tau_2
\Big[
\int^{\tau_2}
d\tau_3
(-\tau_3)^{-\beta -3}
\Big]
\times
\nn\\
&
\times
\bigg[
(-\tau_2)^{\frac{3}{2}}
H_{\beta +\frac{3}{2}}^{(1)}(-k\tau_2 )
\int^{\tau_2}
 d\tau_1
(-\tau_1) ^{\beta +\frac{3}{2}}
\Big(
\big[M_{S\,l}^{,\,l}(\tau_1)\big]_{k}
-E_{S\, k}^{\,\prime}(\tau_1)
\Big)
H_{\beta +\frac{3}{2}}^{(2)}(-k \tau_1)
\nn\\
&
-(-\tau_2)^{\frac{3}{2}}
H_{\beta +\frac{3}{2}}^{(2)}(-k\tau_2 )
\int^{\tau_2}
 d\tau_1
(-\tau_1)^{\beta +\frac{3}{2}}
\Big(
\big[M_{S\,l}^{,\,l}(\tau_1)\big]_{ k}
-E_{S\,  k}^{\,\prime}(\tau_1)
\Big)
 H_{\beta +\frac{3}{2}}^{(1)}(-k \tau_1)
\bigg]
\nn\\
&
-\int^\tau d\tau_1 \Big[\nabla^{-2}M_{S\,l}^{,\,l}\Big]_{ k} \ ,
\el
where $\Big[\nabla^{-2}M_{S\,l}^{,\,l}\Big]_{k}$
 is the Fourier mode of $\nabla^{-2}M_{S\,l}^{,\,l}$ given by  (\ref{MSkkScalar}),
$E_{S\, k}$ is the $k$ mode of $E_S$ given by (\ref{ES1RD}).
With this result,
we can give the separate
solutions of  $\phi^{(2)}_{S}$ and $\chi^{||(2)}_{S}$.
Writing (\ref{Ein2th003RD}) in  $k$-space and
using \eqref{SephiChiZeta}, we have
\bl\label{SeEin1st003}
\phi^{(2)'}_{S\,  k}
=&
\frac{k^2 }{6(\beta +1)} \tau\zeta^{(2)}_{S{\, k}}
-\frac{1}{M_{\rm Pl}}\frac{\sqrt{2(\beta +1)
(\beta +2)}}{6(\beta +1)}\varphi^{(2)'}_{S\, k}
\nn\\
&
+  \frac{1}{M_{\rm Pl}}\frac{\sqrt{2(\beta +1)
  (\beta +2)}(2\beta+1)}{6(\beta +1) \tau}\varphi^{(2)}_{S\, k}
-\frac{\tau}{6(\beta +1)} E_{S\, k}
.
\el
where the functions on the rhs are known.
By  integration we  obtain
\bl\label{Sephi1kk222}
&
\phi^{(2)}_{S \,  k}
= \bigg\{
-\frac{\pi}{48  i }
\frac{(\beta +2) }{(\beta +1)}
k^2\tau ^2
\Big[
\int^{\tau} d\tau_3
(-\tau_3)^{-\beta -3}
\Big]
+\frac{\pi}{48 i }
\frac{(\beta +2) }{(\beta +1)}k^2
\frac{1}{\beta (-\tau )^{\beta}}
\nn\\
&
+\frac{\pi}{24 i }
\frac{ (\beta +2)}{(\beta +1)}
(-\tau)^{-\beta -2}
+\frac{\pi}{24 i }
\frac{(\beta +2)(2 \beta +1)}{(\beta +1)}
\Big[
\int^\tau d\tau_3
(-\tau_3)^{-\beta -3}
\Big]
\bigg\}
\times
\nn\\
&
\times
\bigg[
\int^{\tau }
  d\tau_2
(-\tau_2)^{\frac{3}{2}}
H_{\beta +\frac{3}{2}}^{(1)}(-k\tau_2 )
\int^{\tau_2}
 d\tau_1
(-\tau_1) ^{\beta +\frac{3}{2}}
\Big(
\big[M_{S\,l}^{,\,l}(\tau_1)\big]_{ k}
-E_{S\,  k}^{\,\prime}(\tau_1)
\Big)
H_{\beta +\frac{3}{2}}^{(2)}(-k \tau_1)
\nn\\
&
-\int^{\tau }
 d\tau_2
(-\tau_2)^{\frac{3}{2}}
H_{\beta +\frac{3}{2}}^{(2)}(-k\tau_2 )
\int^{\tau_2}
 d\tau_1
(-\tau_1)^{\beta +\frac{3}{2}}
\Big(
\big[M_{S\,l}^{,\,l}(\tau_1)\big]_{ k}
-E_{S\,  k}^{\,\prime}(\tau_1)
\Big)
 H_{\beta +\frac{3}{2}}^{(1)}(-k \tau_1)
\bigg]
    \nn\\
    &
+\Big[
\frac{\pi}{48 i}
\frac{(\beta +2) }{(\beta +1)}
k^2\tau ^2
-\frac{\pi}{24 i}
\frac{(\beta +2)(2 \beta +1)}{(\beta +1)}
\Big]\int^{\tau }
  d\tau_2
\bigg\{
\Big[
\int^{\tau_2} d\tau_3
(-\tau_3)^{-\beta -3}
\Big]
\times
\nn\\
&
\times
\bigg[
(-\tau_2)^{\frac{3}{2}}
H_{\beta +\frac{3}{2}}^{(1)}(-k\tau_2 )
\int^{\tau_2}
 d\tau_1
(-\tau_1) ^{\beta +\frac{3}{2}}
\Big(
\big[M_{S\,l}^{,\,l}(\tau_1)\big]_{ k}
-E_{S\,  k}^{\,\prime}(\tau_1)
\Big)
H_{\beta +\frac{3}{2}}^{(2)}(-k \tau_1)
\nn\\
&
-(-\tau_2)^{\frac{3}{2}}
H_{\beta +\frac{3}{2}}^{(2)}(-k\tau_2 )
\int^{\tau_2}
 d\tau_1
(-\tau_1)^{\beta +\frac{3}{2}}
\Big(
\big[M_{S\,l}^{,\,l}(\tau_1)\big]_{ k}
-E_{S\,  k}^{\,\prime}(\tau_1)
\Big)
 H_{\beta +\frac{3}{2}}^{(1)}(-k \tau_1)
\bigg]
\bigg\}
    \nn\\
    &
-\frac{\pi}{48 i}
\frac{(\beta +2) }{(\beta +1)}k^2
\int^{\tau }
  d\tau_2
\bigg\{
\Big[
\int^{\tau_2} d\tau_3
(-\tau_3)^{-\beta -1}
\Big]
\times
\nn\\
&
\times
\bigg[
(-\tau_2)^{\frac{3}{2}}
H_{\beta +\frac{3}{2}}^{(1)}(-k\tau_2 )
\int^{\tau_2}
 d\tau_1
(-\tau_1) ^{\beta +\frac{3}{2}}
\Big(
\big[M_{S\,l}^{,\,l}(\tau_1)\big]_{  k}
-E_{S\,  k}^{\,\prime}(\tau_1)
\Big)
H_{\beta +\frac{3}{2}}^{(2)}(-k \tau_1)
\nn\\
&
-(-\tau_2)^{\frac{3}{2}}
H_{\beta +\frac{3}{2}}^{(2)}(-k\tau_2 )
\int^{\tau_2}
 d\tau_1
(-\tau_1)^{\beta +\frac{3}{2}}
\Big(
\big[M_{S\,l}^{,\,l}(\tau_1)\big]_{k}
-E_{S\, k}^{\,\prime}(\tau_1)
\Big)
 H_{\beta +\frac{3}{2}}^{(1)}(-k \tau_1)
\bigg]
\bigg\}
\nn\\
&
-\frac{k^2 \tau ^2}{12(\beta +1)}
\int^{\tau} d\tau_1
\Big[
\nabla^{-2}M_{S\,l}^{,\,l}(\tau_1)
\Big]_{ k}
+\frac{k^2 }{12(\beta +1)}
\int^{\tau} d\tau_1
\Big\{
\tau_1 ^2
\Big[
\nabla^{-2}M_{S\, l}^{,\,l}(\tau_1)
\Big]_{ k}
\Big\}
\nn\\
&
-\frac{1}{6(\beta +1)} \int^\tau d\tau_1
\, \tau_1 E_{S\, k}(\tau_1)
.
\el
Then by \eqref{SephiChiZeta}, we obtain
{\allowdisplaybreaks
\bl\label{Sechi1kkfinal22}
&
\chi^{||(2)}_{S \,  k}
= \bigg\{
-\frac{\pi}{8 i}
\frac{(\beta +2) }{(\beta +1)}
\tau ^2
\Big[
\int^{\tau} d\tau_3
(-\tau_3)^{-\beta -3}
\Big]
+\frac{\pi}{8 i}
\frac{(\beta +2) }{(\beta +1)}
 \frac{1}{\beta (-\tau )^{\beta}}
\nn\\
&
+\frac{\pi}{4 i}
\frac{ (\beta +2)}{(\beta +1)}
k^{-2}(-\tau)^{-\beta -2}
-\frac{\pi}{4 i}
\frac{(\beta +2)^2}{(\beta +1)}k^{-2}
\Big[
\int^\tau d\tau_3
(-\tau_3)^{-\beta -3}
\Big]
\bigg\}
\times
\nn\\
&
\times
\bigg[
\int^{\tau }
  d\tau_2
(-\tau_2)^{\frac{3}{2}}
H_{\beta +\frac{3}{2}}^{(1)}(-k\tau_2 )
\int^{\tau_2}
 d\tau_1
(-\tau_1) ^{\beta +\frac{3}{2}}
\Big(
\big[M_{S\,l}^{,\,l}(\tau_1)\big]_{  k}
-E_{S\,  k}^{\,\prime}(\tau_1)
\Big)
H_{\beta +\frac{3}{2}}^{(2)}(-k \tau_1)
\nn\\
&
-\int^{\tau }
 d\tau_2
(-\tau_2)^{\frac{3}{2}}
H_{\beta +\frac{3}{2}}^{(2)}(-k\tau_2 )
\int^{\tau_2}
 d\tau_1
(-\tau_1)^{\beta +\frac{3}{2}}
\Big(
\big[M_{S\,l}^{,\,l}(\tau_1)\big]_{  k}
-E_{S\, k}^{\,\prime}(\tau_1)
\Big)
 H_{\beta +\frac{3}{2}}^{(1)}(-k \tau_1)
\bigg]
\nn\\
&
+\Big[
\frac{\pi}{8 i}
\frac{(\beta +2) }{(\beta +1)}
\tau ^2
+\frac{\pi}{4 i}
\frac{(\beta +2)^2}{(\beta +1)}k^{-2}
\Big]\int^{\tau }
  d\tau_2
\bigg\{
\Big[
\int^{\tau_2} d\tau_3
(-\tau_3)^{-\beta -3}
\Big]
\times
\nn\\
&
\times
\bigg[
(-\tau_2)^{\frac{3}{2}}
H_{\beta +\frac{3}{2}}^{(1)}(-k\tau_2 )
\int^{\tau_2}
 d\tau_1
(-\tau_1) ^{\beta +\frac{3}{2}}
\Big(
\big[M_{S\,l}^{,\,l}(\tau_1)\big]_{  k}
-E_{S\,  k}^{\,\prime}(\tau_1)
\Big)
H_{\beta +\frac{3}{2}}^{(2)}(-k \tau_1)
\nn\\
&
-(-\tau_2)^{\frac{3}{2}}
H_{\beta +\frac{3}{2}}^{(2)}(-k\tau_2 )
\int^{\tau_2}
 d\tau_1
(-\tau_1)^{\beta +\frac{3}{2}}
\Big(
\big[M_{S\,l}^{,\,l}(\tau_1)\big]_{  k}
-E_{S\, k}^{\,\prime}(\tau_1)
\Big)
 H_{\beta +\frac{3}{2}}^{(1)}(-k \tau_1)
\bigg]
\bigg\}
\nn\\
&
-\frac{\pi}{8 i}
\frac{(\beta +2) }{(\beta +1)}
\int^{\tau }
  d\tau_2
\bigg\{
\Big[
\int^{\tau_2} d\tau_3
(-\tau_3)^{-\beta -1}
\Big]
\times
\nn\\
&
\times
\bigg[
(-\tau_2)^{\frac{3}{2}}
H_{\beta +\frac{3}{2}}^{(1)}(-k\tau_2 )
\int^{\tau_2}
 d\tau_1
(-\tau_1) ^{\beta +\frac{3}{2}}
\Big(
\big[M_{S\,l}^{,\,l}(\tau_1)\big]_{  k}
-E_{S\, k}^{\,\prime}(\tau_1)
\Big)
H_{\beta +\frac{3}{2}}^{(2)}(-k \tau_1)
\nn\\
&
-(-\tau_2)^{\frac{3}{2}}
H_{\beta +\frac{3}{2}}^{(2)}(-k\tau_2 )
\int^{\tau_2}
 d\tau_1
(-\tau_1)^{\beta +\frac{3}{2}}
\Big(
\big[M_{S\,l}^{,\,l}(\tau_1)\big]_{  k}
-E_{S\, k}^{\,\prime}(\tau_1)
\Big)
 H_{\beta +\frac{3}{2}}^{(1)}(-k \tau_1)
\bigg]
\bigg\}
\nn\\
&
+\big(-\frac{ \tau ^2}{2(\beta +1)}-\frac{3}{k^2}\big)
\int^{\tau} d\tau_1\Big[\nabla^{-2}M_{S\,l}^{,\,l}(\tau_1)\Big]_{  k}
+\frac{1}{2(\beta +1)}
\int^{\tau} d\tau_1
 \tau_1 ^2 \Big[ \nabla^{-2}M_{S\,l}^{,\,l}(\tau_1) \Big]_{k}
\nn\\
&
-\frac{1}{(\beta +1)k^2}\int^\tau d\tau_1
\big[\tau_1 E_{S\,  k}(\tau_1) \big] \, .
\el
}
In the 2nd-order solutions \eqref{Sephi1kk222} \eqref{Sechi1kkfinal22},
 a constant gauge mode has been dropped.
Similar to the 1st order,
$\chi^{||(2)}_{ S}$ is $k^{-2}$ divergent at $k\rightarrow0$,
but the metric perturbation $D_{ij}\chi^{||(1)}_{S}$ itself
is actually convergent.

\section{2nd-order residual gauge modes}

The 2nd-order perturbation solutions in Sect 6 generally
contain some residual gauge modes between synchronous coordinates
which need to be identified.
A general 2nd-order coordinate transformation involves
a 2nd-order transformation vector $\xi^{(2)\mu}$,
and the products of a 1st-order transformation vector $\xi^{(1)\mu}$ as well.
For a general RW spacetime
the synchronous-synchronous  transformations of the metric perturbations
up to the 2nd-order are given in  Ref.~~\cite{WangZhang2017,WangZhang2018}.
For the power-law inflation and  the scalar-scalar coupling,
$\xi^{(1)\mu}$ is given in  (\ref{xi0trans}) (\ref{gi0}),
and  $\xi^{(2)\mu}$ is given
in (\ref{alpha2RD1}) (\ref{beta2RD1}) (\ref{d2RD1})
in Appendix \ref{sec:gauge_transform},
and the transformation
of the 2nd-order metric perturbations are given by the following
{\allowdisplaybreaks
\bl\label{phi2TransRD}
\bar \phi^{(2)}_S  = & \phi^{(2)}_S
-\frac{\beta  (\beta +1)}{(-\tau )^{2 (\beta +2)}}  A^{(1)}A^{(1)}
+(-\tau )^{-2 (\beta +1)}\bigg[
-\frac{2}{3}A^{(1)}\nabla^2A^{(1)}
-\frac{1}{3}A^{(1),\,l}A^{(1)}_{,\,l}
\bigg]
\nn\\
&
+\frac{4(\beta +1)}{(-\tau )^{\beta +2}} \phi^{(1)}A^{(1)}
-2 (-\tau )^{-\beta -1}\phi^{(1)'}A^{(1)}
\nn\\
&
-\frac{(\beta +1)}{(-\tau )^{\beta +2}}
 \l[\int^\tau \frac{d\tau' }{(-\tau')^{\beta +1}}\r]
\bigg[
-\frac{4}{3}A^{(1)} \nabla^2  A^{(1)}
- A^{(1)}_{,\,l}A^{(1),\,l}
\bigg]
\nn\\
&
+ \l[ \int^\tau  \frac{d\tau'}{(-\tau')^{\beta +1}}
    \int^{\tau'}\frac{d\tau'' }{(-\tau'')^{\beta +1}}\r]
\bigg[
\frac{2}{3}A^{(1),\,l}\nabla^2A^{(1)}_{,\,l}
+ \frac{2}{3} A^{(1),\,lm}A^{(1)}_{,\,lm}
\bigg]
\nn\\
&
+\l[\int^\tau \frac{d\tau' }{(-\tau')^{\beta +1}}\r]^2
\bigg[
-\frac{1}{3}A^{(1),\,l}  \nabla^2  A^{(1)}_{,\,l}
-\frac{2}{3}A^{(1)}_{,\,lm} A^{(1),\,lm}
\bigg]
\nn\\
&
+ \l[ \int^\tau \frac{d\tau'}{(-\tau')^{\beta +1}} \r]
\bigg[
\frac{2}{3} A^{(1),\,lm}C^{||(1)}_{,\,lm}
+ \frac{2}{3}A^{(1),\,l}\nabla^2C^{||(1)}_{,\,l}
-\frac{4}{3} \phi^{(1)} \nabla^2  A^{(1)}
\nn\\
&
-2\phi^{(1)}_{,\,l}A^{(1),\,l}
+\frac{2}{3}D_{\,lm}\chi^{||(1)}A^{(1),\,lm}
\bigg]
\nn\\
&
+\nabla^2A^{(1)} \int^\tau\frac{4\phi^{(1)}(\tau',{\bf x})}{3(-\tau')^{\beta +1}}d\tau'
+A^{(1)}_{,\,l}\int^\tau\frac{4\phi^{(1),\,l}(\tau',{\bf x})}{3(-\tau')^{\beta +1}}d\tau'
\nn\\
&
-A^{(1),\,lm}\int^\tau\frac{2D_{\,lm}\chi^{||(1)}(\tau',{\bf x})}{3(-\tau')^{\beta +1}}d\tau'
-A^{(1),\,l}\int^\tau\frac{4\nabla^2\chi^{||(1)}_{,\,l}(\tau',{\bf x})}{9(-\tau')^{\beta +1}}d\tau'
\nn\\
&
-\frac{(\beta +1)}{(-\tau )^{\beta +2}} A^{(2)}
+\frac{1}{3}\nabla^2A^{(2)}\int^\tau \frac{d\tau'}{(-\tau')^{\beta +1}}
+\frac{1}{3}\nabla^2C^{||(2)}
,
\el
}
where $A^{(2)}$ and $C^{||(2)}$  in the last line
are due to the vector $\xi^{(2)\mu}$,
and all $A^{(1)}$ has implicitly absorbed a factor $l_0^{-1}$,
and
{\allowdisplaybreaks
\bl\label{chi||2transRD}
 \bar\chi^{||(2)}_S
&=
  \chi^{||(2)}_S
+ (-\tau )^{-2 (\beta +1)}   \bigg[
A^{(1)}A^{(1)}
+2\nabla^{-2}\big(A^{(1)}\nabla^2A^{(1)}\big)
+3\nabla^{-2}\nabla^{-2}\big(
A^{(1),\,lm}A^{(1)}_{,\,lm}
\nn\\
&
-\nabla^2A^{(1)}\nabla^2A^{(1)}\big)
\bigg]
-4 \frac{(\beta +1)}{(-\tau )^{\beta +2}}
 \l[\int^\tau  \frac{d\tau' }{(-\tau' )^{\beta +1}}\r]
  \bigg[
2\nabla^{-2}\big(A^{(1)}\nabla^2A^{(1)}\big)
\nn\\
&
+3\nabla^{-2}\nabla^{-2}\big(
A^{(1),\,lm}A^{(1)}_{,\,lm}
-\nabla^2A^{(1)}\nabla^2A^{(1)}\big)
\bigg]
\nn\\
&
-\frac{(\beta +1)}{(-\tau )^{\beta +2}}
  \bigg[
8\nabla^{-2}\big(A^{(1)}\nabla^2C^{||(1)}\big)
+\nabla^{-2}\nabla^{-2}\big(
-4A^{(1)}\nabla^2\nabla^2\chi^{||(1)}
\nn\\
&
-6A^{(1),\,lm}D_{\,lm}\chi^{||(1)}
-8 A^{(1),\,l}\nabla^2\chi^{||(1)}_{,\,l}
+12A^{(1),\,lm}C^{||(1)}_{,\,lm}
-12\nabla^2A^{(1)}\nabla^2C^{||(1)}
\big)
\bigg]
\nn\\
&
- (-\tau )^{-\beta -1}   \nabla^{-2} \nabla^{-2}\bigg[
2A^{(1)}\nabla^2\nabla^2\chi^{||(1)'}
+3A^{(1),\,lm}D_{\,lm}\chi^{||(1)'}
+4A^{(1),\,m}\nabla^2\chi^{(1)'}_{,\,m} \bigg]
\nn \\
&
-2\l[ \int^\tau  \frac{d\tau'}{(-\tau' )^{\beta +1}}
  \int^{\tau'}\frac{d\tau'' }{(-\tau'' )^{\beta +1}}  \r]
  A^{(1),\,l}A^{(1)}_{,\,l}
+\l[\int^\tau  \frac{d\tau' }{(-\tau' )^{\beta +1}}\r]^2   \bigg[
2A^{(1)}_{,\,l}A^{(1),\,l}
\nn \\
&
-2\nabla^{-2}\big(A^{(1),\,l}\nabla^2A^{(1)}_{,\,l}\big)
+3\nabla^{-2}\nabla^{-2}\big(
\nabla^2A^{(1),\,l}\nabla^2A^{(1)}_{,\,l}
-A^{(1),\,lmn}A^{(1)}_{,\,lmn}\big)
\bigg]
\nn\\
&
+ \l[\int^\tau \frac{d\tau' }{(-\tau' )^{\beta +1}} \r] \bigg[
2A^{(1),\,l}C^{||(1)}_{,\,l}
+2\nabla^{-2}\big(4\phi^{(1)}\nabla^2A^{(1)}
+A^{(1),\,lm}D_{\,lm}\chi^{||(1)}
\nn\\
&
-2A^{(1),\,l}\nabla^2C^{||(1)}_{,\,l}
\big)
+\nabla^{-2}\nabla^{-2}\big(
-2A^{(1)}_{,l}\nabla^2\nabla^2\chi^{||(1),\,l}
-9A^{(1),\,lmn}D_{\,lm}\chi^{||(1)}_{,n}
\nn\\
&
-8A^{(1),\,lm}\nabla^2\chi^{||(1)}_{,lm}
-4\nabla^2\chi^{||(1)}_{,\,l}\nabla^2A^{(1),\,l}
-6D_{\,lm}\chi^{||(1)}\nabla^2A^{(1),\,lm}
+12\phi^{(1),\,lm}A^{(1)}_{,\,lm}
\nn\\
&
-12\nabla^2\phi^{(1)}\nabla^2A^{(1)}
+6\nabla^2A^{(1),\,l}\nabla^2C^{||(1)}_{,\,l}
-6A^{(1),\,lmn}C^{||(1)}_{,\,lmn}
\big)
\bigg]
\nn\\
&
+   \bigg[ 2C^{||(1)}_{,\,l}C^{||(1),\,l}
+\nabla^{-2}\big(8\phi^{(1)}\nabla^2C^{||(1)}
+2 C^{||(1),\,lm}D_{\,lm}\chi^{||(1)}
-2 C^{||(1),\,l}\nabla^2C^{||(1)}_{,\,l}\big)
\nn\\
&
+\nabla^{-2}\nabla^{-2}\big(
12\phi^{(1),\,lm}C^{||(1)}_{,\,lm}
-12\nabla^2\phi^{(1)}\nabla^2C^{||(1)}
-2C^{||(1)}_{,\,l}\nabla^2\nabla^2\chi^{||(1),\,l}
\nn\\
&
-9 C^{||(1),\,lmn}D_{\,lm}\chi^{||(1)}_{,n}
-8 C^{||(1),\,lm}\nabla^2\chi^{||(1)}_{,\,lm}
-4\nabla^2\chi^{||(1)}_{,\,l}\nabla^2C^{||(1),\,l}
\nn\\
&
-6D_{\,lm}\chi^{||(1)}\nabla^2C^{||(1),\,lm}
+3\nabla^2C^{||(1),\,l}\nabla^2C^{||(1)}_{,\,l}
-3C^{||(1),\,lmn}C^{||(1)}_{,\,lmn}
\big)
\bigg]
\nn\\
&
-4 \nabla^{-2}\bigg[
\nabla^2A^{(1)} \int^\tau\frac{2\phi^{(1)}(\tau',{\bf x})}{(-\tau' )^{\beta +1}}d\tau'
+A^{(1)}_{,\,l} \int^\tau\frac{2\phi^{(1)}(\tau',{\bf x})^{,\,l}}{(-\tau' )^{\beta +1}}d\tau'
\nn\\
&
- A^{(1),\,lm} \int^\tau \frac{D_{\,lm}\chi^{||(1)}(\tau',{\bf x})}{(-\tau' )^{\beta +1}}d\tau'
- A^{(1),\,l}  \int^\tau \frac{2\nabla^2\chi^{||(1),\,l}(\tau',{\bf x})}{3(-\tau' )^{\beta +1}}d\tau'
\bigg]
\nn\\
&
-2 A^{(2)}  \int^\tau \frac{d\tau' }{(-\tau' )^{\beta +1}}
-2C^{||(2)}     ,
\el
}
and
{
\allowdisplaybreaks
\bl\label{chiPerp2TransRD}
\bar\chi^{\perp(2)}_{S\,ij}
&=
\chi^{\perp(2)}_{S\,ij}
+(-\tau )^{-2 (\beta +1)}\bigg[
-2\partial_i\nabla^{-2}\big( A^{(1)}_{,j}\nabla^2A^{(1)} \big)
+\partial_i\partial_j\nabla^{-2}\big(
A^{(1),\,l}A^{(1)}_{,\,l}\big)
\nn\\
&
-2\partial_i\partial_j\nabla^{-2}\nabla^{-2}\big(
A^{(1),\,lm}A^{(1)}_{,\,lm}
-\nabla^2A^{(1)}\nabla^2A^{(1)}
\big)
\bigg]
\nn\\
& - 4 \frac{(\beta +1)}{(-\tau )^{\beta +2}}
 \l[ \int^\tau \frac{d\tau' }{(-\tau' )^{\beta +1}}  \r]
 \bigg[
-2\partial_i\nabla^{-2}\big( A^{(1)}_{,j}\nabla^2A^{(1)} \big)
+\partial_i\partial_j\nabla^{-2}\big(
A^{(1),\,l}A^{(1)}_{,\,l}
\big)
\nn\\
&
-2\partial_i\partial_j\nabla^{-2}\nabla^{-2}\big(
A^{(1),\,lm}A^{(1)}_{,\,lm}
-\nabla^2A^{(1)}\nabla^2A^{(1)}
\big)
\bigg]
\nn\\
& -2 \frac{(\beta +1)}{(-\tau )^{\beta +2}}
 \bigg[
2\partial_i\nabla^{-2}\big( 2A^{(1),\,l}C^{||(1)}_{,\,lj}
-2A^{(1)}_{,j}\nabla^2C^{||(1)} -\chi^{(1),\,l}_{lj}A^{(1)}
-\chi^{(1)}_{lj}A^{(1),\,l} \big)
\nn\\
&
+2\partial_i\partial_j\nabla^{-2}\nabla^{-2}\big(
2\nabla^2A^{(1)}\nabla^2C^{||(1)}
-2A^{(1),\,lm}C^{||(1)}_{,\,lm}
+\chi^{(1),\,lm}_{lm}A^{(1)}
\nn\\
&
+\chi^{(1)}_{lm}A^{(1),\,lm}
+2\chi^{(1),\,l}_{lm}A^{(1),m}
\big)
\bigg]
- (-\tau )^{-\beta -1} \bigg[
2\partial_i\nabla^{-2}\big( \chi^{(1)',\,l}_{lj}A^{(1)}
 +\chi^{(1)'}_{lj}A^{(1),\,l} \big)
\nn \\
&
-2\partial_i\partial_j\nabla^{-2}\nabla^{-2}\big(
\chi^{(1)',\,lm}_{lm}A^{(1)}
+\chi^{(1)'}_{lm}A^{(1),\,lm}
+2\chi^{(1)',\,l}_{lm}A^{(1),m}
\big)
\bigg]
\nn \\
&
+ \l[\int^\tau \frac{d\tau' }{(-\tau' )^{\beta +1}}\r]^2\bigg[
2\partial_i\nabla^{-2}\big( A^{(1)}_{,\,lj}\nabla^2A^{(1),\,l} \big)
-\partial_i\partial_j\nabla^{-2}\big(
A^{(1),\,lm}A^{(1)}_{,\,lm}
\big)
\nn\\
&
+2\partial_i\partial_j\nabla^{-2}\nabla^{-2}\big(
A^{(1),\,lmn}A^{(1)}_{,\,lmn}
-\nabla^2A^{(1),\,l}\nabla^2A^{(1)}_{,\,l}
\big)
\bigg]
\nn \\
&
- \l[\int^\tau \frac{d\tau' }{(-\tau' )^{\beta +1}} \r]\bigg[
2\partial_i\nabla^{-2}\big(
4\phi^{(1)}_{,j}\nabla^2A^{(1)}
-4\phi^{(1),\,l}A^{(1)}_{,\,lj}
+\chi^{(1),\,l}_{lj,n}A^{(1),n}
+2\chi^{(1)}_{lj,n}A^{(1),\,ln}
\nn\\
&
+\chi^{(1),m}_{lm}A^{(1),\,l}_{,j}
+\chi^{(1)}_{lm}A^{(1),\,lm}_{,j}
+\chi^{(1)}_{lj}\nabla^2A^{(1),\,l}
+2A^{(1),\,ln}C^{||(1)}_{,\,lnj}
-2A^{(1)}_{,\,lj}\nabla^2C^{||(1),\,l}
\big)
\nn\\
&
-2\partial_i\partial_j\nabla^{-2}\nabla^{-2}\big(
4\nabla^2\phi^{(1)}\nabla^2A^{(1)}
-4\phi^{(1),\,lm}A^{(1)}_{,\,lm}
+\chi^{(1),\,lm}_{lm,n}A^{(1),n}
+3\chi^{(1)}_{lm,n}A^{(1),\,lmn}
\nn\\
&
+4\chi^{(1),\,l}_{lm,n}A^{(1),\,mn}
+2\chi^{(1),\,l}_{lm}\nabla^2A^{(1),m}
+2\chi^{(1)}_{lm}\nabla^2A^{(1),\,lm}
+2A^{(1),\,lmn}C^{||(1)}_{,\,lmn}
\nn\\
&
-2\nabla^2A^{(1),\,l}\nabla^2C^{||(1)}_{,\,l}
\big)
\bigg]
+\bigg[
-2\partial_i\nabla^{-2}\big(
4\phi^{(1)}_{,j}\nabla^2C^{||(1)}
-4\phi^{(1),\,l}C^{||(1)}_{,\,lj}
+\chi^{(1),\,l}_{lj,n}C^{||(1),n}
\nn\\
&
+2\chi^{(1)}_{lj,n}C^{||(1),\,ln}
+\chi^{(1),m}_{lm}C^{||(1),\,l}_{,j}
+\chi^{(1)}_{lm}C^{||(1),\,lm}_{,j}
+\chi^{(1)}_{lj}\nabla^2C^{||(1),\,l}
+C^{||(1),\,ln}C^{||(1)}_{,\,lnj}
\nn\\
&
-C^{||(1)}_{,\,lj}\nabla^2C^{||(1),\,l}
\big)
+2\partial_i\partial_j\nabla^{-2}\nabla^{-2}\big(
4\nabla^2\phi^{(1)}\nabla^2C^{||(1)}
-4\phi^{(1),\,lm}C^{||(1)}_{,\,lm}
+\chi^{(1),\,lm}_{lm,n}C^{||(1),n}
\nn\\
&
+4\chi^{(1),\,l}_{lm,n}C^{||(1),\,mn}
+3\chi^{(1)}_{lm,n}C^{||(1),\,lmn}
+2\chi^{(1),\,l}_{lm}\nabla^2C^{||(1),m}
+2\chi^{(1)}_{lm}\nabla^2C^{||(1),\,lm}
\nn\\
&
+C^{||(1),\,lmn}C^{||(1)}_{,\,lmn}
-\nabla^2C^{||(1),\,l}\nabla^2C^{||(1)}_{,\,l}
\big)
\bigg]
+\partial_i\bigg[
A^{(1),\,l}\int^\tau \frac{2D_{lj}\chi^{||(1)}(\tau',{\bf x})}{(-\tau' )^{\beta +1}}d\tau'
\nn \\
&
-A^{(1)}_{,j}\int^\tau \frac{4\phi^{(1)}(\tau',{\bf x})}{(-\tau' )^{\beta +1}}d\tau'
\bigg]
+\partial_i\partial_j\nabla^{-2}\bigg[
\nabla^2A^{(1)} \int^\tau \frac{4\phi^{(1)}(\tau',{\bf x})}{(-\tau' )^{\beta +1}}d\tau'
\nn\\
&
+A^{(1)}_{,\,l}\int^\tau \frac{4\phi^{(1),\,l}(\tau',{\bf x})}{(-\tau' )^{\beta +1}}d\tau'
-A^{(1),\,lm}\int^\tau \frac{2D_{lm}\chi^{||(1)}(\tau' ,{\bf x})}{(-\tau' )^{\beta +1}}d\tau
\nn\\
&
-A^{(1),\,l}\int^\tau \frac{4\nabla^2\chi^{||(1)}_{,\,l}(\tau' ,{\bf x})}{3(-\tau' )^{\beta +1}}d\tau' \bigg]
  -C^{\perp(2)}_{i,j}
\nn\\
&
+(i \leftrightarrow j ) \ ,
\el
}
and
{\allowdisplaybreaks
\bl\label{chiT2transRD}
\bar\chi^{\top(2)}_{S\,ij}
&=
\chi^{\top(2)}_{S\,ij}
+(-\tau )^{-2 (\beta +1)}\bigg[
\delta_{ij}\nabla^{-2}\big(
A^{(1),\,lm}A^{(1)}_{,\,lm}
-\nabla^2A^{(1)}\nabla^2A^{(1)}
\big)
+4\nabla^{-2}\big(
A^{(1)}_{,ij}\nabla^2A^{(1)}
\nn\\
&
-A^{(1),\,l}_{,i}A^{(1)}_{,\,lj}
\big)
+\partial_i\partial_j\nabla^{-2}\nabla^{-2}\big(
A^{(1),\,lm}A^{(1)}_{,\,lm}
-\nabla^2A^{(1)}\nabla^2A^{(1)}
\big)
\bigg]
\nn\\
&
-4 \frac{(\beta +1)}{(-\tau )^{\beta +2}}
 \l[\int^\tau  \frac{d\tau' }{(-\tau' )^{\beta +1}}\r]\bigg[
\delta_{ij}\nabla^{-2}\big(
A^{(1),\,lm}A^{(1)}_{,\,lm}
-\nabla^2A^{(1)}\nabla^2A^{(1)}
\big)
\nn\\
&
+4\nabla^{-2}\big(
A^{(1)}_{,ij}\nabla^2A^{(1)}
-A^{(1),\,l}_{,i}A^{(1)}_{,\,lj}
\big)
+\partial_i\partial_j\nabla^{-2}\nabla^{-2}\big(
A^{(1),\,lm}A^{(1)}_{,\,lm}
-\nabla^2A^{(1)}\nabla^2A^{(1)}
\big)
\bigg]
\nn\\
& -2 \frac{(\beta +1)}{(-\tau )^{\beta +2}}
\bigg[
-\delta_{ij}\nabla^{-2}\big(
\chi^{(1),\,lm}_{lm}A^{(1)}
+\chi^{(1)}_{lm}A^{(1),\,lm}
+2\chi^{(1),\,l}_{lm}A^{(1),m}
+2\nabla^2A^{(1)}\nabla^2C^{||(1)}
\nn\\
&
-2A^{(1),\,lm}C^{||(1)}_{,\,lm}
\big)
-2\chi^{(1)}_{ij}A^{(1)}
+2\partial_i\nabla^{-2}\big(
\chi^{(1),\,l}_{lj}A^{(1)}
+\chi^{(1)}_{lj}A^{(1),\,l}
\big)
+2\partial_j\nabla^{-2}\big(
\chi^{(1),\,l}_{li}A^{(1)}
\nn\\
&
+\chi^{(1)}_{li}A^{(1),\,l}
\big)
+4\nabla^{-2}\big(
A^{(1)}_{,ij}\nabla^2C^{||(1)}
+C^{||(1)}_{,ij}\nabla^2A^{(1)}
-A^{(1),\,l}_{,i}C^{||(1)}_{,\,lj}
-A^{(1),\,l}_{,j}C^{||(1)}_{,\,li}
\big)
\nn\\
&
-\partial_i\partial_j\nabla^{-2}\nabla^{-2}\big(
\chi^{(1),\,lm}_{lm}A^{(1)}
+\chi^{(1)}_{lm}A^{(1),\,lm}
+2\chi^{(1),\,l}_{lm}A^{(1),m}
+2A^{(1),\,lm}C^{||(1)}_{,\,lm}
\nn\\
&
-2\nabla^2A^{(1)}\nabla^2C^{||(1)}
\big)
\bigg]
- (-\tau )^{-\beta -1} \bigg[
\delta_{ij}\nabla^{-2}\big(\chi^{(1)',\,lm}_{lm}A^{(1)}
+\chi^{(1)'}_{lm}A^{(1),\,lm}
+2\chi^{(1)',\,l}_{lm}A^{(1),m}
\big)
\nn\\
&
+2\chi^{(1)'}_{ij}A^{(1)}
-2\partial_i\nabla^{-2}\big(
\chi^{(1)',\,l}_{lj}A^{(1)}
+\chi^{(1)'}_{lj}A^{(1),\,l}
\big)
-2\partial_j\nabla^{-2}\big(
\chi^{(1)',\,l}_{li}A^{(1)}
+\chi^{(1)'}_{li}A^{(1),\,l}
\big)
\nn\\
&
+\partial_i\partial_j\nabla^{-2}\nabla^{-2}\big(
\chi^{(1)',\,lm}_{lm}A^{(1)}
+\chi^{(1)'}_{lm}A^{(1),\,lm}
+2\chi^{(1)',\,l}_{lm}A^{(1),m}
\big) \bigg]
\nn\\
&
-\l[\int^\tau  \frac{d\tau' }{(-\tau' )^{\beta +1}}\r]^2\bigg[
\delta_{ij}\nabla^{-2}\big(
A^{(1),\,lmn}A^{(1)}_{,\,lmn}
-\nabla^2A^{(1),\,l}\nabla^2A^{(1)}_{,\,l}
\big)
\nn\\
&
+2\nabla^{-2}\big(-2A^{(1),\,lm}_{,i}A^{(1)}_{,\,lmj}
+2A^{(1)}_{,\,lij}\nabla^2A^{(1),\,l}
\big)
+\partial_i\partial_j\nabla^{-2}\nabla^{-2}\big(
A^{(1),\,lmn}A^{(1)}_{,\,lmn}
\nn\\
&
-\nabla^2A^{(1),\,l}\nabla^2A^{(1)}_{,\,l}
\big)
\bigg]
+\l[\int^\tau  \frac{d\tau' }{(-\tau' )^{\beta +1}}\r]\bigg[
\delta_{ij}\nabla^{-2}\big(
4\phi^{(1),\,lm}A^{(1)}_{,\,lm}
-4\nabla^2\phi^{(1)}\nabla^2A^{(1)}
\nn\\
&
-\chi^{(1),\,lm}_{lm,n}A^{(1),n}
-4\chi^{(1),\,l}_{lm,n}A^{(1),mn}
+\chi^{(1)}_{lm,n}A^{(1),\,lmn}
-2\chi^{(1),\,l}_{lm}\nabla^2A^{(1),m}
\nn\\
&
+2 A^{(1),\,lm}\nabla^2\chi^{(1)}_{lm}
-2A^{(1),\,lmn}C^{||(1)}_{,\,lmn}
+2\nabla^2A^{(1),\,l}\nabla^2C^{||(1)}_{,\,l}
\big)
-2\chi^{(1)}_{ij,\,l}A^{(1),\,l}
\nn\\
&
-2\chi^{(1)}_{li}A^{(1),\,l}_{,j}
-2\chi^{(1)}_{lj}A^{(1),\,l}_{,i}
-4\nabla^{-2}\big(
2\phi^{(1),\,l}_{,i}A^{(1)}_{,\,lj}
+2\phi^{(1),\,l}_{,j}A^{(1)}_{,\,li}
-2\phi^{(1)}_{,ij}\nabla^2A^{(1)}
\nn\\
&
-2A^{(1)}_{,ij}\nabla^2\phi^{(1)}
-A^{(1),\,lm}_{,i}C^{||(1)}_{,\,lmj}
-A^{(1),\,lm}_{,j}C^{||(1)}_{,\,lmi}
+A^{(1)}_{,\,lij}\nabla^2C^{||(1),\,l}
+C^{||(1)}_{,\,lij}\nabla^2A^{(1),\,l}
\big)
    \nn\\
    &
+2\partial_i\nabla^{-2}\big(
\chi^{(1),\,l}_{lj,m}A^{(1),m}
+2\chi^{(1)}_{lj,m}A^{(1),\,lm}
+\chi^{(1),m}_{lm}A^{(1),\,l}_{,j}
+\chi^{(1)}_{lm}A^{(1),\,lm}_{,j}
+\chi^{(1)}_{lj}\nabla^2A^{(1),\,l}
\big)
    \nn\\
    &
+2\partial_j\nabla^{-2}\big(
\chi^{(1),\,l}_{li,m}A^{(1),m}
+2\chi^{(1)}_{li,m}A^{(1),\,lm}
+\chi^{(1),m}_{lm}A^{(1),\,l}_{,i}
+\chi^{(1)}_{lm}A^{(1),\,lm}_{,i}
+\chi^{(1)}_{li}\nabla^2A^{(1),\,l}
\big)
\nn\\
&
+\partial_i\partial_j\nabla^{-2}\nabla^{-2}\big(
4\phi^{(1),\,lm}A^{(1)}_{,\,lm}
-4\nabla^2\phi^{(1)}\nabla^2A^{(1)}
-\chi^{(1),\,lm}_{lm,n}A^{(1),n}
-4\chi^{(1),\,l}_{lm,n}A^{(1),mn}
\nn\\
&
-7\chi^{(1)}_{lm,n}A^{(1),\,lmn}
-2\chi^{(1),\,l}_{lm}\nabla^2A^{(1),m}
-4\chi^{(1)}_{lm}\nabla^2A^{(1),\,lm}
-2 A^{(1),\,lm}\nabla^2\chi^{(1)}_{lm}
\nn\\
&
-2A^{(1),\,lmn}C^{||(1)}_{,\,lmn}
+2\nabla^2A^{(1),\,l}\nabla^2C^{||(1)}_{,\,l}
\big)
\bigg]
+\bigg[
\delta_{ij}\nabla^{-2}\big(
4\phi^{(1),\,lm}C^{||(1)}_{,\,lm}
\nn\\
&
-4\nabla^2\phi^{(1)}\nabla^2C^{||(1)}
-\chi^{(1),\,lm}_{lm,n}C^{||(1),n}
-4\chi^{(1),\,l}_{lm,n}C^{||(1),mn}
+\chi^{(1)}_{lm,n}C^{||(1),\,lmn}
\nn\\
&
-2\chi^{(1),\,l}_{lm}\nabla^2C^{||(1),m}
+2C^{||(1),\,lm}\nabla^2\chi^{(1)}_{lm}
+\nabla^2C^{||(1),\,l}\nabla^2C^{||(1)}_{,\,l}
-C^{||(1),\,lmn}C^{||(1)}_{,\,lmn}
\big)
\nn\\
&
-2 \chi^{(1)}_{ij,\,l}C^{||(1),\,l}
-2\chi^{(1)}_{li}C^{||(1),\,l}_{,j}
-2\chi^{(1)}_{lj}C^{||(1),\,l}_{,i}
-4\nabla^{-2}\big(2\phi^{(1),\,l}_{,i}C^{||(1)}_{,\,lj}
+2\phi^{(1),\,l}_{,j}C^{||(1)}_{,\,li}
\nn\\
&
-2\phi^{(1)}_{,ij}\nabla^2C^{||(1)}
-2C^{||(1)}_{,ij}\nabla^2\phi^{(1)}
-C^{||(1),\,lm}_{,i}C^{||(1)}_{,\,lmj}
+C^{||(1)}_{,\,lij}\nabla^2C^{||(1),\,l}
\big)
\nn\\
&
+2 \partial_i\nabla^{-2}\big(
\chi^{(1),\,l}_{lj,m}C^{||(1),m}
+2\chi^{(1)}_{lj,m}C^{||(1),\,lm}
+\chi^{(1),m}_{lm}C^{||(1),\,l}_{,j}
+\chi^{(1)}_{lm}C^{||(1),\,lm}_{,j}
\nn\\
&
+\chi^{(1)}_{lj}\nabla^2C^{||(1),\,l}
\big)
+2 \partial_j\nabla^{-2}\big(
\chi^{(1),\,l}_{li,m}C^{||(1),m}
+2\chi^{(1)}_{li,m}C^{||(1),\,lm}
+\chi^{(1),m}_{lm}C^{||(1),\,l}_{,i}
\nn\\
&
+\chi^{(1)}_{lm}C^{||(1),\,lm}_{,i}
+\chi^{(1)}_{li}\nabla^2C^{||(1),\,l}
\big)
-\partial_i\partial_j\nabla^{-2}\nabla^{-2}\big(
4\nabla^2\phi^{(1)}\nabla^2C^{||(1)}
-4\phi^{(1),\,lm}C^{||(1)}_{,\,lm}
\nn\\
&
+\chi^{(1),\,lm}_{lm,n}C^{||(1),n}
+4\chi^{(1),\,l}_{lm,n}C^{||(1),mn}
+7\chi^{(1)}_{lm,n}C^{||(1),\,lmn}
+2\chi^{(1),\,l}_{lm}\nabla^2C^{||(1),m}
\nn\\
&
+4\chi^{(1)}_{lm}\nabla^2C^{||(1),\,lm}
+2 C^{||(1),\,lm}\nabla^2\chi^{(1)}_{lm}
+C^{||(1),\,lmn}C^{||(1)}_{,\,lmn}
-\nabla^2C^{||(1),\,l}\nabla^2C^{||(1)}_{,\,l}
\big)
\bigg]
   .
\el
}
In the above
the transformation of vector modes $ \chi^{\perp(2)}_{ij}$
involves  $\xi^{(1)\mu}$
and only $C^{\perp(2)}_{(i,j)}$ of $\xi^{(2)\mu}$,
the transformation of tensor modes $ \chi^{\top(2)}_{ij}$
involves  only  $\xi^{(1)\mu}$,
but not $\xi^{(2)\mu}$.
The  transformation of the 2nd-order perturbed  scalar field is
{\allowdisplaybreaks
\bl\label{rho2TransInflation}
\bar \varphi^{(2)}_S
=&
\varphi^{(2)}_S
-\beta  (\beta +1) (-\tau )^{-2 (\beta +2)}\varphi^{(0)'}A^{(1)}A^{(1)}
+(-\tau )^{-2 (\beta +1)}\varphi^{(0)''}A^{(1)}A^{(1)}
\nn\\
&
+(-\tau )^{-\beta -1}
\bigg[
-2\varphi^{(1)'}A^{(1)}
+\varphi^{(0)'} A^{(1)}_{,\, l}C\, ^{ ||(1) ,\,l}
\bigg]
\nn\\
&
+(-\tau )^{-\beta -1}\l[\int^\tau \frac{d\tau' }{(-\tau' )^{\beta +1}}\r]
\varphi^{(0)'}A^{(1)}_{,\, l}A^{(1),\,l}
-2\l[\int^\tau \frac{d\tau' }{(-\tau')^{\beta +1}}\r]
\varphi^{(1)}_{,\, l}A^{(1),\,l}
\nn\\
&
-2\varphi^{(1)}_{,\, l}C\, ^{ ||(1) ,\,l}
- (-\tau)^{-\beta-1}\varphi^{(0)}_{,0}A^{(2)}
\,   .
\el }

The above  synchronous-to-synchronous transformations
involve two vector fields $\xi^{(1)}$ and $\xi^{(2)}$.
These lengthy expressions contain many terms due to
the 1st-order parameters $A^{(1)}$,   $C^{||(1)}$, and $C^{\perp(1)}_i$
of the 1st-order vector $\xi^{(1)}$,
and only a few terms of the 2nd-order parameters
$A^{(2)}$, $C^{||(2)}$, and $C^{\perp(2)}_i$  of $\xi^{(2)}$.
As pointed out in Ref.~\cite{WangZhang2017,WangZhang2018},
distinctions should be made between  $\xi^{(2)\mu}$ and   $\xi^{(1)\mu}$.
In applications
we are often interested in the following case:
the 2nd-order solutions  are transformed \cite{Gleiser1996}
when  the 1st-order solutions are fixed.
\be\label{effecttrs}
\xi^{(1)\mu}=0, ~~  {\text{but}} ~~ \xi^{(2)\mu} \ne 0 .
\ee
This is referred to as the effective 2nd-order transformations.
If one would set  \cite{HwangNoh2012}
\be \label{untrs}
\xi^{(2)\mu}=0,  ~~{\text{but}} ~~  \xi^{(1)\mu} \ne 0
\ee
in \eqref{phi2TransRD} $\sim$ \eqref{rho2TransInflation},
one would have no freedom to make $\bar g^{(2)}_{0\mu }=0$ anymore,
as  $\xi^{(1)\mu}$ has been already fixed
in ensuring  $\bar g^{(1)}_{0\mu }=0$,
\eqref{untrs}   will not be a true 2nd-order gauge transformation.

For the effective 2nd-order transformations,
the  transformation vector
(\ref{alpha2RD1}) $\sim$ (\ref{d2RD1})  reduce to
\be
\alpha^{(2)}(\tau,\mathbf x)
=\frac{ A^{(2)}(\mathbf x)}{(-\tau)^{1+\beta}} \, ,
\ee
\be\label{beta2RD2}
\beta^{(2)}(\tau,\mathbf x)=
A^{(2)}(\mathbf x) \int^\tau \frac{d\tau'}{(-\tau')^{1+\beta}}
+C^{||(2)}(\mathbf x)
\,,
\ee
\be
\label{d2RD2}
d^{(2)}_i(\mathbf x)
=C^{\perp(2)}_i ({\bf x})   ,
\ee
and
(\ref{phi2TransRD})  $\sim$ (\ref{rho2TransInflation})
reduce to
\be\label{phi2TransRD2}
\bar \phi^{(2)}_S (\tau,\mathbf x) =
\phi^{(2)}_S(\tau,\mathbf x)
-\frac{(\beta +1)}{(-\tau )^{\beta +2}}
 A^{(2)}({\bf x})
+\frac{1}{3}\nabla^2A^{(2)}({\bf x})\int^\tau \frac{d\tau'}{(-\tau')^{\beta +1}}
+\frac{1}{3}\nabla^2C^{||(2)}({\bf x})
,
\ee
\be\label{chi||2transRD2}
 \bar\chi^{||(2)}_S(\tau,\mathbf x)
=
  \chi^{||(2)}_S(\tau,\mathbf x)
-2 A^{(2)}({\bf x})  \int^\tau \frac{d\tau' }{(-\tau' )^{\beta +1}}
-2C^{||(2)}({\bf x})     ,
\ee
\be\label{chiPerp2TransRD2}
\bar\chi^{\perp(2)}_{S\,ij}(\tau,\mathbf x)
=
\chi^{\perp(2)}_{S\,ij}(\tau,\mathbf x)
 -\partial_{j}C^{\perp(2)}_{i}({\bf x})
-\partial_{i}C^{\perp(2)}_{j}({\bf x})\ ,
\ee
\bl\label{chiT2transRD2}
\bar\chi^{\top(2)}_{S\,ij}
=
\chi^{\top(2)}_{S\,ij} ,
\el
\bl\label{rho2TransInflation2}
\bar \varphi^{(2)}_S
=&
\varphi^{(2)}_S
+M_{\rm Pl}\sqrt{2(\beta +1) (\beta +2)}  (-\tau)^{-\beta -2}
A^{(2)} ({\bf x})
\,   ,
\el
which have the same structure as the 1st-order transformations
(\ref{gaugetrphi})-- (\ref{deltarho}).
By (\ref{chiT2transRD2}),
 the 2nd-order tensor is invariant
under the effective 2nd-order  transformation \eqref{effecttrs}
between  synchronous coordinates.
The effective 2nd-order  transformation
(\ref{phi2TransRD2}) (\ref{chi||2transRD2}) (\ref{chiPerp2TransRD2})
(\ref{rho2TransInflation2})
have been used in the last section
to eliminate the effective 2nd-order gauge modes from
the 2nd-order vector solution (\ref{vect2ndsol})
and  the scalar solutions
(\ref{Sevarphi1solv2}) (\ref{Sephi1kk222})  (\ref{Sechi1kkfinal22}).

For the effective 2nd-order transformation
the gauge invariant scalar perturbations
can be constructed as the 1st-order ones in Section 4,
but for a general  2nd-order transformation
the 2nd-order gauge invariant perturbations are not easy to
construct (see Ref.~\cite{AcquavivaBartoloMatarrese2003}).

\section{   Conclusion and discussion}
\label{sec:Conclusion}

We present a systematic study of the cosmological perturbations
up to the 2nd-order   in a class of models of
the power-law inflation driven by a scalar field
 in synchronous coordinates.
We list the solutions of all the 1st-order metric perturbations
and of the 1st-order perturbed scalar field as well,
and identify the residual gauge modes of the scalar and vector perturbations.
It is seen that
the 1st-order tensorial perturbation as RGW is decoupled from the perturbed scalar field
and is gauge invariant under synchronous-to-synchronous transformations.
The 1st-order vector perturbation is not a wave
and decreases with expansion,
but the scalar metric perturbation
and the perturbed scalar field during inflation
are waves propagating at the speed of light,
analogous to the tensor perturbation.
In comparison, in the MD stage,
the scalar metric perturbation is not a wave
and does not propagate \cite{Matarrese98,WangZhang2017,ZhangQinWang2017},
and  in the RD stage
the scalar metric perturbation propagates
at the sound speed $c_s\sim \frac{1}{\sqrt 3}$,
instead of the speed of light \cite{WangZhang2018,WangZhang2019}.
We use the quantum normalization that
each $k$ mode of the gauge invariant scalar field $Q_{\varphi}$
in the Bunch-Davies vacuum has the zero point energy $\frac12 \hbar k$ to determine the coefficients $(d_1, d_2)$ in the 1st-order scalar solutions.
This  determines the amplitude of all four gauge invariant 1st-order scalar perturbations
for a given inflation model.
The 2nd-order perturbed Einstein equation
can be put into a form similar to the 1st-order ones,
the source contains not only the 2nd-order stress tensor of the perturbed scalar field,
but also the products of 1st-order perturbations,
among which the scalar-scalar coupling is considered in this paper.
We have explicitly demonstrated that
the equation of 2nd-order perturbed scalar field
is the same as the 2nd-order covariant energy conservation,
just like  the 1st-order case.
In particular, for both the 1st- and 2nd-order,
the momentum conservation is a trivial identity
because  the inflation scalar field  contains no curl part.
This feature is similar to the stress tensor of dust
in the MD stage  \cite{WangZhang2017,ZhangQinWang2017},
but is different from the stress tensor of a relativistic fluid
in the RD stage in which the velocity contains a curl part
and yields a nontrivial momentum conservation \cite{WangZhang2018,WangZhang2019}.
We have obtained the solutions of all the 2nd-order perturbations,
which contain a homogeneous part having the same structure as the 1st-order solutions,
and an inhomogeneous part consisting of integrations of the effective source.
To identify  the 2nd-order residual gauge modes,
we have presented the 2nd-order synchronous-to-synchronous  transformations.
A general 2nd-order transformation
contains both the 2nd-order transformation vector $\xi^{(2)\mu}$ and
the 1st-order transformation vector $\xi^{(1)\mu}$.
Since the 1st-order solutions are practically often fixed in actual applications,
only the transformations by $\xi^{(2)\mu}$
are effective \cite{WangZhang2017,WangZhang2018},
which have similar structure to the 1st-order transformations.
Using these effective 2nd-order transformations
we are able to identify the residual gauge modes in the 2nd-order solutions.

{

For a comprehensive comparison of
the inflation stage
and  MD stage \cite{Matarrese98,VillaRampf2016,WangZhang2017},
in Table \ref{wavesOrNot}
we give a list on the nature of 1st- and 2nd-order perturbations
generated by the scalar-scalar coupling \cite{WangZhang2017}.
During the MD stage,
only 1st-order tensors are waves  propagating at the speed of light,
while all 1st- and 2nd-order scalars and 2nd-order tensors are not.
The 1st-order scalar perturbations are not waves
since their equations contain no sound speed term
due to the vanishing pressure.
Consequently, the inhomogeneous 2nd-order metric perturbations induced by these scalars
are not waves
because their source terms are the nonwave scalar-scalar coupling.
These results for the MD stage agree with those in Ref.~\cite{VillaRampf2016}.
During the inflation stage
all perturbations (1st- and 2nd-order)
are waves except for the decaying 1st-order vectors.
The 1st-order scalar field and the 1st-order scalar metric perturbations
are now waves,
so is the scalar-scalar coupling.
Consequently, the inhomogeneous 2nd-order perturbations
are waves driven by these coupling.
\begin{table}[htbp]
\caption{
The nature of the  1st- and 2nd-order perturbations
during the inflation  and MD stages.
}
\begin{center}
\label{wavesOrNot}
\begin{tabular}{|c|c|c|c|c|c|c|}
\hline \multicolumn{7}{|c|}{ 1st-order perturbations } \\
\hline & \multicolumn{3}{|c|}{ Inflation stage } & \multicolumn{3}{c|}{ MD stage } \\
\hline Perturbations & Scalar & Vector & Tensor & Scalar & Vector & Tensor \\
\hline Waves or not? & Yes & Decay & Yes & No & Decay & Yes \\
\hline  \multicolumn{7}{|c|}{   } \\
\hline  \multicolumn{7}{|c|}{ 2nd-order perturbations by scalar-scalar coupling } \\
\hline & \multicolumn{3}{|c|}{ Inflation stage } & \multicolumn{3}{c|}{ MD stage } \\
\hline Perturbations & Scalar & Vector & Tensor & Scalar & Vector & Tensor \\
\hline Waves or not? & Yes & Yes & Yes & No & No & No
\\
\hline
\end{tabular}
\end{center}
\end{table}

}

The perturbation solutions of the inflation stage derived in this paper
can be joined with those of the RD and MD stages
\cite{WangZhang2017,ZhangQinWang2017,WangZhang2018,WangZhang2019}
for broad applications in cosmology.
The 2nd-order solutions will be helpful
to explore many aspects due to nonlinearity,
such as the non-Gaussianity,
the energy and momentum transfer between $k$-modes,
 and among the tensor, vector, and scalar modes.

\

\textbf{Acknowledgements}

This work is supported in part by National Key RD Program of China \seqsplit{(2021YFC2203100)},  by  the Fundamental Research Funds for the Central Universities under Grant, NSFC Grants No.12003029, 12261131497, 11675165, 11961131007, 12261131497, 11875113, 11833005, 12192224, by 111 Project for ``Observational and Theoretical Research on Dark Matter and Dark Energy" (B23042), by CAS project for young scientists in basic research (YSBR-006), CAS Young Interdisciplinary Innovation Team (JCTD-2022-20).

\
\\
{\large{\bf Author Contributions} Bo Wang completed the main part of the calculation and prepared figures 1-2. Yang Zhang did some calculations and revised the figures. Bo Wang and Yang Zhang wrote the manuscript together and of course all authors reviewed the manuscript.}

\
\\
{\large{{\bf Data availability} We declare that this is a theoretical work with no data associated to it.}

\

\
\\
{\bf \Large Declarations}

\
\\
{\large{{\bf Conflict of interest} The authors declare that they have no conflict of interest.}

\appendix

\section{Perturbed stress tensor }
\label{sec:perturbedGT}

The nonvanishing  {Christoffel} symbols, the perturbed Ricci tensor,
the perturbed Einstein tensors up to 2nd order,
are listed in Ref.~\cite{WangZhang2018}
for a general flat RW spacetime in synchronous coordinate.
In the following we list the stress tensor and the conservation equation
of the inflaton scalar field with a general potential up to the 2nd order.

The stress tensor of the scalar field is
\be\label{Tuvphi}
T_{\mu \nu} = \varphi_{, \mu} \varphi_{, \nu}
  -g_{\mu \nu}\left[\frac{1}{2}
  g^{\alpha \beta} \varphi_{, \alpha} \varphi_{, \beta}
   + V(\varphi)\right] ,
\ee
the energy density $\rho = - T^{0}\, _{0}$
and the pressure $p=\frac13  T^{i}\, _{i}$.
The 0th-order is
\bl \label{T000th}
T^{(0)}_{00}
=&
\frac12 \varphi^{(0)'}\varphi^{(0)'}  +a^2 V(\varphi^{(0)}) ,
\\
T^{(0)}_{0i} =& 0,
\\
T^{(0)}_{ij}  = & \big[
\frac12 \varphi^{(0)'}\varphi^{(0)'}
  - a^2  V(\varphi^{(0)}) \big]\delta_{ij},
\el
the 0th-order energy density $\rho = - T^{(0)0}\,_{0}
= \frac12 a^{-2} \varphi^{(0)'}\varphi^{(0)'} + V(\varphi^{(0)})$,
and the 0th-order pressure $p=\frac13  T^{(0)i}\, _{i}
= \frac12 a^{-2} \varphi^{(0)'}\varphi^{(0)'} - V(\varphi^{(0)})$ .
The  1st-order perturbed stress tensor   is
{\allowdisplaybreaks
\bl
T^{(1)}_{00}
=&
\varphi^{(0)'}\varphi^{(1)'}
+a^2\varphi^{(1)} V(\varphi^{(0)})_{,\varphi}
,\label{T001stre}
\\
T^{(1)}_{0i}
=&
\varphi^{(0)'} \varphi^{(1)}_{, i}
,\label{T0i1stre}
\\
T^{(1)}_{ij}
=& \left[ \varphi^{(0)'}\varphi^{(1)'}
-a^2\varphi^{(1)} V(\varphi^{(0)})_{,\varphi}
\right]\delta_{ij}
\nn \\
& + \left[\frac{1}{2} \left(\varphi^{(0)'}\right)^{2}
-a^2 V(\varphi^{(0)})\right] (-2 \phi^{(1)}) \delta_{ij}
\nn\\
&
+\left[\frac{1}{2}\left(\varphi^{(0)'}\right)^{2}
-a^2 V(\varphi^{(0)})\right] D_{ij}\chi^{||(1)}
\nn\\
&
+\left[\frac{1}{2}\left(\varphi^{(0)'}\right)^{2}
-a^2 V(\varphi^{(0)})\right] \chi^{\perp(1)}_{ij}
\nn \\
&
+\left[\frac{1}{2}\left(\varphi^{(0)'}\right)^{2}
-a^2 V(\varphi^{(0)})\right]\chi^{\top(1)}_{ij}
.\label{Tij1stre}
\el
}
Raising up an index of \eqref{Tij1stre} gives
\bl\label{Tij1}
T^{(1)i}\, _j & \simeq  a^{-2}(\delta^{i\mu} -\gamma^{(1)i\mu}) T_{\mu j}
 \nn \\
& =  \left[ a^{-2} \varphi^{(0)'}\varphi^{(1)'}
    - \varphi^{(1)} V(\varphi^{(0)})_{,\varphi}
   \right]   \delta^i_j \, ,
\el
agreeing  with the expression  (25) of Ref.~\cite{Grishchuk1994}.
The 1st-order perturbed energy density and pressure are
\bl \label{rho1}
\rho^{(1)} = \frac12 a^{-2} \varphi^{(0)'}\varphi^{(1)'}
       +\varphi^{(1)} V(\varphi^{(0)})_{\, ,\varphi} \, ,
 \\
p^{(1)}  = \frac12 a^{-2} \varphi^{(0)'}\varphi^{(1)'}
       - \varphi^{(1)} V(\varphi^{(0)})_{\, ,\varphi} \, .
   \label{p1}
\el
Notice that $\rho^{(1)}=0=p^{(1)}$
for the  exact de Sitter inflation,
since $\varphi^{(0)'}=0= V(\varphi^{(0)})_{\, ,\varphi}$.

The 2nd-order  perturbed stress tensor   is
\bl
T^{(2)}_{00}
=&
\frac{1}{2}\varphi^{(0)'} \varphi^{(2)'}
+\frac{1}{2}a^{2} V(\varphi^{(0)})_{,\varphi} \varphi^{(2)}
\nn\\
&
+\frac{1}{2}\varphi^{(1)'}\varphi^{(1)'}
+\frac{1}{2}\varphi^{(1),l}\varphi^{(1)}_{,l}
+\frac{1}{2}a^{2}\varphi^{(1)}\varphi^{(1)}
  V(\varphi^{(0)})_{,\varphi \varphi }
,
\label{T002_1app}
\\
T^{(2)}_{0i}
=&
\frac{1}{2}M_{\rm Pl}\frac{\sqrt{2(\beta +1) (\beta+2)}}{\tau}\varphi^{(2)}_{, i}
+\varphi^{(1)}_{, i}\varphi^{(1)'},
\label{T0iIflationapp}
\\
T^{(2)}_{ij}
=&
\Big[    -\frac{1}{2} ( \varphi^{(0)'})^2
+a^2 V(\varphi^{(0)})
\Big]\phi^{(2)}\delta_{ij}
+\Big[
     \frac{1}{2}\varphi^{(0)'}\varphi^{(2)'}
-\frac{1}{2}a^2V(\varphi^{(0)})_{,\, \varphi}\varphi^{(2)}
\Big]\delta_{ij}
\nn\\
&
+  \Big[    \frac{1}{4}   ( \varphi^{(0)'})^2
-\frac{1}{2}a^2 V(\varphi^{(0)})
\Big]\chi_{ij}^{(2)}
\nn\\
&
+\Big[  \frac{1}{2}\varphi^{(1)'}\varphi^{(1)'}
        -  \frac{1}{2}  \varphi^{(1),k}\varphi^{(1)}_{, k}
-\frac{1}{2}a^2 V(\varphi^{(0)})_{,\, \varphi \varphi} \varphi^{(1)}\varphi^{(1)}
\nn\\
&
-2 \varphi^{(0)'}\varphi^{(1)'}\phi^{(1)}
+2a^2 V(\varphi^{(0)})_{,\,  \varphi}\varphi^{(1)}\phi^{(1)}
\Big]\delta_{ij}
\nn\\
&
+\varphi^{(1)}_{, i}\varphi^{(1)}_{, j}
-a^2 V(\varphi^{(0)})_{,\,  \varphi} \varphi^{(1)}\chi_{ij}^{(1)}
+\varphi^{(0)'}\varphi^{(1)'}\chi_{ij}^{(1)}
.
\label{Tij2orighinapp}
\el
The 2nd-order perturbed stress tensor are generally nonzero
even for the de Sitter inflation.
These stress tensors generally
will change under residual gauge transformations,
and we shall not discuss in details.
See Ref.~\cite{ChoGongOh2022} for relevant discussions
for a general metric.

The  equations  of covariant conservation are  given by (\ref{covcons}).
By calculation,
we get the 1st-order energy conservation $[T^{ 0\nu}\,_{; \, \nu}]^{(1)}=0$
as the following
\bl \label{equvarphi1st3}
 \varphi^{(0)'}\varphi^{(1)''}
+ 2\frac{a'}{a} \varphi^{(0)'}\varphi^{(1)'}
-\varphi^{(0)'}\nabla^2\varphi^{(1)}
+a^{2}V(\varphi^{(0)})_{,\, \varphi \varphi} \varphi^{(0)'} \varphi^{(1)}
-3 (\varphi^{(0)'} )^2\phi^{(1)'} =0
,
\el
which is equivalent to
the equation of the 1st-order perturbed scalar field  \eqref{e1stord}.
The 1st-order momentum conservation
$[T^{\,  i\nu}\,_{; \, \nu}]^{(1)}=0$ leads to an identity $0=0$
because the scalar field $\varphi$ contains no curl part.
This property of the stress tensor of the scalar field $\varphi$
is similar to
that of a dust \cite{WangZhang2017,ZhangQinWang2017},
but in contrast to the relativistic fluid
for which the momentum conservation
is not an trivial identity \cite{WangZhang2018,WangZhang2019}.

The 2nd-order energy conservation $[T^{ 0\nu}\,_{; \, \nu}]^{(2)}=0$  gives
\bl\label{Enconsv2nd0}
0=
&
\frac{1}{2}a^{-2} \varphi^{(0)'} \varphi^{(2)''}
+a'a^{-3}\varphi^{(0)'}\varphi^{(2)'}
-\frac{1}{2}a^{-2}\varphi^{(0)'}\nabla^2\varphi^{(2)}
+\frac{1}{2}V_{,\, \varphi \varphi} \varphi^{(0)'}\varphi^{(2)}
\nn\\
&
+\frac{1}{4}a^{-2}(\varphi^{(0)'})^{2}\gamma^{(2)'k}_k
+\frac{1}{2}V_{,\varphi  \varphi \varphi }
    \varphi^{(0)'}   \varphi^{(1)}\varphi^{(1)}
+\frac{1}{2}a^{-2}\varphi^{(0)'}\varphi^{(1)'}\gamma^{(1)'k}_{k}
\nn\\
&
+a^{-2}\varphi^{(0)'}\gamma^{(1)kl}_{,l}\varphi_{,k }^{(1)}
+a^{-2}\varphi^{(0)'}\gamma^{(1)kl}\varphi_{,kl }^{(1)}
-\frac{1}{2}a^{-2}\varphi^{(0)'}\gamma^{(1)k}_{k,l}\varphi^{(1),\,l}
\nn\\
&
-\frac{1}{2}a^{-2}(\varphi^{(0)'})^{2}\gamma^{(1)kl}\gamma^{(1)'}_{kl} ,
\el
which can be  written as
\bl\label{enerConsev2}
&
\varphi^{(2)''}_S
+2 \frac{a'}{a} \varphi^{(2)'}_S
-\nabla^2\varphi^{(2)}_S
+a^2 V(\varphi^{(0)})_{,\, \varphi \varphi  }\varphi^{(2)}_S
= - A_S  + 3\varphi^{(0)'}\phi^{(2)'}_S  ,
\el
where the subindex $`` S"$   indicates the scalar-scalar coupling, and
\bl\label{AS2}
A_S
\equiv &
a^2V_{,\varphi^{(0)}\varphi^{(0)}\varphi^{(0)}}\varphi^{(1)}\varphi^{(1)}
-6\varphi^{(1)'}\phi^{(1)'}
+2\varphi_{,k}^{(1)}\phi^{(1),k}
-4\phi^{(1)}\nabla^2\varphi^{(1)}
\nn\\
&
-12\varphi^{(0)'}\phi^{(1)} \phi^{(1)'}
+\frac{4}{3}\varphi_{,k }^{(1)}\nabla^2\chi^{||(1),k}
+2\varphi_{,kl}^{(1)}\chi^{||(1),kl}
\nn\\
&
-\frac{2}{3}\nabla^2\varphi^{(1)}\nabla^2\chi^{||(1)}
-\varphi^{(0)'}\chi^{||(1),kl}\chi^{||(1)'}_{,kl}
+\frac{1}{3}\varphi^{(0)'}\nabla^2\chi^{||(1)}\nabla^2\chi^{||(1)'}.
\el
We have also checked that
the 2nd-order momentum conservation $[T^{\,  i\nu}\,_{; \, \nu}]^{(2)}=0$
is also a trivial identity,  $0=0$, as the 1st-order one.

\section{Gauge transformations  from Synchronous to Synchronous}
\label{sec:gauge_transform}

The formulae  for the  transformations
between two general coordinates  for a flat RW spacetime,
as well as between two synchronous coordinates,
have been given in Appendix C in Ref.~~\cite{WangZhang2017}.
Here we add the transformation of the scalar field $\varphi$.

First consider the 1st-order transformation.
In the synchronous coordinate,
for the power-law inflation,
the transformation vector is the following
\be \label{xi0trans}
\xi^{(1)0}(\tau, {\bf x}) = \frac{A^{(1)}({\bf x})}{a(\tau)}
= \frac{A^{(1)}({\bf x})}{(-\tau)^{\beta+1}},
\ee
\bl  \label{gi0}
 \xi^{(1)i} (\tau, {\bf x})
 & = A^{(1)}({\bf x})^{,i}
 \int^\tau \frac{d\tau' }{a(\tau')} + C^{(1)i}({\bf x})
 \nn \\
&  = A^{(1)}({\bf x})^{,i}
 \int^\tau \frac{d\tau' }{(-\tau')^{\beta+1}} + C^{(1)i}(\bf x) ,
\el
where $A^{(1)}$ and $C^{(1)i}$
are small, arbitrary functions depending on $\bf x$ only,
and the constant $l_0^{-1}$ from $a(\tau)$  has been absorbed into  $A^{(1)}$
for notational simplicity.
 $C^{(1)i}$ can be decomposed into
\be\label{C1decomp}
C ^{(1)i} ({\bf x}) = C\, ^{ ||(1) ,\,i}({\bf x})
+  C\, ^{ \bot(1) i}(\bf x),
\ee
where the transverse part satisfies
$ \partial_i C^{ \bot(1)\, i}=0$.
Corresponding to \eqref{xi0trans} \eqref{gi0},
the  residual gauge transform
of the metric perturbations between  the synchronous coordinates is
\bl
\bar\phi^{(1)} & =   \phi^{(1)}
  + \frac{1}{3}\nabla^2 A^{(1)} \int^\tau \frac{d\tau }{(-\tau)^{\beta+1}}
   + \frac{1}{3}\nabla^2 C^{||(1)}
  +  \frac{a'}{a^2}   A^{(1)} ,
\label{gaugetrphi}
\\
 \bar \chi^{||(1)} & =
 \chi^{||(1)}  - 2 A^{(1)} \int^\tau \frac{d\tau }{(-\tau)^{\beta+1}}
   -2C^{||(1)},
 \label{gaugetrchi}
\\
\label{gaugePerpchi}
\bar\chi^{\perp(1)}_{ij} &  =  \chi^{\perp(1)}_{ij}
-C^{\perp(1)}_{i,j}
-C^{\perp(1)}_{j,i},
\\
\bar\chi^{\top(1)}_{ij} & =  \chi^{\top(1)}_{ij} ,
\label{gaugeGW}
\el
and the combined  scalar $\zeta$ of \eqref{phiChiZeta} transforms as
\bl
\bar\zeta
= &
\zeta
  -2 \frac{a'}{a^2}  A^{(1)} .
\el
We see that the 1st-order tensor modes remains invariant under the residual transformation.
The 1st-order perturbed scalar field
 transforms between  the synchronous coordinates as
\bl \label{deltarho}
\bar   \varphi ^{(1)}
& =  \varphi ^{(1)} - \varphi^{(0)'} \frac{A^{(1)}}{a}
 \\
& =  \varphi ^{(1)}
+M_{\rm Pl}\sqrt{2(\beta +1) (\beta +2)} A^{(1)} (-\tau)^{-\beta -2} ,
\label{deltarho2}
\el
It is seen that
$\varphi ^{(1)}_{gi}$ of \eqref{ginvvarphi} is invariant
under the transformation  \eqref{gaugetrchi} \eqref{deltarho},
$\Psi$ of \eqref{Psigauginv} is invariant
under the transformation  \eqref{gaugetrphi}   \eqref{gaugetrchi},
and $Q_\varphi$ of \eqref{Qvarphi}   and
${\cal R}$ of \eqref{scalarcurvtpertOur} are also invariant
as combinations of $\varphi ^{(1)}_{gi}$ and $\Psi$.

Now  we  list
the 2nd-order synchronous-to-synchronous  gauge transformations.
(See Appendix C in Ref.~\cite{WangZhang2018}  for a general RW spacetime.)
For the power-law inflation and  the scalar-scalar coupling,
 we give the 2nd-order transformation vector
{\allowdisplaybreaks
\bl  \label{alpha2RD1}
\xi^{(2)0}
 & =\frac{ A^{(2)}(\mathbf x)}{(-\tau)^{1+\beta}} \, ,
 \\
  \xi^{(2)i } & = \partial^i \beta^{(2)} + d^{(2)i},
\el
with
\bl\label{beta2RD1}
\beta^{(2)}=&
\nabla^{-2}\Big[
\nabla^2 A^{(1)} \int^\tau\frac{4\phi^{(1)}(\tau',{\bf x})}{(-\tau')^{1+\beta}}d\tau'
+A^{(1)}_{,i}\int^\tau\frac{4\phi^{(1)}(\tau',{\bf x})^{,i}}{(-\tau')^{1+\beta}}d\tau'
\nn\\
&
-A^{(1) ,\, ij}\int^\tau\frac{2\chi^{(1)}_{ij}(\tau',{\bf x})}{(-\tau')^{1+\beta}}d\tau'
-A^{(1) ,\, j}\int^\tau\frac{2\chi^{(1)}_{ji}(\tau',{\bf x})^{,\,i}}{(-\tau')^{1+\beta}}d\tau'
\nn\\
&
+ 2 A^{(1),\, ij}C^{||(1)}_{,\, ij}\int^\tau \frac{d\tau'}{(-\tau')^{1+\beta}}
+ 2 A^{(1) ,\, i}\nabla^2C^{||(1)}({\bf x})_{,i} \int^\tau\frac{d\tau'}{(-\tau')^{1+\beta}}\Big]
\nn\\
&-\frac{1}{2(-\tau)^{2+2\beta}}A^{(1)} A^{(1)}
+ A^{(1),\, i} A^{(1)}_{,\, i}
    \int^\tau  \frac{d\tau'}{(-\tau')^{1+\beta}}
    \int^{\tau'}\frac{d\tau'' }{(-\tau'')^{1+\beta}}
\nn\\
&
+A^{(2)} \int^\tau \frac{d\tau'}{(-\tau')^{1+\beta}}
+C^{||(2)}
\,,
\el
}
{
\bl   \label{d2RD1}
d^{(2)}_i
=&
\partial_i\nabla^{-2}\Big[
-\nabla^2 A^{(1)} \,  \int^\tau \frac{4\phi^{(1)}(\tau',{\bf x})}{(-\tau')^{1+\beta}}d\tau'
-A^{(1)}_{, j}\int^\tau \frac{4\phi^{(1)}(\tau',{\bf x})^{,j}}{(-\tau')^{1+\beta}}d\tau'
\nn\\
&
+2A^{(1) ,\, lj}\int^\tau \frac{ \chi^{(1)}_{lj}(\tau',{\bf x})}{(-\tau')^{1+\beta}}d\tau'
+2A^{(1),\, l}\int^\tau \frac{\chi^{(1)}_{ lj }(\tau',{\bf x})^{,\, j}}{(-\tau')^{1+\beta}}d\tau'
\nn\\
&
- 2 A^{(1),\, lj }C^{||(1)}({\bf x})_{,\, lj }\int^\tau \frac{d\tau'}{(-\tau')^{1+\beta}}
- 2 A^{(1),\, j} \nabla^2C^{||(1)}_{,j }  \int^\tau \frac{d\tau'}{(-\tau')^{1+\beta}}\Big]
\nn\\
&
+ 4A^{(1)}_{,\, i}\int^\tau  \frac{\phi^{(1)}(\tau',{\bf x})}{(-\tau')^{1+\beta}}d\tau'
-2A^{(1)\, ,\, l}\int^\tau  \frac{\chi^{(1)}_{li}(\tau',{\bf x})}{(-\tau')^{1+\beta}}d\tau'
\nn\\
&+ 2 A^{(1)\, ,\, l}C^{||(1)}_{,\, li}\int^\tau \frac{d\tau'}{(-\tau')^{1+\beta}}
+ C^{\perp(2)}_i  .
\el
}
where  $A^{(2)}$ is an arbitrary function of 2nd order,
$ C^{(2)}_i $  is an arbitrary 3-vector of  2nd order
and can be decomposed into
$C^{(2)}_i   =   C^{||(2)}_{,\,i}   +C^{\perp(2)}_{\,i}$.
(See also Ref.~~\cite{WangZhang2017}.)
From these we obtain
the residual transformations of the 2nd-order metric  perturbations
(\ref{phi2TransRD}) $\sim$  \eqref{chiT2transRD},
and of the 2nd-order perturbed scalar field  \eqref{rho2TransInflation}
in the context,
where we have  omitted  the 1st-order curl vector.

\section{1st-order perturbations}
\label{sec:appendix_1stSolution}

Here we show detailed calculations for the 1st-order perturbations.

The $(00)$ component of 1st-order perturbed Einstein equation
  gives the 1st-order energy constraint,
\be \label{energyconstr1st}
-6\frac{a'}{a}\phi^{(1)'}
          +2\nabla^2\phi^{(1) }
          +\frac{1}{3}\nabla^2\nabla^2\chi^{||(1)}
=8\pi G
\Big[
\varphi^{(0)'}\varphi^{(1)'}
+a^2 V(\varphi^{(0)})_{,\varphi} \varphi^{(1)}
\Big] \, ,
\ee
and for the power-law inflation,
\bl\label{Ein1st00}
&
-6\frac{\beta +1}{\tau }\phi^{(1)'}
          +2\nabla^2\phi^{(1) }
          +\frac{1}{3}\nabla^2\nabla^2\chi^{||(1)}
\nn\\
=&
\frac{1}{M_{\rm Pl}}\frac{\sqrt{2(\beta +1) (\beta +2)}}{\tau}\varphi^{(1)'}
-\frac{1}{M_{\rm Pl}}\frac{ \sqrt{2(\beta +1) (\beta +2)}
(2 \beta +1)}{\tau^{2}} \varphi^{(1)}
,
\el
which contains no second order time derivative.
The  $(0i)$ component of  1st-order perturbed Einstein
 gives the 1st-order momentum constraint
\be \label{momentconstr1RD}
2\phi^{(1)' }_{,i} + \frac{1}{2} D_{ij}\chi^{||(1)',j  }
+ \frac{1}{2} \chi^{\perp(1)',j  }_{ij}
 = 8\pi G \varphi^{(0)'} \varphi^{(1)}_{, i},
\ee
which, for the power-law inflation, is
\be\label{momentconstr1RDinf}
2\phi^{(1)' }_{,i} + \frac{1}{2} D_{ij}\chi^{||(1)',j  }
+ \frac{1}{2} \chi^{\perp(1)',j  }_{ij}
 = \frac{1}{M_{\rm Pl}}\frac{\sqrt{2(\beta +1)
 (\beta +2)}}{\tau}\varphi^{(1)}_{, i} .
\ee
The vector mode is formed from a curl vector as \eqref{chiVec0}
and can be separated out from \eqref{momentconstr1RDinf}
\be\label{vector1stequ}
\chi^{\perp(1)', \, j  }_{ij}=0,
\ee
and  the scalar part in \eqref{momentconstr1RDinf}
 can be rewritten as (by dropping the spatial differentiation)
\be \label{drmomentconstr1RD}
2\phi^{(1)' }  + \frac{1}{3} \nabla^2\chi^{||(1)'}
 = 8\pi G \varphi^{(0)'} \varphi^{(1)} ,
\ee
(which can be written as
$ \varphi^{(1)} =\frac{1}{8\pi G \varphi^{(0)'} }
( 2\phi^{(1)' }  + \frac{1}{3} \nabla^2\chi^{||(1)'} )
=  - \frac{1}{8\pi G \varphi^{(0)'} }  \zeta'$.)
i.e.,
\be\label{Ein1st0i}
2\phi^{(1)' } + \frac{1}{3} \nabla^2\chi^{||(1)'}
 = \frac{1}{M_{\rm Pl}}\frac{ \sqrt{2(\beta +1)
 (\beta +2)} }{\tau} \varphi^{(1)} .
\ee

The $(ij)$ component of 1st-order perturbed
Einstein equation  gives the 1st-order evolution equation
\allowdisplaybreaks
\bl  \label{evoEq1stRD2}
&
2\phi^{(1)''} \delta_{ij}
+4\frac{a'}{a}\phi^{(1)'}\delta_{ij}
+\phi^{(1) }_{,ij}
-\nabla^2\phi^{(1) }\delta_{ij}
+\l[4\frac{a''}{a}-2(\frac{a'}{a})^2\r]\phi^{(1)}\delta_{ij}
\nn\\
&
+\frac{1}{2} D_{ij}\chi^{||(1)''}
+\frac{a'}{a}D_{ij}\chi^{||(1)'}
+\l[(\frac{a'}{a})^2-2\frac{a''}{a}\r]D_{ij}\chi^{||(1)}
\nn\\
&
+\frac{1}{6}\nabla^2D_{ij}\chi^{||(1)}
-\frac{1}{9}\delta_{ij}\nabla^2\nabla^2\chi^{||(1) }
\nn\\
&
+\frac{1}{2} \chi^{\perp(1)''}_{ij}
+\frac{a'}{a}\chi^{\perp(1)'}_{ij}
+\l[(\frac{a'}{a})^2-2\frac{a''}{a}\r]\chi^{\perp(1)}_{ij}
\nn \\
&
+\frac{1}{2} \chi^{\top(1)''}_{ij}
+\frac{a'}{a} \chi^{\top(1)'}_{ij}
+\l[(\frac{a'}{a})^2 -2\frac{a''}{a}\r]\chi^{\top(1)}_{ij}
- \frac{1}{2}\nabla^2\chi^{\top(1) }_{ij}
\nn\\
=&
8\pi G \Big\{
 \left[-\left(\varphi^{(0)'}\right)^{2}
+2a^2 V(\varphi^{(0)})\right]\phi^{(1)}\delta_{ij}
+\left[
 \varphi^{(0)'}\varphi^{(1)'}
-a^2\varphi^{(1)} V(\varphi^{(0)})_{,\varphi^{(0)}}
\right]\delta_{ij}
\nn\\
&
+\left[\frac{1}{2}\left(\varphi^{(0)'}\right)^{2}
-a^2 V(\varphi^{(0)})\right]D_{ij}\chi^{||(1)}
\nn\\
&
+\left[\frac{1}{2}\left(\varphi^{(0)'}\right)^{2}
-a^2 V(\varphi^{(0)})\right]\chi^{\perp(1)}_{ij}
\nn\\
&
+\left[\frac{1}{2}\left(\varphi^{(0)'}\right)^{2}
-a^2 V(\varphi^{(0)})\right]\chi^{\top(1)}_{ij}
\Big\}
\el
which is valid for a general RW spacetime with a general $a(\tau)$.
Using \eqref{Ein0thij1} to cancel some terms on both sides,
the  vector  and tensor parts of \eqref{evoEq1stRD2} are the following
\bl  \label{vt1stRD2}
 \chi^{\perp(1)''}_{ij}
+ 2 \frac{a'}{a}\chi^{\perp(1)'}_{ij} =0 ,
\\
 \chi^{\top(1)''}_{ij}
+2 \frac{a'}{a} \chi^{\top(1)'}_{ij}
-  \nabla^2\chi^{\top(1) }_{ij} =0 .
  \label{t1stRD2}
\el
The scalar part  of \eqref{evoEq1stRD2} is the following
\bl
&
2\phi^{(1)''} \delta_{ij}
+4\frac{a'}{a}\phi^{(1)'}\delta_{ij}
+D_{ij}\phi^{(1)} +\frac13 \delta_{ij} \nabla^2 \phi^{(1)}
-\nabla^2\phi^{(1) }\delta_{ij}
-\frac{1}{9}\delta_{ij}\nabla^2\nabla^2\chi^{||(1) }
\nn\\
&
+\frac{1}{2} D_{ij}\chi^{||(1)''}
+\frac{a'}{a}D_{ij}\chi^{||(1)'}
+\frac{1}{6}\nabla^2D_{ij}\chi^{||(1)}
\nn\\
=&
8\pi G \left[ \varphi^{(0)'}\varphi^{(1)'}
-a^2\varphi^{(1)} V(\varphi^{(0)})_{,\varphi^{(0)}}
\right]\delta_{ij} ,
\el
which is also split  into the trace part
\bl  \label{evoEq1stD21}
&
2\phi^{(1)''}
+4\frac{a'}{a}\phi^{(1)'}
- \frac23   \nabla^2 \phi^{(1)}
-\frac{1}{9} \nabla^2\nabla^2\chi^{||(1) }
= 8\pi G \left[ \varphi^{(0)'}\varphi^{(1)'}
-a^2 V(\varphi^{(0)})_{,\varphi^{(0)}} \varphi^{(1)}  \right]  ,
\el
and the  traceless part
\bl
&
D_{ij}\phi^{(1)}
+\frac{1}{6}\nabla^2D_{ij}\chi^{||(1)}
+\frac{1}{2} D_{ij}\chi^{||(1)''}
+\frac{a'}{a}D_{ij}\chi^{||(1)'}
= 0 .
\el
Dropping off $D_{ij}$ in the traceless further gives
\be \label{trcls}
\phi^{(1)}
+\frac{1}{6}\nabla^2\chi^{||(1) }
+\frac{1}{2} \chi^{||(1)''}
+\frac{a'}{a}\chi^{||(1)'}
=0 .
\ee
For the power-law inflation
the above  equations are
\bl  \label{evoEq1stRD3}
&
2\phi^{(1)''} \delta_{ij}
+4\frac{\beta +1}{\tau }\phi^{(1)'}\delta_{ij}
+\phi^{(1) }_{,ij}
-\nabla^2\phi^{(1) }\delta_{ij}
\nn\\
&
+\frac{1}{2} D_{ij}\chi^{||(1)''}
+\frac{\beta +1}{\tau }D_{ij}\chi^{||(1)'}
+\frac{1}{6}\nabla^2D_{ij}\chi^{||(1)}
-\frac{1}{9}\delta_{ij}\nabla^2\nabla^2\chi^{||(1) }
\nn \\
&
+\frac{1}{2} \chi^{\perp(1)''}_{ij}
+\frac{\beta +1}{\tau } \chi^{\perp(1)'}_{ij}
\nn \\
&
+\frac{1}{2} \chi^{\top(1)''}_{ij}
+\frac{\beta +1}{\tau } \chi^{\top(1)'}_{ij}
- \frac{1}{2}\nabla^2\chi^{\top(1) }_{ij}
\nn\\
=&
 \frac{1}{M_{\rm Pl}}\frac{\sqrt{2(\beta +1) (\beta +2)}}{\tau}\varphi^{(1)'}\delta_{ij}
+\frac{1}{M_{\rm Pl}}
\frac{\sqrt{2(\beta +1) (\beta +2)} (2 \beta +1)}{\tau^2}\varphi^{(1)}\delta_{ij}.
\el
which
involves  all the scalar, vector and tensor modes.
The trace part of  (\ref{evoEq1stRD3}) is
\bl\label{evoEq1stRDtrace}
&
\phi^{(1)''}
+2\frac{\beta +1}{\tau }\phi^{(1)'}
-\frac{1}{3}\nabla^2\phi^{(1) }
-\frac{1}{18}\nabla^2\nabla^2\chi^{||(1) }
\nn\\
=&
 \frac{1}{M_{\rm Pl}}\frac{\sqrt{2(\beta +1) (\beta +2)}}{2\tau}\varphi^{(1)'}
+\frac{1}{M_{\rm Pl}}
\frac{\sqrt{2(\beta +1) (\beta +2)} (2 \beta +1)}{2\tau^2}\varphi^{(1)},
\el
which is an inhomogeneous equation
with  $\varphi^{(1)}$  as a source.
The traceless part of  (\ref{evoEq1stRD3}) is
\bl  \label{evoEq1stTraceless}
&
D_{ij}\phi^{(1) }
+\frac{1}{2} D_{ij}\chi^{||(1)''}
+\frac{\beta +1}{\tau }D_{ij}\chi^{||(1)'}
+\frac{1}{6}\nabla^2D_{ij}\chi^{||(1) }
\nn \\
&
+\frac{1}{2} \chi^{\perp(1)''}_{ij}
+\frac{\beta +1}{\tau } \chi^{\perp(1)'}_{ij}
\nn \\
&
+\frac{1}{2} \chi^{\top(1)''}_{ij}
+\frac{\beta +1}{\tau } \chi^{\top(1)'}_{ij}
- \frac{1}{2}\nabla^2\chi^{\top(1) }_{ij}
=0,
\el
which is a homogeneous equation,
and can be further decomposed into scalar, vector, tensor parts as
\be  \label{evoEq1stTracelessScalar0}
D_{ij}\phi^{(1)}
+\frac{1}{6}\nabla^2D_{ij}\chi^{||(1)}
+\frac{1}{2} D_{ij}\chi^{||(1)''}
+\frac{\beta +1}{\tau }D_{ij}\chi^{||(1)'}
=0,
\ee
\be  \label{evoEq1stTracelessVector}
\chi^{\perp(1)''}_{ij}
+\frac{2\beta +2}{\tau } \chi^{\perp(1)'}_{ij}
=0,
\ee
\be  \label{evoEq1stTracelessTensor}
 \chi^{\top(1)''}_{ij}
+\frac{2\beta +2}{\tau } \chi^{\top(1)'}_{ij}
-\nabla^2 \chi^{\top(1) }_{ij}
=0.
\ee
Taking off $D_{ij}$ from  (\ref{evoEq1stTracelessScalar0}),  we  have
\be \label{evoEq1stTracelessScalar}
\phi^{(1)}
+\frac{1}{6}\nabla^2\chi^{||(1) }
+\frac{1}{2} \chi^{||(1)''}
+\frac{\beta +1}{\tau }\chi^{||(1)'}
=0 .
\ee
Thus,  the 1st-order   vector modes evolve independently,
so do the 1st-order tensor modes,
and their solutions are simple to give.
The evolution equation \eqref{evoEq1stTracelessVector}
of the vector mode is not a hyperbolic equation,
so the vector mode is not a  wave, and does not propagate.
The vector equation \eqref{evoEq1stTracelessVector}  is not hyperbolic
and its solution is of the form
\be\label{vect1stsol}
\chi^{\perp(1)}_{ij}
 \propto \tau^{1-2(\beta+1)},  ~   \tau^0
\ee
which is  not an oscillator.
So the vector perturbation is not a wave.
Moreover, the constant mode $\propto  \tau^0 $
is a gauge mode and can be removed
by the gauge transformation \eqref{gaugePerpchi},
and the  mode $ \tau^{1-2(\beta+1)} \sim  a^{-3}$ for $\beta\sim -2$
is decaying away during inflation.
Thus,  we can set
\be\label{novectapp}
\chi^{\perp(1)}_{ij}=0 ,
\ee
 like in the dust and fluid models
\cite{WangZhang2017,ZhangQinWang2017,WangZhang2018,WangZhang2019}.

The  equation  (\ref{evoEq1stTracelessTensor}) of the  tensor perturbation
is a hyperbolic 2nd-order  partial differential equation
which describes the 1st-order gravitational wave propagating
at  the speed of light.
The solution of  (\ref{evoEq1stTracelessTensor}) is the RGW
and has been studied in Refs.~\cite{Grishchuk,Allen1988,zhangyang05,ZhangWang2016,ZhangWang2018b}.
It is written as
\be  \label{Fourierapp}
\chi^{\top(1)}_{ij}  ( {\bf x},\tau)= \frac{1}{(2\pi)^{3/2}}
\int d^3k   e^{i \,\bf{k}\cdot\bf{x}}
\sum_{s={+,\times}} {\mathop \epsilon
\limits^s}_{ij}(k) ~ {\mathop h\limits^s}_k(\tau),
\ee
where ${\mathop \epsilon  \limits^s}_{ij}(k) $
are the polarization tensors,
${\mathop h\limits^s}_k(\tau)$ with  $s= {+,\times}$
are two modes of RGW
and can be assumed to be statistically equivalent,
so that the superscript $s$ can be dropped,
\bl \label{GWmodeapp}
h_k(\tau ) = &
\frac{1}{a(\tau )}\sqrt{\frac{\pi }{2}} \sqrt{\frac{-\tau }{2}}
\left[b_1 H_{\beta +\frac{1}{2}}^{(1)}(-k \tau )
+b_2 H_{\beta +\frac{1}{2}}^{(2)}(-k \tau )\right]
,
\el
where  $H^{(1)}_{\beta +\frac{1}{2}}$, $H^{(2)}_{\beta +\frac{1}{2}}$
are the Hankel functions,
and  the constant coefficients $b_1$ and $b_2$
can be fixed by  the initial condition.
At high $k$,  $H_{\beta +\frac{1}{2}}^{(1)}(-k\tau ) \simeq
\sqrt{\frac{2}{\pi }} \frac{1}{\sqrt{- k\tau }}
 e^{ -i k\tau -\frac{i \pi (\beta+1) }{2}}$ and
\be\label{hiktapp}
h_k(\tau ) \sim \frac{1}{a(\tau )}
\big[ b_1 \frac{1}{ \sqrt{2k} } e^{ -i k\tau -\frac{i \pi  (\beta+1)  }{2}}
+  b_2 \frac{1}{ \sqrt{2k} } e^{ i k\tau +\frac{i \pi (\beta+1) }{2}} \big] .
\ee
For the Bunch-Davies vacuum state during inflation
and assuming the  quantum  normalization condition
that for  each $k$ mode and each polarization of tensor,
there is a zero point energy $\frac12 \hbar k$ in high frequency limit,
we obtain  $b_1=  \frac{2}{M_{\rm Pl}} e^{  \frac{i \pi (\beta+1) }{2}}$,
 $b_2=0$.
The primordial spectrum of tensor perturbation
is $\Delta^2_t \simeq 0.5 \frac{H^2}{M_{Pl}} k^{2\beta+4}$
at  low $k|\tau|\ll 1 $
\cite{zhangyang05,ZhangWang2018b,ZhangWang2016}.

We also need the 1st-order perturbed  scalar field.
By  expanding \eqref{fequvarphi},
we get the 1st-order perturbed field equation
\bl \label{e1stord}
  \varphi^{(1)'' }  +  2 \frac{a'}{a}  \varphi^{(1)'}
    -     \nabla^2  \varphi^{(1)}
           + a^2  V(\varphi^{(0)})_{,\varphi \varphi} \,  \varphi^{(1)}
  =     3  \varphi^{(0)'}   \phi^{(1)'} ,
\el
which is not closed with the metric perturbation
$\phi^{(1)'}$ on the rhs.
(The equation \eqref{e1stord} is equivalent to
the 1st-order energy conservation \eqref{equvarphi1st3} in  Appendix \ref{sec:perturbedGT}.)
For the power-law inflation, \eqref{e1stord}
becomes the following
\be\label{Enconsv1st}
\varphi^{(1)''}
+\frac{2(\beta +1)}{\tau}\varphi^{(1)'}
+\frac{ 2  (2 \beta +1)(\beta +2)}{\tau^{2}}\varphi^{(1)}
-\nabla^2\varphi^{(1)}
= 3M_{\rm Pl}\frac{\sqrt{2(\beta+1)(\beta+2)}}{\tau} \phi^{(1)'}
.
\ee
Combining \eqref{Enconsv1st}
and the constraints (\ref{Ein1st00}) (\ref{Ein1st0i}),
we get a third-order, closed equation
\bl \label{phi1d3}
\varphi ^{(1)'''}
& +\frac{3 \beta  +4  }{\tau}\varphi^{(1)''}
+\frac{2 \beta ^2+2\beta-2 }{\tau ^2}\varphi^{(1) '}
\nn \\
&
-\frac{4 \beta ^2+10 \beta+4}{\tau ^3} \varphi^{(1)}
 -\nabla^2 \varphi^{(1) '}
- \frac{ \beta+2}{\tau} \nabla^2 \varphi^{(1)}
=0 .
\el
It  has a solution
\be \label{phi1d3sol}
\varphi^{(1)}(\tau,\mathbf x) \propto  \tau ^{-\beta -2},
\ee
which  is a gauge mode and can be dropped
by the gauge transform (\ref{deltarho}).
We look for other two solutions  of (\ref{phi1d3}).
Introduce
\be\label{phiBarTrans}
  \hat{\varphi}^{(1)}
= (-\tau) ^{\beta +1}( \varphi^{(1) '}
+\frac{ \beta+2}{\tau}\varphi^{(1)} ) .
\ee
Then (\ref{phi1d3}) becomes
the following  second-order wave equation
\be
  \hat{\varphi}^{(1)}\,  ^{''}
-\frac{(\beta+2)(\beta+1)}{\tau^2}  \hat{\varphi}^{(1)}
-\nabla^2  \hat{\varphi}^{(1)} =0 ,
\ee
which has the solutions
\be\label{FourierEx}
 \hat{\varphi}^{(1)} ({\bf x},\tau)
\equiv  \frac{1}{(2\pi)^{3/2}} \int d^3 k
  e^{-i \,\bf{k}\cdot\bf{x}}  \hat{\varphi}^{(1)} _k (\tau),
\ee
with the Fourier modes
\be\label{varphiTild}
 \hat{\varphi}^{(1)} _k (\tau)
=
\sqrt{-\tau }\Big[
d_1 H^{(1)}_{\beta +\frac{3}{2}}(-k \tau )
+d_2  H^{(2)}_{\beta +\frac{3}{2}}(-k \tau )
\Big]
,
\ee
where  $d_1$, $d_2$ are the constant coefficients,
and will be determined by the initial conditions later.
From  (\ref{phiBarTrans}) \eqref{FourierEx},
we obtain the solution of 1st-order perturbed scalar field
as the following
\bl\label{varphi1solv2app}
\varphi^{(1) }_k (\tau)
= &
k^{-1}(-\tau) ^{-\beta -\frac{1}{2}}
\Big[ d_1 H_{\beta +\frac{1}{2}}^{(1)}(-k\tau )
      +d_2 H_{\beta +\frac{1}{2}}^{(2)}(-k\tau ) \Big]
\nn\\
&
+(\beta +2)k^{-1} (-\tau) ^{-\beta -2}
 \int^\tau d\tau_1 \sqrt{-\tau_1 }
 \Big[d_1H_{\beta +\frac{1}{2}}^{(1)}(-k\tau_1 )
     +d_2 H_{\beta +\frac{1}{2}}^{(2)}(-k\tau_1 )  \Big],
\el
which contains no explicit residual gauge mode.
The perturbed scalar field $\varphi^{(1)}$
is also waves propagating with the speed of light,  like the tensor modes.

Now we  derive the scalar metric perturbations $\phi^{(1)}$ and $\chi^{||(1)}$.
Denote
\be\label{phiChiZeta}
\zeta\equiv
-2\phi^{(1)}
-\frac{1}{3}\nabla^2 \chi^{||(1)} .
\ee
From the  constraints  \eqref{energyconstr1st}  \eqref{drmomentconstr1RD}
and the equations \eqref{Ein0th001} \eqref{Ein0thij1}  \eqref{EnergyCons0th},
we obtain the third-order, closed equation
\bl
\zeta ''' &
    + \big(a H - \frac{H'}{H} -\frac{H''}{H'} \big)\zeta''
    + \big(\frac{H''}{H}-\frac{a H H''}{H'}
    +\frac{\left(H''\right)^2}{\left(H'\right)^2} -\frac{H ''' }{H'} \big) \zeta'
\nn \\
&   - \nabla^2  \zeta'  + \frac{a^2  H' }{a'} \nabla^2 \zeta =0 \, ,
 \label{zteq2}
\el
which is valid  for a general $a(\tau)$.
For the power-law inflation,  \eqref{zteq2} reduces to
\bl \label{2zeta3rd}
 \zeta '''  + \frac{3 (\beta +2)}{\tau}  \zeta ''
+ \frac{ 2 (\beta +1) (\beta +3)}{\tau^2} \zeta '
-  \nabla^2  \zeta '
- \frac{(\beta +2)}{ \tau }\nabla^2 \zeta
=0.
\el
It  has a solution
\[
\zeta(\tau,\mathbf x)  \propto  \tau^{-\beta-2} \propto  \frac{ a'}{a^2},
\]
which is a gauge mode associated with the gauge mode \eqref{phi1d3sol}
 and can be dropped.
Other two solutions  can be derived in the same procedure as for \eqref{phi1d3}.
Actually the moment constraint \eqref{Ein1st0i} gives a simple relation
\be \label{zeta}
\zeta'  = - \frac{1}{M_{\rm Pl}}\frac{\sqrt{2(\beta +1)
 (\beta +2)}}{\tau} \varphi^{(1)} ,
\ee
Since $\varphi^{(1)}$ is known by \eqref{varphi1solv2app},
 integrating \eqref{zeta} gives
\be\label{zetasolfinal2}
 \zeta_{ k}  =
\frac{\sqrt{2(\beta+1)(\beta+2)}}{M_{\rm Pl}\,k}(-\tau) ^{-\beta -2}
 \int^{\tau }
\sqrt{-\tau_1 }\Big[d_1 H^{(1)}_{\beta +\frac{1}{2}}(-k \tau_1 )
+d_2  H^{(2)}_{\beta +\frac{1}{2}}(-k \tau_1 )   \Big] \, d\tau_1 .
\ee
where $d_1$ and $d_2$ are the same constants in \eqref{varphi1solv2app},
and  represent  two independent solutions.

The separate solutions of  $\phi^{(1)}$ and $\chi^{||(1)}$
can be given as follows.
By use of the relation \eqref{zeta}
we write   (\ref{Ein1st00}) in the $k$-space as
\be\label{phi1p}
\phi^{(1)'}_{ k}
=   \frac{k^2 }{6(\beta +1)} \tau\zeta_{ k}
-\frac{1}{6 M_{\rm Pl}}\frac{\sqrt{2(\beta +1)
 (\beta +2)}}{(\beta +1)}\varphi^{(1)'}_{ k}
-\frac{2 \beta +1}{6(\beta +1)}\zeta_{ k}^{\,'}  .
\ee
Since   $\zeta_{k}$  and $\varphi^{(1)}_{k}$ are known,
integrating \eqref{phi1p} gives
\bl\label{phi1kkfinalapp}
\phi^{(1)}_{ k}
=&
-\frac{\sqrt{2(\beta +1) (\beta +2)}}{(\beta +1)}
\frac{(-\tau)^{-\beta -\frac{1}{2}}}{6 M_{\rm Pl}k}
\Big[
d_1 H_{\beta +\frac{1}{2}}^{(1)}(-k\tau )
+d_2 H_{\beta +\frac{1}{2}}^{(2)}(-k\tau )
\Big]
\nn\\
&
-\frac{1}{M_{\rm Pl}}\frac{\sqrt{2(\beta +1) (\beta +2)}}{(\beta +1)}
\Big( \frac{\beta +1}{2 \,k} (-\tau) ^{-\beta -2}
 + \frac{k}{6 \beta} (- \tau)^{-\beta}  \Big)
\nn\\
&
\times\int^{\tau} d\tau_1 \sqrt{-\tau_1 }\Big[
d_1 H^{(1)}_{\beta +\frac{1}{2}}(-k \tau_1 )
+d_2 H^{(2)}_{\beta +\frac{1}{2}}(-k \tau_1 )   \Big]
\nn\\
&
+\frac{\sqrt{2(\beta +1) (\beta +2)}}{(\beta +1)\beta}
\frac{k}{6 M_{\rm Pl}}
\int^\tau d\tau_1  (-\tau_1)^{-\beta+ \frac12 }
\nn\\
&
\times
\Big[
d_1 H^{(1)}_{\beta +\frac{1}{2}}(-k \tau_1 )
+d_2 H^{(2)}_{\beta +\frac{1}{2}}(-k \tau_1 )   \Big]
 .
\el
Then by the relation \eqref{phiChiZeta} we get
\bl\label{chi1kkfinalapp}
&
\chi^{||(1)}_{ k}
=
-\frac{\sqrt{2(\beta +1)(\beta +2)}}{ M_{\rm Pl}(\beta +1)
k^3(-\tau) ^{\beta +\frac{1}{2}}}
\Big[
d_1 H_{\beta +\frac{1}{2}}^{(1)}(-k\tau )
+d_2 H_{\beta +\frac{1}{2}}^{(2)}(-k\tau )
      \Big]
\nn\\
&
-\frac{\sqrt{2(\beta +1) (\beta +2)}}{ M_{\rm Pl}(\beta +1)k}
 \frac{1}{\beta}(-\tau )^{-\beta}
\int^\tau d\tau_1 (-\tau_1)^{\frac{1}{2}}
\Big[d_1H_{\beta +\frac{1}{2}}^{(1)}(-k\tau_1 )
     +d_2 H_{\beta +\frac{1}{2}}^{(2)}(-k\tau_1 ) \Big]
\nn\\
&
+\frac{\sqrt{2(\beta +1) (\beta +2)}}{ M_{\rm Pl}(\beta +1)k}
  \frac{1}{\beta}  \int^{\tau}d\tau_1
 (-\tau_1)^{-\beta+ \frac12  }
\Big[
d_1 H_{\beta +\frac{1}{2}}^{(1)}(-k\tau_1 )
+d_2 H_{\beta +\frac{1}{2}}^{(2)}(-k\tau_1 )
\Big]
.
\el
A constant gauge mode has been dropped
from \eqref{phi1kkfinalapp} \eqref{chi1kkfinalapp}.
In absence of an anisotropic stress tensor,
the two scalars $\phi^{(1)}$ and $\chi^{||(1)}$ are not independent.
Moreover,
$\phi^{(1)'}$, $\chi^{||(1)}$ and $\varphi^{(1)}$ are all related,
and one  set of coefficients $(d_1, d_2)$
fixes all the scalar perturbations.

\section{The gauge invariant 1st-order scalar perturbations}
\label{sec:appendix_gaugeInvariant}

The above 1st-order scalar metric  perturbations and
perturbed scalar field are generally subject to
changes under the transformations
between synchronous coordinates
(see \eqref{gaugetrphi} $\sim $  \eqref{deltarho} in Appendix \ref{sec:gauge_transform}.)
Therefore, the gauge-invariant 1st-order scalar perturbations
are often used in cosmology, and mostly expressed
in the Poisson coordinates with off-diagonal metric perturbations
 \cite{Bardeen1980,KodamaSasaki1984,Mukhanov1992,Hwang1993,Sasaki1986,
Mukhanov1988,Mukhanov2005}.
 Here in the synchronous coordinates
we give a complete treatment of four  gauge-invariant scalars as the following
\be\label{Psigauginv}
\Psi = \phi^{(1)} +  \frac16 \nabla^2 \chi^{||(1)}
         +  \frac{ a H}{2}  \chi^{||(1)'}  \, ,
\ee
\be
Q_\varphi   = \varphi^{(1)}  + \frac{\varphi^{(0)'} }{H a}
             \big(\phi^{(1)} +\frac16\nabla^2 \chi^{||(1)} \big) ,
\label{Qvarphiapp}
\ee
\be \label{ginvvarphi}
 \varphi ^{(1)}_{gi} =  \varphi ^{(1)} - \frac12 \varphi^{(0)'}\chi^{||(1)'}
 = Q_\varphi - \frac{\varphi^{(0)'}}{a H } \Psi ,
\ee
\be \label{scalarcurvtpertOur}
{\cal R}   =  \frac{a H }{\varphi^{(0)'}} Q_\varphi
    \,  .
\ee
These scalars are invariant
under the 1st-order gauge transformation \eqref{xi0trans} \eqref{gi0}.
$\Psi$ and   $\varphi ^{(1)}_{gi}$  correspond to
respectively  (3.13) (6.8) in  Ref.~\cite{Mukhanov1992}.
$Q_\varphi$ is often used
in the presence of a scalar field  \cite{Hwang1993,TaruyaNambu1998}.
$\cal R$  amounts to a rescaling of $Q_\varphi$
and is called the comoving  curvature perturbation
\cite{GordonWandsBassettMaartens2000,BassettTsuijikawa2006,Lukash1980,Lyth1985}.
Their  equations for a general inflationary model
in synchronous coordinates are given by
\bl
\Psi''  + \big( a H - \frac{H''}{H'} \big) \Psi'
   + \big( a^2 H^2 + 2 a H' -  \frac{a  H  H''}{H'} \big) \Psi
   - \nabla^2 \Psi  =0 ,
    \label{Psieq}
\el
\bl
 Q_\varphi ''
&   +2  a H Q_\varphi ' - \nabla^2  Q_\varphi
 +  \Big[ \frac{5 a^2 H^2}{4}+ \frac32  a H'
 -\frac{2 \left(H'\right)^2}{H^2}
 - \frac{a H H''}{2 H'}
\nn \\
&   +\frac{2 H''}{H}
+\frac{\left(H''\right)^2}{4 \left(H'\right)^2}
- \frac12 \frac{H'''}{H'}
 \Big] Q_\varphi     = 0  \, ,
\label{eqQvarphi}
\el
\bl \label{3red8}
 \varphi^{(1)'' }_{gi} &  + 2 a H  \varphi^{(1)'}_{gi}
        - \nabla^2 \varphi^{(1)}_{gi}
    + \Big[ \frac{5 a^2 H^2}{4}
-\frac{5 a H'}{2}
-\frac{a H H''}{2 H'}
\nn \\
& +\frac{\left(H''\right)^2}{4 \left(H'\right)^2}
-\frac{H '''}{2 H'} \Big] \varphi^{(1)}_{gi}
=  \varphi^{(0)'} \Big[ 4 \Psi'
  + \big(5 a H  + \frac{H''}{H'} \big) \Psi \Big]    ,
\el
\bl \label{Requgen}
{\cal R}''
 + \big(  a H -2 \frac{H'}{H}  + \frac{H''}{H'} \big) {\cal R}'
  -\nabla^2 {\cal R} =0 \, .
\el

For the power-law inflation, these equations reduce to
\bl
\Psi_k \, ''   + \frac{2\beta+4}{\tau}   \Psi_k \, ' + k^2 \Psi_k =0 ,
    \label{Psieqinfl}
\el
\bl
Q\,  '' _{\varphi \, k}
   + \frac{2(\beta+1)}{\tau} Q\,  ' _{\varphi \, k}
   + k^2 Q_{\varphi \, k}     =0 \,   ,  \label{Qeqplawinf}
\el
\bl \label{3redpwl}
\varphi^{(1)'' }_{gi}   +  &  \frac{2(\beta+1)}{\tau}  \varphi^{(1)'}_{gi}
  +k^2 \varphi^{(1)}_{gi}
  + \frac{2 (\beta+2) (2 \beta +1)}{\tau^2}  \varphi^{(1)}_{gi}
  \nn \\
= & M_{\rm Pl}\sqrt{2(\beta +1) (\beta +2)} \tau^{-1}
        \big( 4\Psi \, ' + \frac{4 \beta +2}{\tau} \Psi \big) ,
\el
\be \label{Requ}
{\cal R}''_k
+\frac{2 (\beta +1)}{\tau} {\cal R}'_k + k^2 {\cal R}_k=0 .
\ee
The equation \eqref{3redpwl} of $\varphi^{(1)}_{gi\, k}$ is not closed
and contains  inhomogeneous terms.
Observe that Eq.\eqref{Psieqinfl} of $\Psi$
differs  from Eq.\eqref{evoEq1stTracelessTensor}
of the tensor  $\chi^{\top(1) }_{ij}$.
The solutions  are given by
\bl\label{Psiex}
\Psi_{k}
=&
\frac{\sqrt{2 (\beta +1) (\beta +2)}}{2 k^2 M_{\text{Pl}}}
(-\tau )^{-\beta -\frac{3}{2}}
\Big[
d_1 H^{(1)}_{\beta +\frac{3}{2}}(-k \tau )
+d_2  H^{(2)}_{\beta +\frac{3}{2}}(-k \tau )
\Big]
,
\el
\bl \label{Qsol}
Q_{\varphi\, k}
& =   k^{-1}(-\tau) ^{-\beta -\frac{1}{2}}
\Big[ d_1 H_{\beta +\frac{1}{2}}^{(1)}(-k\tau )
      +d_2 H_{\beta +\frac{1}{2}}^{(2)}(-k\tau ) \Big] \, ,
\el
\bl  \label{varphiex}
\varphi ^{(1)}_{gi\,k}
 = & k^{-1}(-\tau) ^{-\beta -\frac{1}{2}}
    \Big[ d_1 H_{\beta +\frac{1}{2}}^{(1)}(-k\tau )
         +d_2 H_{\beta +\frac{1}{2}}^{(2)}(-k\tau ) \Big]
\nn\\
&
-k^{-2}(\beta +2) (-\tau )^{-\beta -\frac{3}{2}}
\Big[ d_1 H^{(1)}_{\beta +\frac{3}{2}}(-k \tau )
+d_2  H^{(2)}_{\beta +\frac{3}{2}}(-k \tau ) \Big],
\el
\bl \label{calR}
{\cal R}_k    & = \sqrt{\frac{\beta +1}{2 (\beta+2)}} \frac{1 }{M_{\rm Pl}}
     Q_{\varphi\, k} .
\el
Note that these solutions can be used to solve the 1st-order perturbations through the relation (\ref{Psigauginv})--(\ref{scalarcurvtpertOur}), and the results are the same as those given in Appendix \ref{sec:appendix_1stSolution}.

These gauge-invariant scalar solutions are simple,
contain no integration terms,
and can be also obtained
from the scalar solutions \eqref{varphi1solv2app}
\eqref{phi1kkfinalapp} \eqref{chi1kkfinalapp}.
Among these,
$\Psi$ and $Q_{\varphi}$ are sufficient to describe
the 1st-order gauge  invariant scalar perturbations.

The coefficients $d_1, d_2$ in \eqref{Psiex} -- \eqref{calR}
are the same as those in \eqref{varphi1solv2app}
\eqref{phi1kkfinalapp} \eqref{chi1kkfinalapp},
and can be fixed as the following.
We consider the scalar field $Q_\varphi$ as a quantum field
\bl\label{oprQvarphi}
Q_{\varphi}  ({\bf x},\tau)
& \equiv   \int  \frac{d^3 k}{(2\pi)^{3/2}}
    [ a_k Q_{\varphi \, k } (\tau) e^{-i \,\bf{k}\cdot\bf{x}}
  + a^\dagger_k Q^* _{\varphi \, k } (\tau)e^{i \,\bf{k}\cdot\bf{x}}] .
\el
where the  mode $Q_{\varphi \, k }$ is the positive frequency part
of \eqref{Qsol} and  $d_2= 0$ is taken,
and $a_{\bf k}$ and $a_{\bf k}^{\dag}$ are the annihilation and creation operators
\be
 [ a_{\bf k},a_{\bf k'}^{\dag}  ] = \delta^3_{\mathbf k, \mathbf k'}  \,  .
\ee
At high $k$,
 $Q_\varphi$ is approximately described as  a harmonic oscillator
with the energy density
\be\label{Qor}
\rho_Q   =  \frac{1}{2 a^2} \big[  (Q' _{\varphi})^2
+ (\nabla Q_{\varphi})^2  \big],
\ee
(see the expression \eqref{Tuvphi}
with the potential $V$ being neglected at high $k$.)
Consider the Bunch-Davies vacuum state \cite{BunchDavies1978}
during inflation
\[
a_k|0\rangle=0  .
\]
The energy density in the vacuum state is found to be
\bl
\langle 0| \rho_Q  |0\rangle  = \int \frac{d k }{k} \rho_{Q\, k} ,
\el
where  the  spectral energy density is
\bl \label{sprener}
 \rho_{Q\, k} & = \frac{ k^3 }{4\pi^2  a^2} \frac{1}{L^3}
 \Big( | Q' _{\varphi \, k }|^2  + k^2  |Q_{\varphi \, k }|^2  \Big)
\nn \\
&  \simeq  \frac{ k^3 }{4\pi^2  a^2} \frac{1}{L^3} \Big|d_1  \frac{2 l_0}{k \sqrt{\pi} }
  e^{ -\frac{i \pi  \beta }{2} -i\pi} \Big|^2 \frac{1}{a^2} k
  ~~~\text{ at high $k$} .
\el
where $L^3$ is a normalization volume.
By the requirement of quantum normalization that
each $k$ mode in the vacuum
contributes a zero point energy $\frac{k}{2 a(\tau)}$ (with the unit $ \hbar =1$),
the  spectral energy density is given by
$ \rho_{Q\, k}=  \frac{ k^4 }{4\pi^2 L^3 a^4}$,
and the coefficient is determined by
\be \label{determd1app}
d_1 =   \frac{ \sqrt{\pi} k}{2 l_0}  e^{\frac{i \pi  \beta }{2}+i\pi} ,
\ee
(which also leads to
$Q_{\varphi\, k} \simeq \frac{1}{a(\tau)} \frac{1}{\sqrt{2k} } e^{-i k\tau}$
at high $k$).
Thus, all the scalar  modes  \eqref{Psiex} -- \eqref{calR} are fixed,
and we can give their  power spectra.
Consider the auto-correlation function of $Q_\varphi$ in the BD vacuum
\bl
 \langle 0|Q_\varphi(\tau,{\bf x}) Q_\varphi (\tau,{\bf x}) |0\rangle
 = \int^\infty_0 \Delta_Q^2 \frac{d k }{k},
\label{spectrQ}
\el
where  the vacuum power spectrum of $Q_\varphi$ is
\bl
\Delta_Q^2(k, \tau)
& =\frac{k^3}{2 \pi^2}\left|{Q}_{\varphi \,  k}(\tau)\right|^2 .
\el
We are more interested in the spectrum at the low $k$ range
\bl \label{Qspectr}
\Delta_Q^2(k, \tau)   = \Big| \frac{\pi +i \sin (\pi  \beta )
\Gamma\left(-\beta -\frac{1}{2}\right)
\Gamma\left(\beta +\frac{3}{2}\right)}
{2^{\beta + \frac{3}{2}} \pi^{\frac{1}{2}} \Gamma
\left(\beta +\frac{3}{2}\right)}\Big|^2
   l_0^{-2} k^{2\beta+4},
\el
which is independent of time (``conserved" on large scales).
Using  $l_0^{-2}\simeq H^2 /k_*^{2\beta+4}$ where $k_*= a H$
evaluated  at the horizon exit ($k_* |\tau|=1$),
\eqref{Qspectr} is written as
\bl
\Delta_Q^2(k, \tau) = A_Q (\frac{k}{k_*})^{2\beta+4}  ,
\el
with the  amplitude
\be
A_Q = \Big| \frac{\pi +i \sin (\pi  \beta )
\Gamma\left(-\beta -\frac{1}{2}\right)
\Gamma\left(\beta +\frac{3}{2}\right)}
{2^{\beta + \frac{3}{2}} \pi^{\frac{1}{2}} \Gamma \left(\beta +\frac{3}{2}\right)}\Big|^2
   H^{2}
\simeq \frac12 H^{2}
  ~~~\text{for $\beta \simeq -2$} .
\ee
Analogously,
the spectrum of $\Psi$ at low $k$ is also independent of time,
\bl
\Delta_\Psi^2(k, \tau) = \frac{k^3}{2 \pi^2} \left|\Psi_{k}(\tau)\right|^2
   =  A_{\Psi} (\frac{k}{k_*})^{2\beta+4}
\el
with the dimensionless amplitude
\bl
 A_{\Psi} & =  \frac{|(\beta +1) (\beta +2)| }{ 2^{2\beta + 7}  \pi }
      \Big| \frac{1}{\Gamma \left(\beta +\frac{5}{2}\right)}
   -\frac{i \sin (\pi  \beta ) \Gamma   \left(-\beta -\frac{3}{2}\right) }{\pi } \Big|^2
   \big(\frac{H}{M_{\text{Pl}}}\big)^2
\nn \\
& \simeq  \frac{|\beta+2|}{8\pi^2} \big(\frac{H}{M_{\text{Pl}}}\big)^2
 ~~~~ \text{for $\beta \simeq -2$} .
\el
The spectrum of $\varphi_{gi}$ is
\bl
\Delta_{\varphi_{gi} }^2(k, \tau) =
 \frac{k^3}{2 \pi^2} \left|\varphi_{gi\, k}(\tau)\right|^2
 \simeq \Delta_Q^2(k, \tau)
  ~~~~ \text{for $\beta \simeq -2$}  ,
\el
which is  approximately equal to  $\Delta_Q^2$,
since $\varphi_{gi\, k}$ is dominated by the first part of \eqref{varphiex}
for $\beta \simeq -2$.
And the spectrum of ${\cal R}$ at low $k$
(by the relation \eqref{calR}) is
\bl
\Delta_R^2(k, \tau)
& \simeq A_R  \big(\frac{k}{k_*}\big)^{2\beta+4}
\label{Rspectr},
\el
with the  dimensionless amplitude
\be
 A_R  \simeq   \frac{1}{4|\beta+2|}  \big(\frac{H}{M_{\text{Pl}}}\big)^2
  ~~~~ \text{for $\beta \simeq -2$} .
\ee
These spectra at low $k$ have a common power index $(2\beta+4)$,
but with different amplitudes.
$\Delta_\Psi^2$  and $\Delta_R^2$ are dimensionless,
whereas $\Delta_Q^2$  and $\Delta_{\varphi_{gi} }^2$
have the dimension of square mass.
Fig.\eqref{spectra} shows that
 $\Delta_Q^2$ and $\Delta_{\varphi_{gi}}^2$ are almost overlapping,
$\Delta_R^2$ has a shape similar to $\Delta_Q^2$ and $\Delta_{\varphi_{gi}}^2$,
and these three spectra are $\propto k^2$ at high $k$.
But  $\Delta_\Psi^2$ is flat $\propto k^0$ at high $k$.
For the inflation  $\beta=-2.02$,
the ratio  $\Delta_R^2/\Delta_\Psi^2 \sim 2.7 \times 10^3$.
We  enlarge $\Delta_\Psi^2$ in Fig.\eqref{spectrumPsi}
to show its slope $(2\beta+4)$ at low $k$.
\begin{figure}
\centering
\includegraphics[width=0.7 \textwidth]{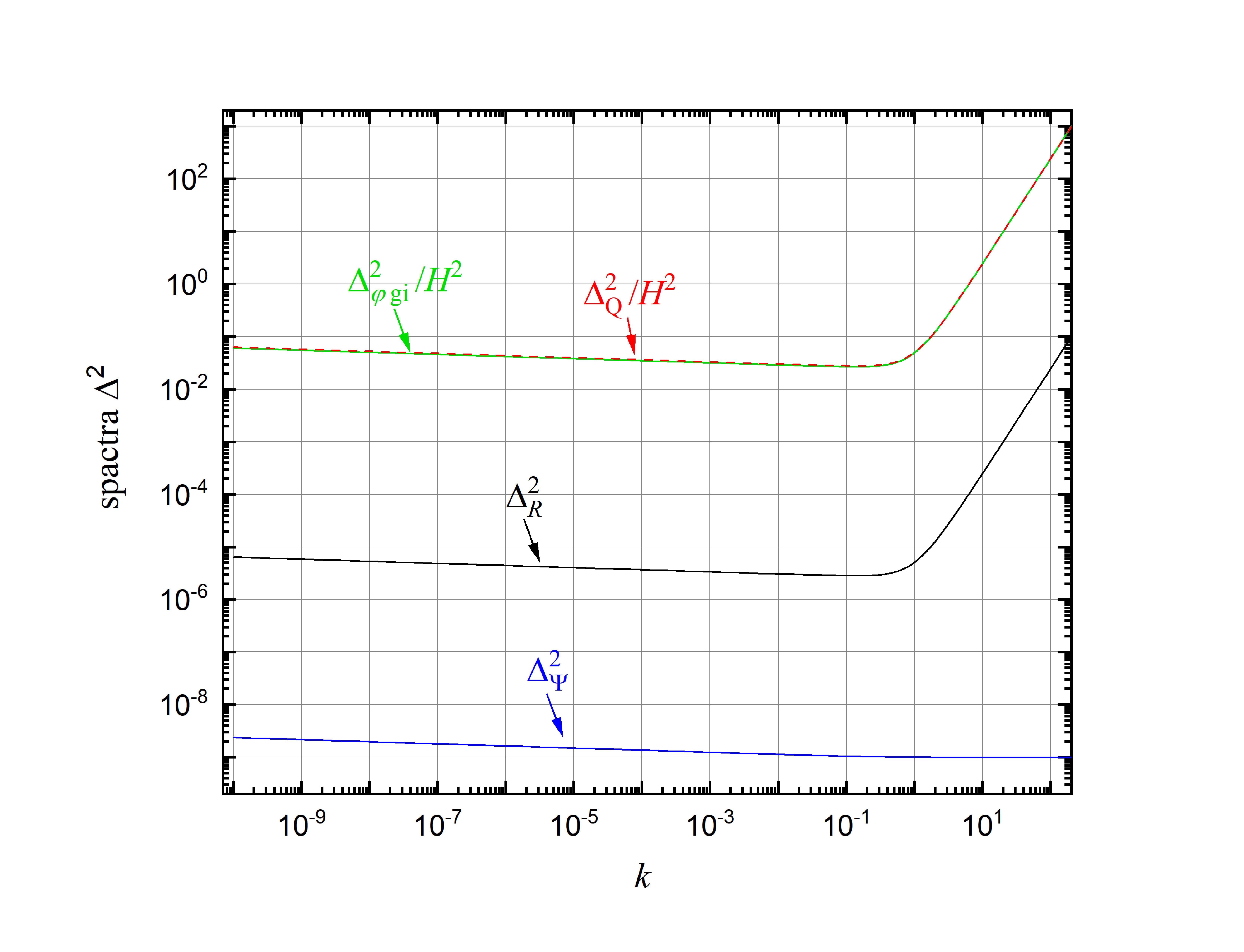}
\caption{ The four primordial spectra $\Delta_Q^2 /H^2$,
$\Delta_{\varphi_{gi} }^2 /H^2$,
$\Delta_R^2$, $\Delta_\Psi^2$ are shown,
where $\Delta_Q^2$ and $\Delta_{\varphi_{gi}}^2$ have been rescaled by $1/H^2$
for a clear illustration.
$\Delta_R^2$ is about  2700 higher than $\Delta_\Psi^2$.
The model $\beta=-2.02$ and $H/M_{Pl}=2\times 10^{-3}$,
and at a fixed time $|\tau|=1$ for illustration.
}
\label{spectra}
\end{figure}
\begin{figure}
\centering
\includegraphics[width=0.7 \textwidth]{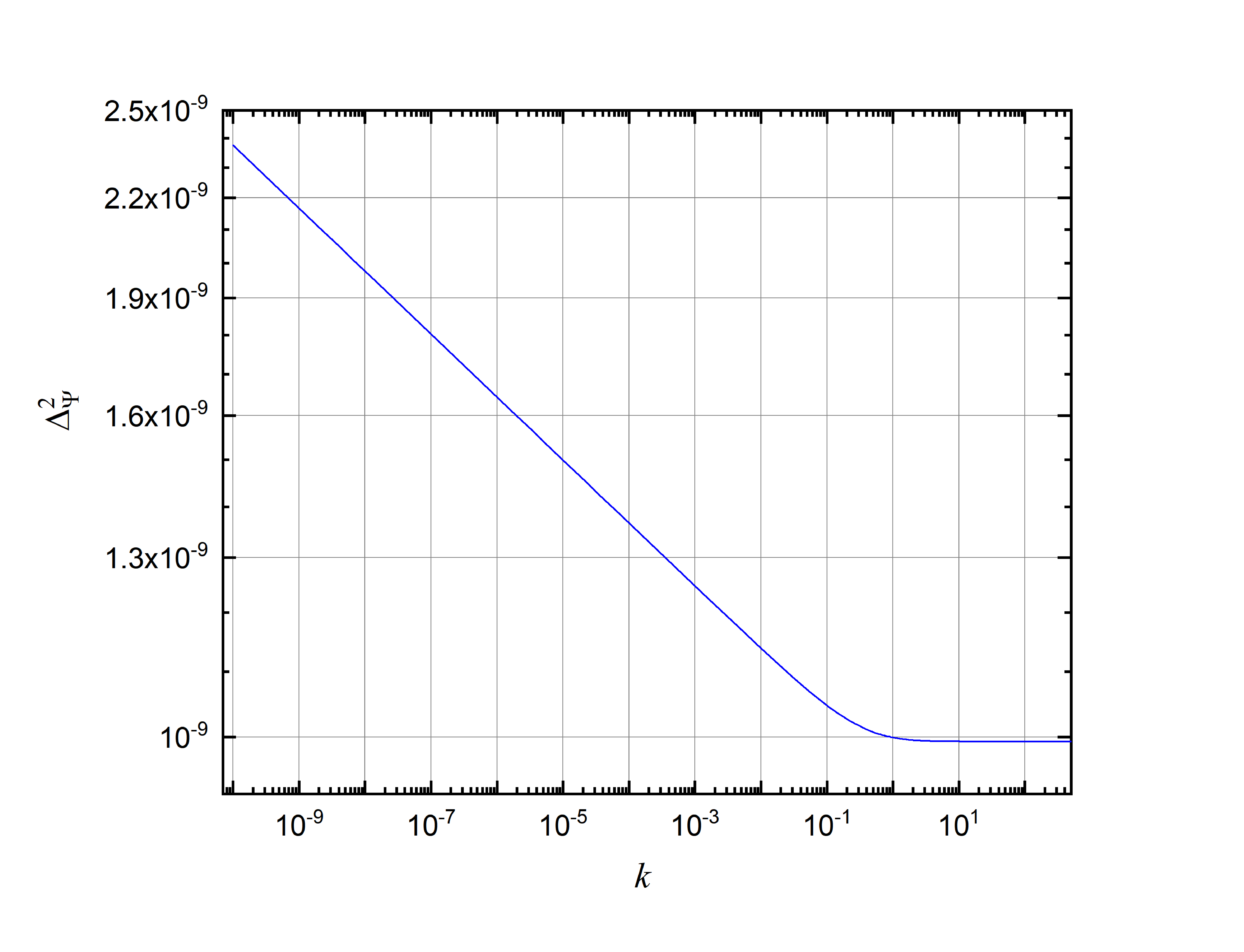}
\caption{ The spectrum  $\Delta_\Psi^2$ of Fig.\ref{spectra}
is enlarged to show the slope at low $k$.}
\label{spectrumPsi}
\end{figure}
It should be mentioned that these spectra are actually UV divergent at high $k$.
The UV divergences can be removed by appropriate regularization,
and we shall not discuss this issue here.
See Refs.~\cite{Parker2007,ZhangWang2016,ZhangWang2018b,ZhangYeWang2020}.
In literature,
the spectrum $\Delta_R^2$ is sometimes  used to
compare with the CMB observation \cite{BassettTsuijikawa2006}.
However,
$\cal R $ is made of the scalar field and the metric perturbation,
and $\Psi$ is the scalar metric perturbations
which directly generates CMB anisotropies,
and is more appropriate
to compare with the CMB observation \cite{AkramiArrojaAshdown2018}.
This  will allow for a higher $H$ of inflation
for a given set of observational date of CMB.
For an accurate description,
one should solve the Boltzmann equation  \cite{ZhaoZhang2006,CaiZhang2012}
of CMB photons that contains
the Sachs-Wolfe term $\gamma \, ' _{ij}e^i e^j$,
where $\gamma_{ij}$ is the metric perturbations
and $e^i$ is the direction vector of photon propagation.

The above results of 1st-order gauge invariant scalar perturbations
are valid for a general  power-law inflation model.
In the special case $\beta =-2$ of exact de Sitter inflation,
the  scalar metric perturbations  $\phi^{(1)}$ and  $\chi^{||(1)}$
in \eqref{phi1kkfinalapp} \eqref{chi1kkfinalapp}
and their gauge invariant
combination $\Psi$ in \eqref{Psiex} are zero,
and the 1st-order perturbed energy density and pressure,
\eqref{rho1} \eqref{p1}, are also zero consistently,
as also  mentioned in Ref.~\cite{Grishchuk1994}.
The perturbed scalar field $\varphi ^{(1)}$ of \eqref{varphi1solv2app},
$Q_{\varphi}$ of \eqref{Qsol}, and $\varphi ^{(1)}_{gi}$
of \eqref{varphiex} are nonzero,
whereas  the comoving  curvature perturbation
 $\cal R$ of \eqref{calR} becomes singular
and is improperly defined
for the  de Sitter inflation.
Nevertheless,  the tensor perturbation \eqref{GWmodeapp}
 is regular for the de Sitter inflation
\cite{zhangyang05,ZhangWang2018b,ZhangWang2016}.


\begin{thebibliography}{34}




\bibitem{Lifshitz1946} E. M. Lifshitz,
          Zh. Eksp. Theor. Fiz \textbf{16},   587 (1946).
	

\bibitem{LifshitzKhalatnikov1963} E. M. Lifshitz and I. M. Khalatnikov,
          Adv. Phys. {\bf 12} 185 (1963).

	

\bibitem{PressVishniac1980} W. H. Press and E. T. Vishniac,
 Astrophys. J. \textbf{239},   1 (1980).
	

\bibitem{Bardeen1980} J.  M.  Bardeen,     Phys.  Rev.  D \textbf{22},     1882 (1980).


\bibitem{KodamSasaki1984} H.  Kodama and  M.  Sasaki,
Prog. Theor. Phys. Suppl. \textbf{78},   1 (1984).
	
\bibitem{BrandenbergerKahnPress1983}
R. Brandenberger, R. Kahn and W. H. Press,
Phys. Rev. D \textbf{28},   1809 (1983).
	




\bibitem{Grishchuk1994} L. P. Grishchuk,
     Phys.Rev.D {\bf 50},  7154-7172 (1994) ;
      arXiv:gr-qc/9410025.


\bibitem{Peebles1980}
P. J. E. Peebles, {\it The Large-Scale Structure of the Universe}
         (Princeton University Press, Princeton, NJ, 1980).

\bibitem{BaskoPolnarev1984} M.M. Basko and A.G. Polnarev,
                   Mon. Not. R. Astron. Soc. \textbf{191},   207 (1980);
                   Sov. Astron. \textbf{24},   268 (1980).


\bibitem{Polnarev1985} A.G. Polnarev, Sov. Astron. {\bf29}, 607  (1985).


\bibitem{MaBertschinger1995} C. P. Ma and E. Bertschinger,
Astrophys. J.  {\bf 455}, 7 (1995).



\bibitem{Bertschinger}
      E. Bertschinger,   arXiv:astro-ph/9503125.


\bibitem{ZaldarriagaHarari1995} M. Zaldarriaga and D. D.Harari,
                   Phys. Rev. D {\bf 52}, 3276 (1995);

            M. Zaldarriaga and U. Seljak,
                  Phys. Rev. D {\bf55},  1830 (1997).


\bibitem{Kosowsky1996} A. Kosowsky, Ann. Phys. (N.Y.) {\bf 246},  49 (1996).



\bibitem{Kamionkowski1997} M. Kamionkowski and A.  Kosowsky, A. Stebbins,
                  Phys. Rev. D {\bf55}, 7368 (1997).


\bibitem{KeatingTimbie1998}  B. Keating, P.  Timbie, A. Polnarev, and J. Steinberger,
Astrophys. J. {\bf 495} 580 (1998).
	

\bibitem{ZhaoZhang2006} W. Zhao and Y. Zhang,
               Phys. Rev. D {\bf74},  083006 (2006);

 T.Y. Xia and Y. Zhang,  Phys. Rev. D {\bf78},  123005 (2008);

 T.Y. Xia and Y. Zhang,  Phys. Rev. D {\bf 79}, 083002  (2009).


\bibitem{CaiZhang2012}    Z. Cai and Y. Zhang,
    Classical Quantum Gravity {\bf 29},  105009 (2012).


\bibitem{Baskaran} D. Baskaran,  L. P. Grishchuk,  and A. G. Polnarev,
                      Phys. Rev. D {\bf 74}, 083008 (2006).

\bibitem{Polnarevmiller2008} A. G. Polnarev, N. J. Miller, and B. G. Keating,
           Mon. Not. R. Astron. Soc. {\bf386}, 1053 (2008).


\bibitem{Grishchuk}L. P. Grishchuk, Sov. Phys. JETP {\bf40}, 409 (1975);
 Classical Quantum Gravity {\bf14} 1445 (1997);
           Lect. Notes Phys. {\bf 562}, 167 (2001).




\bibitem{Starobinsky}A. A. Starobinsky, JETP Lett. {\bf 30}, 682 (1979).


\bibitem{Allen1988}B. Allen, Phys. Rev. D {\bf37}, 2078 (1988);
    B. Allen and S. Koranda, Phys. Rev. D {\bf 50}, 3713  (1994).


\bibitem{FordParker1977GW}  L. H. Ford and L. Parker, Phys. Rev. D {\bf16}, 1601 (1977).



\bibitem{Rubakov}   V. A. Rubakov, M. V. Sazhin,  and  A. V. Veryaskin,
           Phys. Lett. {\bf115B}, 189 (1982 ).

\bibitem {Fabbri}  R. Fabbri and M.D. Pollock, Phys. Lett. {\bf125B},  445 (1983).

\bibitem {AbbottWise1984}  L.F. Abbott and M.B. Wise,  Nucl. Phys. B {\bf 224}, 541 (1984).


\bibitem{Giovannini}     M. Giovannini,
            Phys. Rev. D {\bf 60}, 123511 (1999);
             PMC Phys. {\bf A 4}, 1 (2010).


\bibitem{Tashiro}       H. Tashiro,  T. Chiba and  M. Sasaki,
       Classical Quantum Gravity {\bf 21}, 1761 (2004).


\bibitem{Morais2014} J. Morais, M. Bouhmadi-Lopez, and A. B. Henriques,
           Phys. Rev. D \textbf{89}, 023513 (2014).

\bibitem{zhangyang05} Y. Zhang {\it et al}.,
                Classical Quantum Gravity {\bf22}, 1383 (2005);

       Y. Zhang {\it et al}.,   Classical Quantum Gravity  {\bf23}, 3783 (2006);

       H. X.  Miao   and  Y.  Zhang,
             Phys.  Rev. D {\bf75},    104009 (2007);

       S. Wang, Y. Zhang, T. Y. Xia, and H. X. Miao,
         Phys. Rev. D {\bf 77}, 104016 (2008);

       D.Q. Su and  Y. Zhang, Phys. Rev. D {\bf 85}, 104012  (2012) ;

       B. Wang and Y. Zhang,  Res. Astron. Astrophys. \textbf{19}, 024 (2019).



\bibitem{ZhangWang2016}  D.G. Wang, Y. Zhang and J.W. Chen,
          Phys. Rev. D \textbf{94},   044033 (2016);

\bibitem{ZhangWang2018b} Y. Zhang and B. Wang,
   J. Cosmol. Astropart. Phys. 11 (2018) 006.




\bibitem{BartoloBertaccaMatarrese2020} N. Bartolo, D. Bertacca, S. Matarrese, et al.,
          Phys. Rev. D \textbf{102}, 023527 (2020).


\bibitem{CaiLinWang2021} Y.-F. Cai, C. Lin, B. Wang, S.-F. Yan,
           Phys. Rev. Lett. \textbf{126}, 071303 (2021).


\bibitem{PyneCarroll1996} T. Pyne and S.M. Carroll,
         Phys. Rev. D    {\bf 53}, 2920   (1996).



\bibitem{AcquavivaBartoloMatarrese2003}
V. Acquaviva, N. Bartolo, S. Matarrese, et al., Nucl. Phys. \textbf{B667}, 119 (2003).

\bibitem{Bartolo2010} N. Bartolo, S.  Matarrese,   and   A.  Riotto,
Phys. Rev. D {\bf 69},    043503   (2004);
J. Cosmol. Astropart. Phys.  01  (2004) 003;
J. Cosmol. Astropart. Phys.  10  (2005) 010.




\bibitem{YangZhang2007} Y. Zhang,  Astron. Astrophys. {\bf 464}, 811  (2007);

             Y. Zhang and H.X. Miao,  Res. Astron. Astrophys. {\bf 9}, 501 (2009);

             Y. Zhang and Q. Chen,   Astron. Astrophys.  {\bf 581},  A53 (2015);

             Y. Zhang,  Q. Chen, and S.G. Wu, Res. Astron. Astrophys. {\bf 19}, 53 (2019);

             Y. Zhang and B.C. Li, Phys.Rev.D {\bf 104}, 123513 (2021);

             S.G. Wu and Y. Zhang,  Res. Astron. Astrophys. {\bf 22}, 045015 (2022);

             S.G. Wu and Y. Zhang,  Res. Astron. Astrophys. {\bf 22}, 125001 (2022).



\bibitem{AnandaClarksonWands2007} K. N. Ananda, C. Clarkson and D. Wands,
        Phys. Rev. D {\bf 75},  123518 (2007).



\bibitem{JeongKomatsu2006}  D. Jeong and E. Komatsu,
               Astrophys. J. {\bf 651}, 619  (2006);

       M. Shoji and E. Komatsu,
            Astrophys. J. {\bf 700}:  705 (2009).

\bibitem{Baumann2007} D. Baumann, P. Steinhardt, K. Takahashi, and K. Ichiki,
                Phys. Rev. D \textbf{76},   084019 (2007).
	

\bibitem{Matarrese2007} S. Matarrese and M. Pietroni,
       J. Cosmol. Astropart. Phys.  06  (2007) 026.


\bibitem{Pietroni2008} M. Pietroni,
        J. Cosmol. Astropart. Phys.  10  (2008) 036.



\bibitem{Matsubara2008}  T. Matsubara,
          Phys. Rev. D {\bf 78}, 083519 (2008).


\bibitem{Hwang2017GW}J.-c. Hwang, D. Jeong and H. Noh,
Astrophys. J. \textbf{842}, 46 (2017).

\bibitem{PeresRosen1959}  A. Peres and N. Rosen,
Physical Review \textbf{115}, 1085 (1959).

\bibitem{Tomita1967}   K. Tomita,
     Prog. Theor. Phys. \textbf{37},   831 (1967);	
  \textbf{45},   1747 (1971);	
 \textbf{47},   416 (1972).
	

\bibitem{MatarresePantanoSa'ez1994} S. Matarrese, O. Pantano and D. Saez,
             Phys. Rev. Lett. {\bf 72}, 320 (1994);
              Mon. Not. R. Astron. Soc. {\bf 271}, 513      (1994);

              S.  Matarrese and D. Terranova,
              Mon. Not.  R. Astron.  Soc. {\bf 283}, 400 (1996).

\bibitem{Russ1996} H. Russ, M. Morita, M. Kasai, and G. Borner,
              Phys. Rev. D {\bf 53}  6881 (1996).

\bibitem{Salopek}  D. S. Salopek, J. M. Stewart, and K. M. Croudace,
             Mon. Not. R. Astron. Soc. {\bf 271}, 1005 (1994).

\bibitem{Bruni97} M. Bruni, S. Matarrese, S. Mollerach and S. Sonego,
            Classical Quantum Gravity \textbf{14}, 2585 (1997).

\bibitem{Matarrese98} S.  Matarrese, S.  Mollerach,  and M.  Bruni,
                  Phys.  Rev.  D {\bf 58}, 043504 (1998).

\bibitem{MollerachHarariMatarrese2004} S. Mollerach, D. Harari, and S. Matarrese,
      Phys. Rev. D {\bf 69}, 063002 (2004).

\bibitem{Lu2008} T.  H.-C. Lu, K.  Ananda, C.  Clarkson,
                          Phys. Rev. D {\bf 77} 043523, (2008);

    T.  H.-C. Lu, K.  Ananda, C.  Clarkson, and R.  Maartens,
         J. Cosmol. Astropart. Phys.  02  (2009) 023.


\bibitem{Baumann2012} D. Baumann, A. Nicolis, L. Senatore, and M.  Zaldarriaga,
        J. Cosmol. Astropart. Phys.  07 (2012) 051.

{
\bibitem{VillaRampf2016} E. Villa and C. Rampf, J. Cosmol. Astropart. Phys. 01 (2016) 030; J. Cosmol. Astropart. Phys. 05 (2018) E01.
}

\bibitem{Brilenkov&Eingorm2017} R.  Brilenkov and M. Eingorm,
        Astrophys. J. {\bf 845}, 153 (2017).

\bibitem{NohHwang2004} H. Noh and J.-c. Hwang, Phys. Rev. D {\bf 69}, 104011   (2004);
             Classical Quantum Gravity \textbf{22}, 3181 (2005);

    J.-c. Hwang and  H. Noh, Phys. Rev. D \textbf{72} 044011   (2005); Phys. Rev. D {\bf73}, 044021 (2006); Phys. Rev. D {\bf 76},  103527 (2007);

    J.-c. Hwang, D. Jeong, H. Noh,
             Mon. Not. R. Astron. Soc. \textbf{459},  1124 (2016);

 J.-O. Gong, J.-c. Hwang, H. Noh, et al.,
 J. Cosmol. Astropart. Phys.,  10 (2017) 027.

\bibitem{Noh2014} H. Noh, J. Cosmol. Astropart. Phys.  07 (2014) 037.

\bibitem{Sikora2023} S. Sikora, Classical Quantum Gravity {\bf 40}, 025002 (2023).

\bibitem{HwangNohPark2016} J. C. Hwang, H. Noh and C.G. Park,
   Mon. Not. R. Astron. Soc. \textbf{461}, 3239 (2016).


\bibitem{Vernizzi2005} F. Vernizzi, Phys. Rev. D \textbf{71}, 061301 (2005).


\bibitem{UgglaWainwright2019} C. Uggla and J. Wainwright,
Phys. Rev. D \textbf{100}, 023544 (2019).


\bibitem{Nakamura2003}   K. Nakamura,  Prog. Theor. Phys. \textbf{110},   723  (2003);
	 \textbf{113},   481  (2005);
           Phys. Rev. D {\bf 74},  101301  (2006);
     {\bf 80}, 124021 (2009).


\bibitem{Domenech&Sasaki2017} G. Dom\`{e}nech and M. Sasaki,
      Phys. Rev. D {\bf97}, 023521 (2018).


\bibitem{MalikWands2004} K. A.  Malik  and  D.   Wands,
        Classical Quantum Gravity \textbf{21}, L65 (2004).







\bibitem{WangZhang2017} B. Wang and Y. Zhang,
        Phys. Rev. D \textbf{96},   103522 (2017).
	
\bibitem{ZhangQinWang2017} Y. Zhang, F. Qin and B. Wang,
         Phys. Rev. D \textbf{96},   103523 (2017).

\bibitem{WangZhang2018} B. Wang and Y. Zhang,
        Phys. Rev. D \textbf{98}, 123019 (2018).

\bibitem{WangZhang2019} B. Wang and Y. Zhang,
        Phys. Rev. D \textbf{99}, 123008 (2019).


\bibitem{ChoGongOh2020} I. Cho, J.-O. Gong and S. H. Oh,
Phys. Rev. D \textbf{102}, 043531 (2020).


\bibitem{Lucchin1985}
     F. Lucchin,  S. Matarrese,  Phys. Rev. D {\bf 32}, 1316 (1985).


\bibitem{Guth1981} A. H. Guth, Phys. Rev. D \textbf{23}, 347 (1981).

\bibitem{Starobinsky1980} A. A. Starobinsky, Phys. Lett. B \textbf{91}, 99 (1980).

\bibitem{Linde1982} A. D. Linde, Phys. Lett. B \textbf{108}, 389 (1982).


\bibitem{JHwang1993}    J. Hwang, Phys. Rev. D {\bf48}, 3544 (1993).

\bibitem{LythRiotto1999} D. H. Lyth and A. Riotto,
        Physics Reports \textbf{314}, 1 (1999).

\bibitem{Riotto2002} A. Riotto,  arXiv: hep-ph/0210162 (2002).


\bibitem{nongaussiancite}

P. Coles and J. D. Barrow, Mon. Not. R. Astron. Soc. \textbf{228}, 407 (1987);

N. Bartolo, E. Komatsu, S. Matarrese, et al., Physics Reports \textbf{402}, 103 (2004);

A. P. S. Yadav and B. D. Wandelt, Advances in Astronomy \textbf{2010}, 565248 (2010);


\bibitem{primordialBH}  Ya.  B.  Zeldovich,
        Adv.  Astron. Apstrophys. {\bf 3},  241 (1965);

     S. W. Hawking,
                    Mon. Not. R. Astron. Soc. {\bf152}, 75 (1971);

	        Y.-F. Cai, X. Tong, D.-G. Wang, S.-F. Yan,
        Phys. Rev. Lett. \textbf{121}, 081306 (2018).

        G. Franciolini, A. Kehagias, S. Matarrese, et al.,
        J. Cosmol. Astropart. Phys.  03 (2018) 016.


\bibitem{Rigopoulos2004} G. Rigopoulos,
Class. Quantum Grav. \textbf{21}, 1737 (2004);

G. I. Rigopoulos and E.P.S. Shellard,
J. Cosmol. Astropart. Phys.  10 (2005) 006;

  G. I. Rigopoulos, E. P. S. Shellard and B. J. W. van Tent,
  Phys. Rev. D \textbf{73}, 083521 (2006).


\bibitem{ChoGongOh2022} I. Cho, J.-O. Gong and S. H. Oh,
Phys. Rev. D \textbf{106}, 084027 (2022).


\bibitem{Mukhanov1992}  V. F. Mukhanov,  H.A. Feldman,  and R. H. Brandenberger,
       Phys. Rep. \textbf{215}, 203-333, (1992).

\bibitem{Weinberg2008} S. Weinberg,
{\it Cosmology} (Oxford University Press, 2008).

\bibitem{Planck2018} N. Aghanim, Y. Akrami, M. Ashdown, et al.,
Astronomy \& Astrophysics \textbf{641}, A6 (2020).



\bibitem{Sasaki1986}  M. Sasaki, Prog. Theor. Phys. {\bf 76}, 1036 (1986).

\bibitem{Mukhanov1988} V. F. Mukhanov, Sov. Phys. JETP {\bf 67}, 1297 (1988).



\bibitem{GordonWandsBassettMaartens2000}
    C.  Gordon,  D.  Wands,  B. A.  Bassett,  and  R.  Maartens,
            Phys. Rev. D {\bf63}, 023506 (2000).

\bibitem{BassettTsuijikawa2006} B. A.  Bassett, S. Tsujikawa, D.  Wands, Rev. Mod. Phys. {\bf 78}, 537 (2006).

\bibitem{Baumann2009} D. Baumann, TASI lectures on inflation, arXiv:0907.5424.


\bibitem{BunchDavies1978}T. S. Bunch and P. C. Davies,
Proc. Roy. Soc. Lond. A. \textbf{360}, 117 (1978).


\bibitem{KodamaSasaki1984}  H. Kodama and M. Sasaki,
Prog. Theor. Phys. Suppl. {\bf 78}, 1 (1984).

\bibitem{Mukhanov2005} V. F. Mukhanov,  Physical   Foundations    of  Cosmology
          (Cambridge University Press, Cambridge, 2005).

\bibitem{Hwang1993} J. Hwang,
Phys. Rev. D {\bf48}, 3544 (1993).


\bibitem{Gleiser1996} R. J. Gleiser, C. O. Nicasio, R. H. Price and J. Pullin,
           Classical Quantum Gravity \textbf{13},   L117 (1996).


\bibitem{HwangNoh2012}  J.-c. Hwang,   H. Noh, and J.-O. Gong,
              Astrophys. J. \textbf{752},   50 (2012).


\bibitem{TaruyaNambu1998}  A. Taruya and Y. Nambu,
          Phys. Lett. B {\bf 428}, 37 (1998).


\bibitem{Lukash1980} V. Lukash,
Zh. Eksp. Teor. Fiz \textbf{79}, 1601 (1980)
(Sov. Phys. JETP {\bf 52}, 807 (1980)).

\bibitem{Lyth1985} D. H. Lyth, Phys. Rev. D \textbf{31}, 1792 (1985).



\bibitem{ZhangYeWang2020}Y. Zhang, X. Ye and B. Wang,
           Science China. PMA. {\bf63},  250411  (2020);

  Y. Zhang,  B. Wang, and X. Ye,
       Chinese Physics C, {\bf 44} (9) 095104 (2020);

  X. Ye, Y. Zhang, and B. Wang,  J. Cosmol. Astropart. Phys. 09, 020  (2022);

  Y. Zhang and X. Ye,  Phys. Rev. D {\bf 106}, 065004 (2022).

\bibitem{Parker2007} L. Parker,
     arXiv:hep-th/0702216.



\bibitem{AkramiArrojaAshdown2018}
Y. Akrami, F. Arroja, M. Ashdown, et al.,
 Astronomy \& Astrophysics \textbf{641}, A10 (2020).



\end{thebibliography}
\end{document}